\hoffset=0.1truecm
\voffset=0.1truecm
\vsize=23.0truecm
\hsize=16.25truecm


\parskip=0.2truecm
\def\pp{\parshape 2 0.0truecm 16.25truecm 2truecm 14.25truecm}
\def\mpl{{ M_{\rm Pl} }}
\def\sigbar{ {\langle \sigma \rangle}}
\def\sfr{ {SFR} } 
\def\gp{ \Gamma_P } 
\def\rhovac{ {\rho_{\rm vac} } } 
\def\newpage{\vfill\eject}
\def\bline{{$\ast$\dotfill$\ast$\dotfill$\ast$\dotfill$\ast$\dotfill$\ast$
\dotfill$\ast$\dotfill$\ast$\dotfill$\ast$\dotfill$\ast$\dotfill$\ast$}}
%
%
%
\centerline{\bf A DYING UNIVERSE:} 
\centerline{\bf The Long Term Fate and Evolution of Astrophysical Objects} 
\bigskip
\centerline{\bf Fred C. Adams and Gregory Laughlin}
\bigskip
\medskip
\centerline{\sl Physics Department, University of Michigan}
\centerline{\sl Ann Arbor, MI 48109, USA}
\bigskip
\centerline{fca@umich.edu \quad and \quad gpl@boris.physics.lsa.umich.edu}
\vskip 0.25truein
\centerline{submitted to {\it Reviews of Modern Physics}: 21 June 1996} 
\centerline{revised: 27 Sept. 1996; accepted 15 Oct. 1996} 
\medskip
\bigskip 
\centerline{\bf Abstract} 
\medskip 

This paper outlines astrophysical issues related to the long term fate
of the universe.  We consider the evolution of planets, stars, stellar
populations, galaxies, and the universe itself over time scales which
greatly exceed the current age of the universe. Our discussion starts
with new stellar evolution calculations which follow the future
evolution of the low mass (M type) stars that dominate the stellar
mass function.  We derive scaling relations which describe how the
range of stellar masses and lifetimes depend on forthcoming increases
in metallicity.  We then proceed to determine the ultimate mass
distribution of stellar remnants, i.e., the neutron stars, white
dwarfs, and brown dwarfs remaining at the end of stellar evolution; this
aggregate of remnants defines the ``final stellar mass function''.  
At times exceeding $\sim$1--10 trillion years, the supply of
interstellar gas will be exhausted, yet star formation will continue
at a highly attenuated level via collisions between brown dwarfs.
This process tails off as the galaxy gradually depletes its stars by
ejecting the majority, and driving a minority toward eventual
accretion onto massive black holes. As the galaxy disperses, stellar
remnants provide a mechanism for converting the halo dark matter into
radiative energy. Posited weakly interacting massive particles are
accreted by white dwarfs, where they subsequently annihilate with each
other.  Thermalization of the decay products keeps the old white
dwarfs much warmer than they would otherwise be.  After accounting for
the destruction of the galaxy, we consider the fate of the expelled
degenerate objects (planets, white dwarfs, and neutron stars) within
the explicit assumption that proton decay is a viable process.  The
evolution and eventual sublimation of these objects is dictated by the
decay of their constituent nucleons, and this evolutionary scenario is
developed in some detail.  After white dwarfs and neutron stars have
disappeared, galactic black holes slowly lose their mass as they emit
Hawking radiation.  This review finishes with an evaluation of
cosmological issues that arise in connection with the long-term
evolution of the universe. We devote special attention to the relation
between future density fluctuations and the prospects for continued
large-scale expansion. We compute the evolution of the background
radiation fields of the universe. After several trillion years, 
the current cosmic microwave background will have redshifted into 
insignificance; the dominant contribution to the radiation background
will arise from other sources, including stars, dark matter 
annihilation, proton decay, and black holes. Finally, we consider 
the dramatic possible effects of a non-zero vacuum energy density. 

\newpage 
\bigskip
\noindent 
{\bf TABLE OF CONTENTS} 
\bigskip

\bigskip
\noindent
{\bf I. INTRODUCTION} \dotfill 4

\bigskip
\bigskip
\noindent
{\bf II. THE END OF CONVENTIONAL STELLAR EVOLUTION} \dotfill 5 

{\bf A. Lifetimes of Main Sequence Stars} 

{\bf B. Forthcoming Metallicity Effects}

{\qquad 1. Stellar Lifetimes vs Metallicity} 

{\qquad 2. Stellar Masses vs Metallicity} 

{\bf C. The Fate of the Earth and the Sun} 

{\bf D. Continued Star Formation in the Galaxy} 

{\bf E. The Final Mass Function} 

\bigskip
\bigskip
\noindent
{\bf III. DEATH OF THE GALAXY} \dotfill 12

{\bf A. Dynamical Relaxation of the Galaxy} 

{\bf B. Gravitational Radiation and the Decay of Orbits} 

{\bf C. Star Formation through Brown Dwarf Collisions} 

{\qquad 1. Collision Time Scales} 

{\qquad 2. Collision Cross Sections} 

{\qquad 3. Numerical Simulations and Other Results} 

{\bf D. The Black Hole Accretion Time} 

{\bf E. Annihilation and Capture of Halo Dark Matter}

{\bf F. The Fate of Planets during Galactic Death}  

\bigskip
\bigskip
\noindent
{\bf IV. LONG TERM FATE OF DEGENERATE STELLAR OBJECTS} \dotfill 21 

{\bf A. Proton Decay} 

{\bf B. White Dwarfs Powered by Proton Decay} 

{\bf C. Chemical Evolution in White Dwarfs}

{\bf D. Final Phases of White Dwarf Evolution} 

{\bf E. Neutron Stars Powered by Proton Decay} 

{\bf F. Higher Order Proton Decay} 

{\bf G. Hawking Radiation and the Decay of Black Holes} 

{\bf H. Proton Decay in Planets} 

\newpage  
\bigskip
\noindent
{\bf V. LONG TERM EVOLUTION OF THE UNIVERSE} \dotfill 32 

{\bf A. Future Expansion of a Closed Universe} 

{\bf B. Density Fluctuations and the Expansion of a Flat or Open Universe} 

{\bf C. Inflation and the Future of the Universe} 

{\bf D. Background Radiation Fields} 

{\bf E. Possible Effects of Vacuum Energy Density} 

{\qquad 1. Future Inflationary Epochs}

{\qquad 2. Tunneling Processes}

{\bf F. Speculations about Energy and Entropy Production in the Far Future}

{\qquad 1. Continued Formation and Decay of Black Holes} 

{\qquad 2. Particle Annihilation in an Open Universe}

{\qquad 3. Formation and Decay of Positronium}

\bigskip
\bigskip
\noindent
{\bf VI. SUMMARY AND DISCUSSION} \dotfill 44 

{\bf A. Summary of Results} 

{\bf B. Eras of the Future Universe}

{\bf C. Experimental and Theoretical Implications} 

{\bf D. Entropy and Heat Death} 

\bigskip
\bigskip
\noindent
{\bf Acknowledgments} \dotfill 48 

\bigskip
\bigskip
\noindent
{\bf References} \dotfill 49

\newpage 
\bigskip 
\noindent{\bf I. INTRODUCTION} 
\medskip 

The long term future of the universe and its contents is a topic of
profound scientific and philosophical importance.  With our current
understanding of physics and astrophysics, many of the questions
regarding the ultimate fate of the universe can now be quantitatively
addressed.  Our goal is to summarize and continue the development of a
quantitative theory of the future.

Investigations of the early universe at both accessible and
inaccessible energies have become commonplace, and a great deal of
progress within this discipline has been made (see, e.g., Weinberg, 
1972, 1977; Kolb \& Turner, 1990; Linde, 1990; Peebles, 1993; 
Zuckerman \& Malkan, 1996).  On the other hand, relatively
little work has focused on the future of the universe. The details of
the fiery denouement in store for a closed universe have been outlined
by Rees (1969), whereas an overview of the seemingly more likely
scenario in which the universe is either open or flat, and hence
expands forever, was set forth in the seminal paper {\it Time Without
End} (Dyson, 1979). The development of an open universe was also
considered in detail by Islam (1977, 1979).  The spirit of Rees,
Islam, and Dyson's work inspired several follow-up studies (see also
Rees, 1981).  The forthcoming evolution of very low mass stars has been
discussed in general terms by Salpeter (1982).  The effects of matter
annihilation in the late universe were studied (Page \& McKee, 1981ab),
and some aspects of proton decay have been explored (Dicus et
al., 1982; Turner, 1983).  Finally, the possibility of self-reproducing
inflationary domains has been proposed (Linde, 1988).  In general,
however, the future of the universe has not been extensively probed
with rigorous calculations.

Because the future of the universe holds a great deal of intrinsic
interest, a number of recent popular books have addressed the subject
(e.g., Davies, 1994; Dyson, 1988; Barrow \& Tipler, 1986; Poundstone,
1985).  Authors have also grappled with the intriguing prospects for
continued life, both human and otherwise, in far future (e.g., Dyson, 
1979; Frautschi, 1982; Barrow \& Tipler, 1986; Linde, 1988, 1989; Tipler, 
1992; Gott, 1993; Ellis \& Coule, 1994).  Our aim, however, is to
proceed in as quantitative a manner as possible.  We apply known
physical principles to investigate the future of the universe on
planetary, stellar, galactic, and cosmic scales.  The issue of life,
however alluring, is not considered here.

In standard Big Bang Cosmology, evolutionary epochs are usually 
expressed in terms of the redshift. When considering the far future,
however, time itself is often the more relevant evolutionary measure.  
The immense dynamic range of time scales $\tau$ involved in the subject 
suggests a convenient logarithmic unit of time $\eta$, defined by 
$$\eta \equiv \log_{10} \Bigl [ {\tau \over (1 {\rm yr})} 
\Bigr] \, . \eqno(1.1)$$ 
We refer to a particular integer value of $\eta$ as a ``cosmological 
decade''.  For example, the current age of the 
universe corresponds to $\eta \approx 10$. 

The article of faith inherent in our discussion is that the laws of
physics are constant in time, at least over the range of time scales
$10 < \eta < 100$ under consideration. There is no general guarantee 
that this assumption holds. Nevertheless, modern cosmology suggests 
that physical laws have held constant from the Planck time to the 
present, i.e., over cosmological decades spanning the range 
$-50 \le \eta \le 10$, and there is little reason to expect that 
they will not continue to do so.  We also implicitly assume that all
of the relevant physics is known (with full awareness of the fact that
our version of the future will be subject to revision as physical
understanding improves).

This paper is organized in roughly chronological order, moving from
events in the relatively near future to events in the far future.  
In section \S II, we discuss physical processes that affect
conventional stellar evolution; these processes will take place in the
time range $10 < \eta < 15$.  In \S III, we discuss events which lead to
the disruption and death of the galaxy; these processes unfold over a
time range $15 < \eta < 25$.  Marching further into time, in \S IV, 
we discuss the fate of stellar objects in the face of very long term
processes, including proton decay ($30 < \eta < 40$), and Hawking
radiation ($60 < \eta < 100$).  In \S V, we broaden our scope and
focus on the long term evolution of the universe as a whole.  We
conclude, in \S VI, with a general overview of our results. 
Since physical eschatology remains embryonic, we emphasize 
the major unresolved issues and point out possible avenues 
for further research.

\bigskip 
\bigskip 
\noindent{\bf II. THE END OF CONVENTIONAL STELLAR EVOLUTION} 
\nobreak 
\medskip 

At the present epoch, stars are the cornerstone of astrophysics. 
Stars mediate the appearance and evolution of galaxies, stars are
responsible for evolving the chemical composition of matter, and stars
provide us with much of the information we have regarding the current
state of the universe.

For the next several thousand Hubble times, conventionally evolving
stars will continue to play the central role. We thus consider the
forthcoming aspects of our current epoch, which we term the 
{\sl Stelliferous Era}. In particular, the fact that the majority of
stars have barely begun to evolve motivates an extension of standard
stellar evolution calculations of very low mass stars to time scales
much longer than the current age of the universe. We also discuss
continued star formation within the galaxy, and the final mass
distribution of stellar remnants.

\bigskip 
\noindent{\bf A. Lifetimes of Main Sequence Stars} 
\medskip 

Low mass stars are by far the most commonplace (e.g., Henry, 
Kirkpatrick, \& Simons, 1994), and they live for a long time.
To a working approximation, the main sequence 
(core hydrogen burning) lifetime of a star depends on its 
mass through the relation
$$\tau_\ast = 10^{10}  {\rm yr} \Bigl[ {M_\ast \over 1 M_\odot} 
\Bigr]^{-\alpha} \, , \eqno(2.1{\rm a})$$
where the index $\alpha \approx 3-4$ for stars of low mass. 
In terms of cosmological decades $\eta$, we obtain 
$$\eta_\ast = 10 - \alpha \log_{10} [M_\ast / 1 M_\odot] 
\, . \eqno(2.1{\rm b})$$
Thus, for example, $\eta_\ast \approx 13$ for a small star 
with $M_\ast$ = 0.1 $M_\odot$. Indeed at the present time, only stars
with masses $M_\ast >$ 0.8$M_\odot$ have had time to experience
significant post-main sequence evolution. Hence, a large fraction, 
$$f \equiv { \int_{M_{min}}^{0.8} (dN/dm) dm \over 
\int_{{M}_{min}}^{M_{max}} (dN/dm) dm } 
\sim 80\% \, , \eqno(2.2)$$
of all stars ever formed have yet to experience any significant 
evolution (here, $dN/dm$ is the mass distribution -- see \S II.E). 
We are effectively still in the midst of the transient initial 
phases of the stelliferous epoch. 

Very little consideration has been given to the post-main sequence
development of stars which are small enough to outlive the current age
of the universe.  An essay by Salpeter (1982) contains a qualitative
discussion regarding the evolution of M stars (especially with respect
to $^3$He production) but detailed stellar evolutionary sequences have
not been presented in the literature.  Nevertheless, there is a sizable
collection of papers which discuss the pre-main sequence and main
sequence properties of very low mass stars (e.g., Kumar, 1963;
Copeland, Jensen \& Jorgensen, 1970; Grossman \& Graboske, 1971;
D'Antona \& Mazzitelli, 1985; Dorman, Nelson \& Chau, 1989). The best
comprehensive family of models spanning the M dwarfs and brown dwarfs
is probably that of Burrows et al. (1993). Those authors devote
attention to the formative cooling phases, as well as the exact mass
of the minimum mass star (which for their input physics occurs at
$M_\ast$ = 0.0767 $M_{\odot}$).  Evolution beyond 20 billion years 
was not considered (see also Burrows \& Liebert, 1993). 

The dearth of information regarding the fate of the M dwarfs has
recently been addressed (Laughlin, Bodenheimer, \& Adams, 1996). 
We have performed a detailed series of stellar evolution calculations 
which follow the pre main-sequence through post main-sequence
evolution of late M-dwarfs, yielding the following picture of what
lies in store for the low mass stars.

Newly formed stars containing less mass than $M_\ast \sim 0.25 M_{\odot}$ 
are fully convective throughout the bulk of their structure. The
capacity of these stars to entirely mix their contents has several
important consequences. First, these late M stars maintain access to
their entire initial reserve of hydrogen, greatly extending their
lifetimes in comparison to heavier stars like the sun which see their
fuel supply constricted by stratified radiative cores. Second, as
recognized by Salpeter (1982), full convection precludes the buildup
of composition gradients which are ultimately responsible (in part) 
for a star's ascent up the red giant branch. The lowest mass stars
burn all their hydrogen into helium over an $\eta=13$ time scale, and
then quietly fade from prominence as helium white dwarfs.  This
general evolutionary scenario is detailed in Figure 1 (adapted from
Laughlin et al., 1996), which charts the path in the
Hertzsprung-Russell diagram followed by low mass stars of several
different masses in the range $0.08 M_{\odot}$ 
$\le M_\ast \le$  $0.25 M_{\odot}$.

Upon emerging from its parent cloud core, the lowest mass star 
capable of burning hydrogen ($M_\ast \approx 0.08 M_{\odot}$) 
descends the convective Hayashi track and arrives on the main sequence
with a luminosity $L_\ast \sim 10^{-4} L_{\odot}$.  The main sequence
phase is characterized by gradual prolonged increase in both
luminosity and effective surface temperature as hydrogen is consumed.  
Due to the relatively low prevailing temperature in the stellar core 
($T_{c} \approx 4 \times 10^{6}$ K), the proton-proton nuclear
reaction chain is decoupled from statistical equilibrium, and the
concentration of $^{3}$He increases steadily until $\eta$=12.6, at which
time a maximum mass fraction of 16$\%$ $^{3}$He has been attained.  
As the initial supply of hydrogen is depleted, the star heats up and 
contracts, burns the $^{3}$He, increases in luminosity by a factor of
10, and more than doubles its effective temperature.  After $\sim$11  
trillion years, when the star has become 90\% $^{4}$He by mass, a
radiative core finally develops. The evolutionary time scale begins 
to accelerate, and hydrogen is exhausted relatively quickly in the 
center of the star. When nuclear burning within the modest resulting
shell source can no longer provide the star's mounting energy
requirements, the star begins to contract and cool and eventually 
becomes a helium white dwarf. Stars with masses up to $\sim$0.20
$M_{\odot}$ follow essentially this same evolutionary scenario. As
stellar mass increases, radiative cores develop sooner, and the stars
perform increasingly dramatic blueward excursions in the H-R diagram.

A star with a slightly larger mass, $M_\ast$ = 0.23 $M_{\odot}$, 
experiences the onset of a radiative core when the hydrogen mass
fraction dips below 50\%. The composition gradients which ensue are
sufficient to briefly drive the star to lower effective temperature as
the luminosity increases. In this sense, stars with mass $M_\ast$ = 
$0.23 M_{\odot}$ represent the lowest mass objects that can become 
conventional ``Red Giants''.  At these low masses, however, the full 
giant phase is not completed. Stars with initial mass $M_\ast < 0.5$
$M_{\odot}$ will be unable to generate the high central temperatures 
($T_c \sim 10^8$ K) required for the helium flash; these stars abort 
their ascent up the giant branch by veering to the left in the 
H-R diagram in the manner suggested by Figure 1. 

The steady luminosity increases experienced by aging M dwarfs 
will have a considerable effect on the mass to light ratio of the 
galaxy. For example, as a 0.2 $M_{\odot}$ star evolves, there is a
relatively fleeting epoch (at $\eta \approx$ 12) during which the star
has approximately the same radius and luminosity as the present day
sun. Given that M dwarfs constitute the major fraction of all stars,
the total luminosity of the galaxy will remain respectably large,
$L_{\rm gal} \sim 10^{10} L_\odot$, at this future date.  This
luminosity is roughly comparable to the characteristic luminosity
${\cal L}^\ast$ = 3.4 $\times 10^{10}$ $L_\odot$ displayed by present
day galaxies (Mihalas \& Binney, 1981).

\bigskip 
\noindent{\bf B. Forthcoming Metallicity Effects} 
\nobreak 
\medskip 

The foregoing evolutionary calculations assumed a solar abundance set.
In the future, the metallicity of the galaxy will steadily increase 
as stars continue to process hydrogen and helium into heavy elements. 
It is thus useful to determine the effects of these metallicity 
increases. 

\bigskip 
\noindent 
{1. Stellar Lifetimes vs Metallicity} 
\medskip 

First, it is possible to construct a 
simple scaling relation that clarifies how stellar 
lifetimes $\tau_\ast$ depend on the metallicity $Z$. 
The stellar lifetime is roughly given by amount of fuel
available divided by the rate of fuel consumption, i.e.,
$$\tau_\ast \sim M_\ast \, X \, / L \, , \eqno(2.3)$$
where $M_\ast$ is the stellar mass and $X$ is the hydrogen mass 
fraction.  For relatively low mass stars, the luminosity $L$ 
obeys the scaling relation 
$$L \sim \kappa_0^{-1} \, \mu^{7.5} M_\ast^{5.5} \, , \eqno(2.4)$$
where $\mu$ is the mean molecular weight of the star 
and where $\kappa_0$ is the constant of proportionality 
appearing in the usual opacity relation for stars (Clayton, 1983). 
Thus, for a given stellar mass $M_\ast$, the lifetime scales 
according to 
$$\tau_\ast \sim \kappa_0 \, X \, \mu^{-7.5} \, . \eqno(2.5)$$

To evaluate the stellar lifetime scaling relation, one needs to know
how the parameters $\kappa_0$, $X$, and $\mu$ vary with metallicity. 
The opacity constant $\kappa_0$ is roughly linearly dependent on 
the metallicity, i.e., 
$$\kappa_0 \sim Z \, . \eqno(2.6)$$
The mean molecular weight $\mu$ can be approximately 
written in the form
$$\mu \approx {2 \over (1 + 3 X + Y/2) } \,  , \eqno(2.7)$$
where $Y$ is the helium mass fraction (e.g., see Clayton, 1983). 
By definition, the mass fractions obey the relation
$$X + Y + Z = 1 \, . \eqno(2.8)$$
Finally, for this simple model, we write the helium abundance 
$Y$ in the form 
$$Y = Y_P +  fZ \, , \eqno(2.9)$$
where $Y_P$ is the primordial abundance and the factor {\it f} 
accounts for the increase in helium abundance as the metallicity 
increases.  Big Bang nucleosynthesis considerations indicate that 
$Y_P$ $\approx$ 1/4 (Kolb \& Turner, 1990), whereas $f \approx$ 2 
based on the solar enrichment in $Y$ and $Z$ relative to the 
primordial values.  Combining the above results, we obtain a 
scaling relation for the dependence of stellar lifetimes on 
metallicity, 
$$\tau_\ast \sim Z (1 - a Z) \, (1 - b Z)^{7.5} \, , \eqno(2.10)$$  
where we have defined constants $a \equiv 4 (1 + f)/3 \approx 4$  
and $b \equiv 8/9 + 20 f/27 \approx 64/27$.  This result implies 
that stellar lifetimes have a {\it maximum value}. In particular, 
we find that stars born with metallicity $Z \approx 0.04$ live the 
longest. For larger values of $Z$, the reduction in nuclear fuel 
and the change in composition outweigh the lifetime extending 
decrease in luminosity arising from the increased opacity. 

A recent set of galactic chemical evolution calculations 
(Timmes, 1996) have probed far into the stelliferous epoch. 
The best indications suggest that the galactic abundance 
set will approach an asymptotically constant composition 
($X \sim$ 0.2, $Y \sim$ 0.6, and $Z \sim$ 0.2) over a time 
scale $\eta \sim 12$.  As a consequence, any generations of stars 
formed after $\eta \sim 12$ will suffer significantly shorter 
lifetimes than the theoretical maximum implied by equation [2.10]. 

\bigskip 
\noindent 
{2. Stellar Masses vs Metallicity} 
\medskip 

The maximum stable stellar mass decreases as metallicity increases. 
On the main sequence, the maximum possible mass is reached when the 
star's radiation pressure comes to dominate the thermal (gas) 
pressure within the star.  Here, we introduce the usual ansatz that 
the total pressure at the center of the star can be written in 
the form $P_C = P_R + P_g$, where the thermal gas pressure is
given by the fraction $P_g = \beta P_C$ and, similarly, 
$P_R = (1 - \beta) P_C$.  Using the ideal gas law for the 
thermal pressure and the equation of state for a gas of photons, 
we can write the central pressure in the form 
$$P_C = 
\Bigl[ {3 \over a} {(1-\beta) \over \beta^4} \Bigr]^{1/3} \, 
\Bigl[ {k \rho_C \over \mu m_P}  \Bigr]^{4/3} \, , \eqno(2.11)$$
where $k$ is Boltzmann constant and $a$ is the radiation 
constant. The quantity $\mu$ is again the mean molecular weight 
and can be written in the form of equation [2.7]. 
In hydrostatic equilibrium, the central pressure required 
to support a star of mass $M_\ast$ can be expressed as
$$P_C \approx \Bigl[ {\pi \over 36} \Bigr]^{1/3} 
\, G \, M_\ast^{2/3} \, \rho_C^{4/3} \, , \eqno(2.12)$$ 
where $\rho_C$ is the central density (see Phillips, 1994). 

Equating the above two expressions [2.11] and [2.12], we can solve 
for the mass to find 
$$M_\ast = 
\Bigl[ {108 \over \pi a} {(1-\beta) \over \beta^4} \Bigr]^{1/2} \, 
\Bigl[ {k \over \mu m_P}  \Bigr]^2 \,  G^{-3/2} \, \approx 
40 M_\odot \, \, \mu^{-2} \, , \eqno(2.13)$$ 
where we have set $\beta=1/2$ to obtain the numerical value. 
The maximum stellar mass thus depends somewhat sensitively 
on the mean molecular weight $\mu$, which in turn is a function of 
the metallicity. By applying the approximations [2.7], [2.8], and [2.9], 
one can write the maximum mass in the form 
$$M_\ast = 40 M_\odot \, \Bigl\{ (2 - 5 Y_P /4) - 
(3 + 5 f/2) Z/2 \Bigr\}^2 \, \approx 
114 M_\odot \, (1 - 2.4 Z)^2 \, . \eqno(2.14)$$
Thus, for the expected asymptotic value of the metallicity,
$Z = 0.2$, the maximum mass star is only $M_\ast \approx 30 M_\odot$. 

The continuously increasing metallicity of the interstellar medium will
also have implications for low mass stars. Higher metallicity leads to
more effective cooling, which leads to lower temperatures, which in
turn favors the formation of less massive stars (e.g., see the recent
theory of the initial mass function by Adams \& Fatuzzo, 1996). The IMF
of the future should be skewed even more dramatically in favor of the
faintest stars.

The forthcoming metallicity increases may also decrease the mass of
the minimum mass main sequence star as a result of opacity effects
(cf. the reviews of Stevenson, 1991; Burrows \& Liebert, 1993).  
Other unexpected effects may also occur. For example, when the
metallicity reaches several times the solar value, objects with mass
$M_\ast$ =  0.04 $M_{\odot}$ may quite possibly halt their cooling 
and contraction and land on the main sequence when thick ice clouds 
form in their atmospheres. Such ``frozen stars'' would have an
effective temperature of $T_\ast \approx$ 273 K, far cooler than the
current minimum mass main sequence stars. The luminosity of these
frugal objects would be more than a thousand times smaller than the
dimmest stars of today, with commensurate increases in longevity.

\bigskip 
\bigskip 
\noindent{\bf C. The Fate of the Earth and the Sun} 
\nobreak 
\medskip  

A popular and frequently quoted scenario for the demise of the Earth
involves destruction through evaporation during the Sun's asymptotic
giant branch (AGB) phase.  As the Sun leaves the horizontal branch and
expands to become an AGB star, its outer radius may swell to such an
extent that the photospheric radius overtakes the current orbital 
radius of the Earth.  If this state of affairs comes to pass, 
then two important processes will affect the Earth: 
[1] Evaporation of material due to the extreme heat, and [2]
Orbital decay through frictional drag. This second process drives the
Earth inexorably into the giant sun, thereby increasing the efficacy
of the evaporation process.  Once the earth finds itself {\it inside}
the sun, the time scale for orbital decay is roughly given by the time
required for the expiring Earth to sweep through its mass, $M_E$, in
solar material. This short time interval is given by 
$$\tau = {M_E \over \rho_\odot \, (\pi R_E^2) \, v_{\rm orbit} } 
\, \approx 50 \, {\rm yr} \, , \eqno(2.15)$$ 
where $\rho_\odot$ $\sim 10^{-6}$ g/cm$^3$ is the mass density of
solar material at the photosphere, $R_E \approx$ 6370 km is the 
radius of the Earth, and $v_{\rm orbit} \approx$ 30 km/s is the 
orbital speed. Hence, the demise of the Earth will befall it swiftly, 
even in comparison to the accelerated stellar evolution time scale
inherent to the asymptotic giant branch.  The Earth will be
efficiently dragged far inside the sun and vaporized in the fierce
heat of the stellar plasma, its sole legacy being a small (0.01\%)
increase in the metallicity of the Sun's distended outer envelope.

Recent work suggests, however, that this dramatic scene can be
avoided.  When the sun reaches a luminosity of $\sim 100 L_\odot$ 
on its ascent of the red giant branch, it will experience heavy mass
loss through the action of strong stellar winds. Mass loss results 
in an increase in the orbital radii of the planets and can help the 
Earth avoid destruction. However, the actual amount of mass loss 
remains uncertain; estimates are based largely on empirical 
measurements (see Reimers, 1975), but it seems reasonable that the sun 
will diminish to $\sim 0.70 M_\odot$ when it reaches the tip of 
the red giant branch, and will end its AGB phase as a carbon white 
dwarf with mass $\sim 0.5 M_\odot$.  Detailed stellar evolution 
calculations for the sun have been made by Sackmann, Boothryod, 
\& Kraemer (1993). In their best-guess mass loss scenario, they 
find that the orbital radii for both the Earth and Venus increase
sufficiently to avoid being engulfed during the AGB phase. 
Only with a more conservative mass loss assumption, in which the 
Sun retains $0.83 M_\odot$ upon arrival on the horizontal branch, 
does the solar radius eventually overtake the Earth's orbit.

\bigskip 
\noindent{\bf D. Continued Star Formation in the Galaxy} 
\medskip 

Galaxies can only live as long as their stars. Hence it is useful to
estimate how long a galaxy can sustain normal star formation (see,
e.g., Shu, Adams, \& Lizano, 1987) before it runs out of raw
material. One would particularly like to know when the {\it last}
star forms.

There have been many studies of the star formation history in both 
our galaxy as well as other disk galaxies (e.g., Roberts, 1963; 
Larson \& Tinsley, 1978; Rana, 1991; Kennicutt, Tamblyn, \& Congdon, 
1994; hereafter KTC).  Although many uncertainties arise in these 
investigations, the results can be roughly summarized as follows. 
The gas depletion time $\tau_{R}$ for a disk galaxy is defined 
to be the current mass in gas, $M_{\rm gas}$, divided by the star 
formation rate $\sfr$, i.e., 
$$\tau_R \equiv { M_{\rm gas} \over \sfr } \, . \eqno(2.16)$$
For typical disk galaxies, this time scale is comparable 
to the current age of the universe; KTC cite a range 
$\tau_R \approx$ 5 -- 15 Gyr. The actual time scale for (total) 
gas depletion will be longer because the star formation rate 
is expected to decrease as the mass in gas decreases. For example, 
if we assume that the star formation rate is proportional to the 
current mass in gas, we derive a total depletion time of the form 
$$\tau = \tau_R \, \ln \bigl[ M_0/M_F \bigr] \, , \eqno(2.17)$$
where $M_0$ is the initial mass in gas and $M_F$ is the final 
mass.  For typical disk galaxies, the initial gas mass is 
$M_0 \sim 10^{10} M_\odot$ (see Table 5 of KTC). 
Thus, if we take the extreme case of $M_F$ = 1 $M_\odot$, 
the total gas depletion time is only $\tau \approx 23 \tau_R$ 
$\approx$ 120 -- 350 Gyr.  In terms of cosmological decades, 
the gas depletion time becomes 
$\eta_D = 11.1 - 11.5 \,$. 

Several effects tend to extend the gas depletion time scale
beyond this simple estimate. When stars die, they return a fraction of
their mass back to the interstellar medium.  This gas recycling effect
can prolong the gas depletion time scale by a factor of 3 or 4 (KTC).
Additional gas can be added to the galaxy through infall onto the
galactic disk, but this effect should be relatively small (cf. the
review of Rana, 1991); the total mass added to the disk should not
increase the time scale by more than a factor of 2.  Finally, if the
star formation rate decreases more quickly with decreasing gas mass
than the simple linear law used above, then the depletion time scale
becomes correspondingly larger.  Given these complications, we expect
the actual gas depletion time will fall in the range
$$\eta_D = 12 - 14 \, . \eqno(2.18)$$  
Thus, by the cosmological decade $\eta \approx 14$, essentially all
normal star formation in galaxies will have ceased. Coincidentally,
low mass M dwarfs have life expectancies that are comparable
to this time scale.  In other words, both star formation and stellar
evolution come to an end at approximately the same cosmological decade. 

There are some indications that star formation may turn off
even more dramatically than outlined above.  Once the gas density
drops below a critical surface density, star formation may turn off
completely (as in elliptical and S0 galaxies).  The gas may be heated
entirely by its slow accretion onto a central black hole.

These results indicate that stellar evolution is confined to a 
reasonably narrow range of cosmological decades.  It is presumably 
impossible for stars to form and burn hydrogen before the epoch of 
recombination in the universe (at a redshift $z \sim$ 1000 and hence 
$\eta \sim 5.5$). Thus, significant numbers of stars will exist only 
within the range
$$5.5 < \eta < 14 \, . \eqno(2.19)$$
The current epoch ($\eta \sim 10$) lies
near the center of this range of (logarithmic) time scales. 
On the other hand, if we use a linear time scale, the current 
epoch lies very near the beginning of the stelliferous era. 

\bigskip 
\noindent{\bf E. The Final Mass Function} 
\nobreak 
\medskip 

When ordinary star formation and conventional stellar evolution have 
ceased, all of the remaining stellar objects will be in the form of
brown dwarfs, white dwarfs, neutron stars, and black holes.  One 
way to characterize the stellar content of the universe at this epoch
is by the mass distribution of these objects; we refer to this
distribution as the ``Final Mass Function'' or FMF. Technically, the 
Final Mass Function is not final in the sense that degenerate objects 
can also evolve and thereby change their masses, albeit on vastly 
longer time scales. The subsequent evolution of degenerate objects 
is discussed in detail in \S IV. 

Two factors act to determine the FMF:
[1] The initial distribution of stellar masses (the initial mass 
function [IMF] for the progenitor stars), and [2] The transformation 
between initial stellar mass and the mass of the final degenerate
object.  Both of these components can depend on
cosmological time.  In particular, one expects that metallicity effects 
will tend to shift the IMF toward lower masses as time progresses.

The initial mass function can be specified in terms of a general 
log-normal form for the mass distribution $\psi = dN/d\ln m$, 
$$\ln \psi (\ln m) = A - {1 \over 2 \sigbar^2}
\Bigl\{ \ln \bigl[ m / m_C \bigr] \Bigr\}^2 \, , \eqno(2.20)$$
where $A$, $m_C$, and $\sigbar$ are constants. Throughout this 
discussion, stellar masses are written in solar units, i.e., 
$m \equiv M_\ast / (1M_\odot)$. This general form for the IMF 
is motivated by the both current theory of star formation and 
by general statistical considerations (Adams \& Fatuzzo, 1996; 
Zinnecker, 1984; Larson, 1973; Elmegreen \& Mathieu, 1983). 
In addition, this form is (roughly) consistent with observations
(Miller \& Scalo, 1979), which suggest that the shape parameters 
have the values $\sigbar \approx 1.57$ and $m_C \approx 0.1$ for 
the present day IMF (see also Salpeter, 1955; Scalo, 1986; Rana, 1991).  
The constant $A$ sets the overall normalization of the distribution 
and is not of interest here. 

For a given initial mass function, we must find the final masses 
$m_F$ of the degenerate objects resulting from the progenitor stars 
with a given mass $m$.  For the brown dwarf range of progenitor 
masses, $m < m_H$, stellar objects do not evolve through nuclear 
processes and hence $m_F = m$.  Here, the scale $m_H \approx 0.08$ 
is the minimum stellar mass required for hydrogen burning to take place.  

Progenitor stars in the mass range $m_H \le m \le m_{SN}$ eventually
become white dwarfs, where the mass scale $m_{SN}$ $\approx$ 8 is the
minimum stellar mass required for the star to explode in a supernova 
(note that the mass scale $m_{SN}$ can depend on the metallicity -- 
see Jura, 1986). 
Thus, for the white dwarf portion of the population, we must specify
the transformation between progenitor mass $m$ and white dwarf mass
$m_{WD}$.  The results of Laughlin et al. (1996) indicate that
stars with main sequence masses $m < 0.4$ will undergo negligible 
mass loss in becoming helium white dwarfs. Unfortunately, this 
relationship remains somewhat ill-defined at higher masses, mostly 
due to uncertainties in red giant mass loss rates (e.g., see Wood, 
1992).  For the sake of definiteness, we adopt the following 
transformation between progenitor mass and white dwarf mass,
$$m_{WD} = {m \over 1 + \alpha m} \exp [ \beta m ] 
\, , \eqno(2.21)$$
with $\alpha$ = 1.4 and $\beta$ = 1/15. This formula is 
consistent with the models of Wood (1992) over the appropriate
mass range and approaches the expected form $m_{WD} = m$ in 
the low mass limit. 

Stars with large initial masses, $m > m_{SN}$, end their lives 
in supernova explosions and leave behind a neutron star (although 
black holes can also, in principle, be produced).  The mass of 
the remnant neutron star is expected to be near the Chandrasekhar 
limit $m_{Ch} \approx 1.4$, as confirmed in the case of the binary
pulsar (Manchester \& Taylor, 1977). 

To compute the FMF, one convolves the initial mass function with the
transformations from progenitor stars to white dwarfs and neutron
stars.  The Final Mass Function that results is shown in Figure 2.
For comparison, the initial mass function is also shown (as the dashed
curve). Notice that the two distributions are similar for masses less
than the Chandrasekhar mass ($\sim 1.4 M_\odot$) and completely
different for larger masses.

Once the FMF has been determined, one can estimate the number 
and mass fractions of the various FMF constituents. 
We define ${\cal N}_{BD}$
to be the fraction of brown dwarfs by number and ${\cal M}_{BD}$ 
to be the fraction of brown dwarfs by mass, with analogous 
fractions for white dwarfs (${\cal N}_{WD}$ and ${\cal M}_{WD}$) 
and neutron stars (${\cal N}_{NS}$ and ${\cal M}_{NS}$). 
For an IMF of the form [2.20] with present 
day values for the shape parameters, we 
obtain the following number fractions:
$${\cal N}_{BD} = 0.45 \, , \qquad 
{\cal N}_{WD} = 0.55 \, ,  \qquad 
{\cal N}_{NS} = 0.0026 \, .  \eqno(2.22)$$ 
Similarly, for the mass fractions one finds 
$${\cal M}_{BD} = 0.097 \, , \qquad 
{\cal M}_{WD} = 0.88 \, ,  \qquad 
{\cal M}_{NS} = 0.024 \, .  \eqno(2.23)$$ 
Thus, brown dwarfs are expected to be present in substantial 
numbers, but most of the mass will reside in the form of white 
dwarfs.  Neutron stars will make a relatively small contribution 
to the total stellar population.  The above values for ${\cal N}_{NS}$,
and ${\cal M}_{NS}$ were obtained under the assumption that all stars
$m>m_{SN} \sim 8$ produce neutron stars.  In reality, a portion of 
these high mass stars may collapse to form black holes instead, 
but this complication does not materially affect the 
basic picture described above.

\newpage 
\noindent{\bf III. DEATH OF THE GALAXY} 
\medskip 

We have argued that over the long term, the galaxy will incorporate 
a large fraction of the available baryonic matter into stars. By the 
cosmological decade $\eta = 14-15$, the stellar component of the 
galaxy will be in the form of seemingly inert degenerate remnants. 
Further galactic activity will involve these remnants in phenomena 
which unfold over time scales ranging from $\eta \sim 15-30$. 
This time period is part of what we term the {\sl Degenerate Era}. 

The course of this long term galactic dynamical evolution is dictated
by two generalized competing processes. First, in an isolated physical
system containing any type of dissipative mechanism (for example,
gravitational radiation, or extremely close inelastic encounters
between individual stars), the system must evolve toward a state of
lower energy while simultaneously conserving angular momentum. The net
outgrowth of this process is a configuration in which most of the mass
is concentrated in the center and most of the angular momentum is
carried by small parcels at large radii. (The present day solar system
presents a good example of this process at work.)  Alternatively, a
second competing trend occurs when collisionless relaxation processes
are viable. In a galaxy, distant encounters between individual stars
are effectively collisionless. Over time, stars tend to be evaporated
from the system, the end product of this process is a tightly bound
agglomeration (perhaps a massive black hole) in the center, containing
only a fairly small fraction of the total mass.  Hence, one must
estimate the relative efficiencies of both collisionless and
dissipative processes in order to predict the final state of the
galaxy.  This same competition occurs for physical systems on both
larger scales (e.g., galaxy clusters) and smaller scales (e.g.,
globular clusters).

In addition to gravitational radiation and dynamical relaxation,
occasional collisions between substellar objects -- brown dwarfs --
provide a channel for continued star formation at a very slow
rate. Collisions and mergers involving two white dwarfs will lead 
to an occasional type I supernova, whereas rare impacts involving 
neutron stars will engender even more exotic bursts of energy. 
Such events are impressive today. They will be truly spectacular 
within the cold and impoverished environment of an evolved galaxy. 

\bigskip 
\noindent{\bf A. Dynamical Relaxation of the Galaxy}
\medskip 

A stellar system such as a galaxy relaxes dynamically 
because of stellar encounters.  The characteristic time 
scale associated with this process in the case of purely 
stellar systems is well known and can be written as  
$$\tau_{\rm relax} = {R \over v} {N \over 12 \ln (N/2)} 
\, , \eqno(3.1)$$
where $R$ is the size of the system, $v$ is the typical 
random velocity, and $N$ is the total number of stars
(for further discussion, see Lightman \& Shapiro, 1978; 
Shu, 1982; Binney \& Tremaine, 1987).
The logarithmic factor appearing in the denominator takes into 
account the effects of many small angle deflections of stars 
through distant encounters. The time scale for stars to 
evaporate out of the system is roughly given by 
$$\tau_{\rm evap} = 100 \, \tau_{\rm relax} 
\sim 10^{19} {\rm yr} \, , \eqno(3.2)$$
where we have used $R$ = 10 kpc, $v$ = 40 km/s, and $N = 10^{11}$
to obtain the numerical result. We thus obtain the corresponding 
estimate 
$$\eta_{\rm evap} = 19 + \log_{10} [R/ 10 {\rm kpc} ] 
+ \log_{10} [N/10^{11}] . \, \eqno(3.3)$$ 
Thus, stars escape from the galaxy with a characteristic time 
scale $\eta \approx 19 - 20$ (see also Islam, 1977; Dyson, 1979). 

The stellar dynamical evolution of the Galaxy is more complicated than
the simple picture outlined above.  First, the galaxy is likely to
have an extended halo of dark matter, much of which may be in
non-baryonic form.  Since this dark halo does not fully participate in
the dynamical relaxation process, the halo tends to stabilize the
system and makes the stellar evaporation time scale somewhat longer
than the simple estimate given above. 

Other dynamical issues can also be important.  In globular clusters,
for example, mass segregation occurs long before stellar evaporation
and binary star heating plays an important (actually dominant) role in
the long term evolution.  On the other hand, equation [3.1] is
formally valid only if the stars are not bound into binary or triple
systems.  Binary interaction effects can be important for the long
term evolution of the stellar component of the galaxy.  In particular,
the presence of binaries can increase the effective interaction 
cross section and can lead to a variety of additional types of 
interactions.  Both three-body interactions and binary-binary 
interactions are possible. As a general rule, interactions 
lead to hard binaries becoming harder and wide binaries becoming
softer or even disrupted (``ionized'').  Binaries that become
sufficiently hard (close) can spiral inwards, become mass transfer
systems, and eventually explode as supernovae. These effects are 
just now becoming understood in the context of globular cluster 
evolution (for further discussion of these dynamical issues, 
see, e.g., Chernoff \& Weinberg, 1990; Hut et al., 1992). 

Galaxies in general, and our galaxy in particular, live in 
groups or clusters.  These larger scale systems will also 
undergo dynamical relaxation processes analogous to those 
discussed above.  However, a more immediate issue that can 
affect our galaxy in the relatively near future is the 
possibility of merging with other galaxies in the local group, 
in particular Andromeda (M31). The orbits of nearby galaxies
have been of the subject of much study (e.g., Peebles, 1994), 
but large uncertainties remain.  For the current separation 
between the Milky Way and M31 ($d$ = 0.75 Mpc) and radial 
velocity ($v_r$ = 120 km/s), the two galaxies will experience 
a close encounter at a time $\Delta t$ = $6 \times 10^9$ yr 
in the future (i.e., at $\eta$ = 10.2). Whether this encounter 
will lead to a collision/merger or simply a distant passage 
depends on the tangential velocity component, which is not 
well determined.  The models of Peebles (1994) suggest that 
the distance of closest approach will lie in the range 
20 -- 416 kpc, with more models predicting values near the 
upper end of this range. Thus, more work is necessary to 
determine whether or not the Milky Way is destined to 
collide with M31 in the relatively near future. 

However, even if our galaxy does not collide with M31 on the first
pass, the two galaxies are clearly a bound binary pair. The orbits 
of binary galaxy pairs decay relatively rapidly through dynamical 
friction (e.g., Binney \& Tremaine, 1987; Weinberg, 1989).  Thus, even
if a collision does not occur on the first passing, M31 and the Milky
Way will not survive very long as individual spiral galaxies. 
On a time scale of approximately $\eta = 11 - 12$, the entire 
local group will coalesce into one large stellar system. 

\bigskip 
\noindent{\bf B. Gravitational Radiation and the Decay of Orbits}
\medskip 

Gravitational radiation acts in the opposite direction: it causes 
orbits to lose energy and decay so that the stars move inward. 
We first consider the case of a galaxy and its constituent stars. 
As a given star moves through the potential well of a galaxy, 
its orbit decays through gravitational radiation (e.g., Misner, 
Thorne, \& Wheeler, 1973; Weinberg, 1972).  The rate of energy 
loss is proportional to the square of the quadrapole moment 
of the whole system (see also Ohanian \& Ruffini, 1994). 
For the case in which the galaxy has a large scale 
quadrapole moment (e.g., a bar), the rate of energy loss 
from gravitational radiation can be written in the simple form 
$${ {\dot E} \over E } = 
\Big( {v \over c} \Bigr)^5 \tau^{-1} \, , \eqno(3.4)$$ 
where $\tau = 2 \pi R/v$ is the orbit time.  For a galaxy, 
the rotation curve is almost flat with a nearly constant velocity 
$v \sim 200$ km/s.  The time scale $\tau_{GR}$ for gravitational 
radiation is thus given by 
$$\tau_{GR} = {2 \pi R \over v} 
\Big( {v \over c} \Bigr)^{-5} \approx 10^{24} {\rm yr} 
\Big( {R \over R_0} \Bigr) \, , \eqno(3.5)$$ 
where $R_0$ = 10 kpc is a reference length scale for the 
galaxy. We thus obtain the estimate 
$$\eta_{GR} = 24 + \log_{10} [R/10 {\rm kpc}] \, . \eqno(3.6)$$ 
This time scale corresponds to $\sim 10^{16}$ orbits around the
galactic center.  Notice that if the stars are radiating incoherently
in a nearly smooth potential, the time scale becomes longer by a
factor of $M_{gal} / M_\ast$, where $M_\ast$ is the mass of the star
and $M_{gal}$ is effective galactic mass.  Notice also that
gravitational orbital decay takes substantially longer than stellar
evaporation from the galaxy (see the previous section). Thus, the
evolution of the galaxy will be dominated by the collisionless
process, and hence the majority of stellar remnants will be ejected
into intergalactic space rather than winding up in the galactic core
(see also Islam, 1977; Dyson, 1979; Rees, 1984).

Gravitational radiation also causes 
the orbits of binary stars to lose energy and decay.  
Of particular importance is the decay of binary brown dwarf stars. 
The eventual coalescence of these systems can lead to the formation 
of a new hydrogen burning star, provided that the mass of the entire
system is larger than the hydrogen burning limit $M_H \sim 0.08
M_\odot$. The time scale $\tau_{OD}$ for orbital decay can be written
$$\tau_{OD} = {\pi \over 2} {c^5 R_0^4 \over G^3 M_\ast^3} 
\, , \eqno(3.7)$$
where $M_\ast$ is the mass of the the stars and $R_0$ is the 
initial orbital separation.  Inserting numerical values 
and writing the result in terms of cosmological decades, 
we obtain the result 
$$\eta_{OD} = 19.4 + 4 \log_{10} [ R_0 / (1 {\rm AU}) ] 
- 3 \log_{10} [ M_\ast / (1 M_\odot) ] \, . \eqno(3.8)$$
This result also applies to planetary orbits (see \S III.F below). 

\bigskip 
\noindent{\bf C. Star Formation through Brown Dwarf Collisions} 
\nobreak 
\medskip 

Once all of the interstellar material has been used up, one viable 
way to produce additional stars is through the collisions of brown 
dwarfs.  These objects have masses too small for ordinary hydrogen 
burning to take place and hence their supply of nuclear fuel will 
remain essentially untapped. Collisions between these substellar 
objects can produce stellar objects with masses greater than the 
hydrogen burning limit, i.e., stars of low mass.  We note that 
the search for brown dwarfs has been the focus of much 
observational work (see, e.g., Tinney, 1995) and the existence 
of these objects is now on firm ground 
(e.g., Golimowski et al., 1995; Oppenheimer et al., 1995). 

\bigskip 
\noindent
{1. Collision Time Scales} 
\medskip 

After conventional star formation in the galaxy has ceased, 
the total number of brown dwarfs in the galaxy will be $N_0$.  
Although the value of $N_0$ is uncertain and is currently the 
subject of much current research (e.g., see Alcock et al., 1993; 
Aubourg et al., 1993; Tinney, 1995), we expect 
that $N_0$ is roughly comparable to the number of ordinary stars 
in the galaxy today, $N_0$ $\sim$ $10^{11}$ (see \S II.C).  
The rate $\Gamma$ at which these brown dwarfs collide is given by 
$$\Gamma = {N \sigma v \over V} = - {1 \over N} {dN \over dt} 
\, , \eqno(3.9)$$  
where $N$ is the number of brown dwarfs in a galaxy with volume $V$, 
$\sigma$ is the collision cross section (see below), and $v$ is the 
typical relative velocity. This equation can be integrated to obtain 
$$N(t) = {N_0 \over 1 + t / \tau_C } \, , \eqno(3.10)$$  
where $\tau_C$ is the characteristic time scale 
$$\tau_C = \Gamma^{-1} \sim 10^{22} {\rm yr} , \eqno(3.11)$$ 
or, equivalently, 
$$\eta_C = 22 + \log_{10} \Bigl[ V / (20 {\rm kpc})^3 \Bigr] - 
\log_{10} \Bigl[ v / (200 {\rm km/s}) \Bigr] \, . \eqno(3.12)$$  
To obtain this numerical value for the time scale, we have assumed
that the collision cross section is given by the geometrical cross
section of the brown dwarfs; this assumption is justified below. 
We have also the used numerical values $V \sim (20 {\rm kpc})^3$ 
and $v \sim 200$ km/s which are characteristic of the galactic halo. 

The estimate of collision rates given here is somewhat conservative.
Nearby stellar encounters can lead to the formation of binaries
through tidal excitation of modes on the stars (see Press \& Teukolsky, 
1977; Lee \& Ostriker, 1986).  These binaries can eventually decay and
thereby lead to additional stellar collisions.

The time scale [3.12] is the time required for the halo population of
brown dwarfs to change. Notice that this time scale is larger
than the evaporation time scale calculated in \S III.A. This ordering
makes sense because distant encounters (which lead to evaporation)
must be much more frequent than true collisions.  For $\eta < \eta_C$, 
the collision rate of brown dwarfs for the entire galaxy is given by 
$\Gamma_{\rm tot} = N/\tau_C$ $\sim 10^{-11}$ yr$^{-1}$.  
The typical outcome of a brown dwarf collision will be the production
of a stellar object with mass $M_\ast \sim 0.1 M_\odot$, large enough
to burn hydrogen.  The stellar (main-sequence) lifetime of such a star
is roughly $2.5 \times 10^{13}$ yr. This stellar evolutionary time
scale is longer than the time scale on which stars are forming.  As a
result, the galaxy will produce many stars through this process and 
will contain $\sim 100$ hydrogen burning stars for cosmological 
decades $\eta > 14$. 

Notice that the time scale for producing stars through brown dwarf 
collisions is generally much shorter than the orbit decay time for brown 
dwarf binaries. For orbital decay, equation [3.8] implies that 
$\eta \sim 22.5$ + 4$\log_{10}(R/1 {\rm AU})$. Thus, brown dwarf 
collisions provide the dominant mechanism for continued star 
formation while the galaxy remains intact.  

\bigskip 
\noindent
{2. Collision Cross Sections} 
\medskip 

To complete this argument, we must estimate the cross section for
colliding brown dwarfs. Consider two brown dwarfs with a relative
velocity $v_{rel}$.  For simplicity, we consider the case of equal 
mass brown dwarfs with mass $m$. The orbital angular momentum of 
the system is given by 
$$J = m v_{rel} b \, , \eqno(3.13)$$ 
where $b$ is the impact parameter.  When the two dwarfs
collide and form a composite star of mass $\sim 2m$, the angular 
momentum can be written 
$$I \Omega = f (2m) R^2 \Omega \, , \eqno(3.14)$$ 
where $R$ is the stellar radius, $\Omega$ is the rotation 
rate, and $f$ is a numerical constant of order unity which 
depends on the internal structure of the star.  We next invoke 
the constraint that the rotation rate of the final state must be 
less than the break-up speed, i.e., 
$$\Omega^2 R^2 < {G (2m) \over R} \, . \eqno(3.15)$$ 
Combining the above results, we obtain a bound on the
impact parameter $b$ that can lead to a bound final system. 
We thus obtain 
$$b^2 < {8 f^2 G m R \over v_{rel}^2 } \, , \eqno(3.16)$$ 
which can be used to estimate the cross section, 
$$\sigma \approx \pi b^2 = 
{8 \pi f^2 G m R \over v_{rel}^2 } \, .  \eqno(3.17)$$ 
Using typical numerical values, we find that 
$b \sim R \sim 10^{10}$ cm, which is roughly comparable 
to the radius of the brown dwarf (e.g., Burrows et al., 1993). 

\bigskip 
\noindent
{3. Numerical Simulations and Other Results} 
\medskip 

In order to illustrate the viability of this collision process, 
we have done a set of numerical simulations using smooth particle 
hydrodynamics (SPH). We find that collisions between substellar
objects can indeed form final products with masses greater than the
minimum mass required to burn hydrogen.  Examples of such collisions
are shown in Figure 3.  In these simulations, density structures from
theoretical brown dwarf models (Laughlin \& Bodenheimer, 1993) are
delivered into impact with relative velocity 200 km/s. The
hydrodynamic evolutionary sequences shown are adiabatic. One expects
that the emergent stellar mass object will contract toward the main
sequence on a Kelvin-Helmholtz time scale and then initiate hydrogen
burning.

Finally, we note that white dwarfs will also collide in the galactic
halo.  As outlined in \S II.E, we expect roughly comparable numbers 
of white dwarfs and brown dwarfs at the end of the stelliferous era. 
Although the white dwarfs are actually smaller in radial size, they
are more massive and hence have a larger gravitational enhancement to
their interaction cross section.  As a result, the net cross section
and hence the net interaction rate of white dwarfs should be roughly
comparable to that of brown dwarfs (\S III.C.1).  When white dwarfs
collide with each other, several different final states are possible,
as we discuss below. 

If the two white dwarfs are sufficiently massive, it is possible that
the collision product will have a final mass which exceeds the
Chandrasekhar limit ($M_{Ch} \approx 1.4 M_\odot$) and hence can
explode in a supernova. Using the final mass function (see \S II.E and
Figure 2), we estimate that roughly one third of the white dwarfs will
have masses greater than $0.7 M_\odot$ and hence only about one tenth
of the collisions can possibly result in an object exceeding the
Chandrasekhar mass limit.  The supernova rate from these collisions
can thus be as large as $\Gamma_{SN} \sim 10^{-12}$ yr$^{-1}$,
although it will be somewhat smaller in practice due to
inefficiencies.

The most common type of collision is between two low mass white 
dwarfs -- the final mass function peaks at the mass scale 
$M_\ast \approx 0.13 M_\odot$. These low mass objects will 
have an almost pure helium composition.  If the final product 
of the collision has a mass larger than the minimum mass 
required for helium burning ($M_{\rm He} \approx 0.3 M_\odot$), 
then the product star could land on the helium main-sequence
(see, e.g., Kippenhahn \& Weigert, 1990). In order for the star
to burn helium, the collision must be sufficiently energetic 
to impart enough thermal energy into the star; otherwise, the 
star will become just another helium white dwarf. 
Another possibility exists for collisions between white dwarfs
of slightly larger masses. If the product of the collision 
has a mass smaller than the Chandrasekhar mass and larger than 
the minimum mass to burn carbon 
($0.9 M_\odot \le M_\ast \le 1.4 M_\odot$), the product star
could land on the carbon main sequence.
Thus, this mode of late time star formation can lead to an 
interesting variety of stellar objects. 

\bigskip 
\noindent{\bf D. The Black Hole Accretion Time} 
\nobreak 
\medskip 

Large black holes tend to accrete stars and gas
and thereby increase their mass. The black hole accretion time
is the characteristic time scale for a black hole in the center of a
galaxy to swallow the rest of the galaxy.  If we consider collisions
of the black hole with stars, and ignore the other processes discussed
above (gravitational radiation and stellar evaporation), the time for
the black hole to absorb the stars in the galaxy is given by
$$\tau = { V \over \sigma v} \, , \eqno(3.18)$$ 
where $V = R^3$ is the volume of the galaxy, $v$ is the typical speed
of objects in the galaxy ($v \sim$ 200 km/s), and $\sigma$ is the
effective cross section of the black hole.  As a starting point, 
we write the cross section in the form 
$$\sigma = \Lambda \pi R_S^2 \, , \eqno(3.19)$$  
where $\Lambda$ is a dimensionless enhancement factor due to 
gravitational focusing, and $R_S$ is the Schwarzschild 
radius $R_S$ given by 
$$R_S = G M / c^2 \, . \eqno(3.20)$$ 
We thus obtain the time scale 
$$\tau = 10^{30} {\rm yr} \Bigl[ M/10^6 M_\odot \Bigl]^{-2} 
\Bigl[ R/10 {\rm kpc} \Bigl]^3 \, \Lambda^{-1} 
\, , \eqno(3.21{\rm a})$$
or, equivalently, 
$$\eta_{\rm accrete} = 30 - 2 \log_{10} [ M/10^6 M_\odot ] + 
3 \log_{10} [R/10 {\rm kpc} ] - \log_{10} [ \Lambda ] 
\, . \eqno(3.21{\rm b})$$ 

The time scale $\eta_{\rm accrete} \sim 30$ is much longer than the
time scale for both stellar evaporation and gravitational radiation
(see also the following section).  As a consequence, at these late 
times, all the stars in a galaxy will either have evaporated into 
intergalactic space or will have fallen into the central black hole
via gravitational radiation decay of their orbits. Of course, as the 
black hole mass grows, the accretion time scale decreases. 
Very roughly, we expect $\sim 1-10 \%$ of the stars to fall to the
central black hole and the remainder to be evaporated; the final 
mass of the central black hole will thus be 
$M_{BH} \sim 10^9-10^{10}$ $M_\odot$.

One can also consider this process on the size scale of 
superclusters.  When $\eta \sim 30$, supercluster-sized cosmological
density perturbations of length $R$ will have long since evolved to
nonlinearity, and will be fully decoupled from the Hubble flow. One 
can imagine an ensemble of $\sim 10^9 - 10^{10}$ $M_\odot$ black holes 
which have descended from dead galaxies and are now roaming freely 
and hoovering up an occasional remaining star in the volume $R^{3}$. 
The characteristic time scale for this process is 
$$\eta_{\rm accrete} = 33 - 2 \log_{10} [ M/10^9 M_\odot ] 
+ 3 \log_{10} [R/10 {\rm Mpc} ] \, . \eqno(3.22)$$ 

As for the case of the galaxy, however, this straightforward scenario
is compromised by additional effects. Gravitational radiation will
continuously cause the orbits of the black holes to decay, and some 
of them may eventually merge. Stellar encounters with both other stars
and with the black holes will lead to stellar evaporation from the
supercluster sized system. Over the long term, one expects that the
supercluster will consist of a very large central black hole with a
majority of the stars and many of the original $\sim 10^9 - 10^{10}$ 
$M_\odot$ galactic black holes escaping to large distances. 
In other words, the supercluster-sized system will behave somewhat
analogously to the galaxy, except that it will contain a larger size
scale, a longer time scale, and two widely disparate mass scales
(namely, a stellar mass scale $M_\ast \sim 1 M_{\odot}$, and a black
hole mass scale $M_{BH} \sim 10^9 - 10^{10}$ $M_\odot$). 
Equipartition effects between the two mass scales will come into play,
and will drive the galactic black holes toward the center while
preferentially ejecting the stellar remnants. In principle, this
hierarchy can extend up to larger and larger perturbation length
scales, although the relevant time scales and detailed dynamics become
more uncertain as one proceeds with the extrapolation.

\bigskip
\noindent{\bf E. Annihilation and Capture of Halo Dark Matter} 
\medskip

Galactic halos consist largely of dark matter, much of which may
reside in non-baryonic form.  Although the nature and composition 
of this dark matter remains an important open question, one of the 
leading candidates is Weakly Interacting Massive Particles, usually
denoted as WIMPs. These particles are expected to have masses in the
range $M_W = 10 - 100$ GeV and interact through the weak force and
gravity only (cf. the reviews of Diehl et al., 1995; Jungman,
Kamionkowski \& Griest, 1996; see also the recent proposal of 
Kane \& Wells, 1996).  Many authors have studied the signatures of 
WIMP annihilation, usually with the hope of finding a detectable
signal.  One can apply the results of these studies to estimate 
the time scale for the depletion of WIMPs from a galactic halo. 

We first consider the case of direct particle-particle annihilation. 
Following usual conventions, the rate $\Gamma_W$ for WIMP 
annihilation in the halo can be written in the form  
$$\Gamma_W = n_W \langle \sigma v \rangle \, , \eqno(3.23)$$
where $n_W$ is the number density of WIMPs in the halo 
and $\langle \sigma v \rangle$ is the average value of the 
annihilation cross section times velocity.  If WIMPs make up 
a substantial mass fraction of the galactic halo, their number 
density is expected to be roughly $n_W \sim 1$ cm$^{-3}$. 
The typical velocity of particles in the galactic halo is 
$\sim$200 km/s. Using the most naive dimensional argument, 
we can estimate the interaction cross section as 
$$\sigma \sim M_W^2 G_F^2 \sim 5 \times 10^{-38} 
{\rm cm}^2 \, \Bigl[ {M_W \over 1 {\rm GeV} } \Bigr]^2 
\, , \eqno(3.24)$$
where $M_W$ is the mass of the particle and $G_F$ is the Fermi 
constant. The true cross section has additional factors which take 
into account spin dependences, mixing angles, and other model 
dependent quantities (see Diehl et al., 1995; Jungman et al., 1996); 
the form [3.24] is thus highly approximate, but adequate for our 
purposes.  We also note that the relic abundance of dark matter 
particles is determined by the interaction cross section; 
in order for the abundance to be cosmologically significant, 
the interaction cross section must be of order 
$\sigma \sim 10^{-37}$ cm$^2$ (see Kolb \& Turner, 1990). 

Putting all of the above results together, we can estimate the 
time scale $\tau_W$ for the population of WIMPs to change, 
$$\tau_W = \Gamma^{-1} = 
{1 \over n_W \langle \sigma v \rangle } \, \sim 
3 \times 10^{22} {\rm yr} \, . \eqno(3.25)$$
Thus, in terms of cosmological decades, we obtain the 
annihilation time scale in the form 
$$\eta_W = 22.5 - \log_{10} \Bigl[ {\langle \sigma v \rangle 
\over 10^{-30} \, {\rm cm}^3 \, {\rm s}^{-1} } \Bigr] - 
\log_{10} \Bigl[ { n_W \over 1 {\rm cm}^{-3} } \Bigr] 
\, . \eqno(3.26)$$
It takes a relatively long time for WIMPs to annihilate via 
direct collisions.  In particular, the annihilation time scale 
is much longer than the stellar evaporation time scale (\S III.A). 

Another important related effect is the capture of WIMPs by
astrophysical objects. The process of WIMP capture has been studied
for both the Sun (Press \& Spergel, 1985; Faulkner \& Gilliland, 1985)
and the Earth (Freese, 1986) as a means of helping to detect the dark
matter in the halo (see also Krauss, Srednicki, \& Wilczek, 1986; Gould,  
1987, 1991).  Although WIMP capture by the Sun and the Earth can be
important for dark matter detection, the lifetimes of both (main
sequence) stars and planets are generally too small for WIMP capture
to significantly affect the total population of particles in the
galactic halo.  On the other hand, stellar remnants, in particular
white dwarfs, can be sufficiently long lived to have 
important effects.

In astrophysical objects, WIMPs are captured by scattering 
off of nuclei.  When the scattering event leads to a final 
velocity of the WIMP that is less than the escape speed of 
the object, then the WIMP has been successfully captured. 
For the case of white dwarfs, we can make the following simple 
estimate of the capture process. The mean free path of a WIMP 
in matter with white dwarf densities is generally less than 
the radius of the star.  In addition, the escape speed from 
a white dwarf is large, roughly $\sim 3000$ km/s, which is much 
larger than the velocity dispersion of WIMPs in the halo.  
As a result, to first approximation, most WIMPs that pass 
through a white dwarf will be captured.  The WIMP capture 
rate $\Gamma_{W\ast}$ by a white dwarf is thus given by 
$$\Gamma_{W\ast} = n_W \sigma_{WD} v_{rel} \, , \eqno(3.27)$$
where $\sigma_{WD}$ $\sim 10^{18}$ cm$^2$ is the cross 
sectional area of the white dwarf and $v_{rel}$ $\sim$ 200 km/s 
is the relative velocity.  The capture rate is thus 
$$\Gamma_{W\ast} \sim 10^{25} {\rm s}^{-1} \, . \eqno(3.28)$$
With this capture rate, a white dwarf star can consume its 
weight in WIMPs on a time scale of $\sim 10^{24}$ yr. 
The total mass in WIMPs in the halo is expected to be a factor 
of 1--10 times the mass of stars, which will be mostly in the 
form of white dwarfs at these late times (\S II.E).  As a result, 
the time scale for white dwarfs to deplete the entire halo 
population of WIMPs via capture is roughly given by 
$$\tau \sim 10^{25} {\rm yr} \qquad {\rm or} 
\qquad \eta \sim 25 \, . \eqno(3.29)$$ 
The actual time scales will depend on the fraction of the 
galactic halo in non-baryonic form and on the properties 
(e.g., mass) of the particles; these quantities remain 
unknown at this time.

The annihilation of halo WIMPs has important consequences for both the
galaxy itself and for the white dwarfs. Basically, the galaxy as a
whole loses mass while the white dwarfs are kept hotter than they
would be otherwise.  The population of captured WIMPs inside the star
will build up to a critical density at which the WIMP annihilation
rate is in equilibrium with the WIMP capture rate (see, e.g., Jungman
et al., 1996).  Furthermore, most of the annihilation products will be
absorbed by the star, and the energy is eventually radiated away
(ultimately in photons).  The net result of this process (along with
direct annihilation) is thus to radiate away the mass of the galactic
halo on the time scales given by equations [3.26] and [3.29].  This
process competes with the evaporation of stars through dynamical
relaxation (\S III.A) and the decay of stellar orbits through
gravitational radiation (\S III.B). 

Since the time scale for WIMP evaporation is much longer than the
dynamical time scale, the galaxy will adiabatically expand as the halo
radiates away.  In the outer galaxy, the dark matter in the halo
dominates the gravitational potential well and hence the stars in the
outer galaxy will become unbound as the halo mass is radiated away.
Since WIMPs do not dominate the potential inside the solar circle, 
the corresponding effects on the inner galaxy are relatively weak. 

The white dwarf stars themselves will be kept hot by this WIMP 
capture process with a roughly constant luminosity given by 
$$L_{WD} = {\cal F} m_W \Gamma_{W\ast} = 
{\cal F} m_W n_W \sigma_{WD} v_{rel} \, 
\sim 4 \times 10^{-12} L_\odot \, , \eqno(3.30 )$$
where $\cal F$ is an efficiency factor (expected to be of order 
unity) which takes into account the loss of energy from the star 
in the form of neutrinos.  With this luminosity, the white dwarf 
has a surface temperature $T_\ast \approx$ 63 K, where we have 
assumed a typical white dwarf mass $M_\ast = 0.5 M_\odot$. 
As a reference point, we note that an entire galaxy of such stars
has a total luminosity comparable to that of the sun, 
$L_{\rm gal} \sim 1 L_\odot$.  However, most of the radiation 
will be emitted at infrared wavelengths, $\lambda \sim 50 \mu$m. 

For completeness, we note that axions provide another 
viable candidate for the dark matter in the galactic halo 
(see Chapter 10 of Kolb \& Turner, 1990). 
These particles arise from solutions to the 
strong CP problem in quantum chromodynamics
(see, e.g., Peccei \& Quinn, 1977ab; 
Weinberg, 1978; Wilczek, 1978). The coupling of the 
axion to the photon allows the axion to decay to 
a pair of photons with a lifetime $\tau_a$ given by 
$$\tau_a \approx 2 \times 10^{17} \, {\rm yr} \, 
\bigl( m_a / 1 {\rm eV} \bigr)^{-5} \, , \eqno(3.31)$$
where $m_a$ is the mass of the axion; we have assumed 
representative values for the remaining particle physics 
parameters. Relic axions with sufficient numbers to contribute 
to the dark matter budget of the universe have masses in the 
range $10^{-6}$ eV $< m_a < $ $10^{-3}$ eV, where the value 
depends on the production mechanism.  Using these mass values, 
we obtain an allowed range of axion decay time scales, 
$$32 \le \eta_a \le 47 . \eqno(3.32)$$

\bigskip 
\noindent{\bf F. The Fate of Planets during Galactic Death}
\nobreak  
\medskip  
\nobreak 

Planets can be loosely defined as objects that are small enough 
(in mass) to be supported by ordinary Coulomb forces rather than by
degeneracy pressure.  Over the long term, planets suffer from
several deleterious processes.  They can be vaporized by their evolving
parent stars, and their orbits can either decay or be disrupted.
Barring these more imminent catastrophes, planets will evaporate
as their protons decay (see \S IV.H). 

The theory of general relativity indicates that planetary orbits 
slowly decay via emission of gravitational radiation (see \S III.B). 
To fix ideas, consider a planet orbiting a star of mass $M_\ast$ 
at an initial orbital radius $R$. Gravitational radiation drives 
orbital decay on a time scale given by 
$$\tau = { 2 \pi R \over v} \Bigr( {v \over c} \Bigl)^{-5} \, = 
2.6 \times 10^{19} {\rm yr} \Bigl( {R \over 1 {\rm AU} } \Bigr)^4 
\Bigl( {M_\ast \over 1 M_\odot } \Bigr)^{-3} \, , \eqno(3.33)$$
or, in terms of cosmological decades, 
$$\eta = 19.4 + 4 \log_{10} [R/1 {\rm AU}] 
- 3 \log_{10} [M_\ast / 1 M_\odot] \, . \eqno(3.34)$$

In the interim, planets can be dislodged from their parent stars
during encounters and collisions with interloping stars. The time
scale for these dislocations is given by the time interval required 
to produce a sufficiently close encounter with another star. Very 
roughly, if a perturbing star intrudes within a given planet's orbit,
then the planet is likely to be entirely ejected from the system. 
This time scale is given by 
$$\tau = { 1 \over n \sigma v} \, , \eqno(3.35)$$
where $n$ is the number density of stars ($\sim 0.1$ pc$^{-3}$ 
in our galaxy today), $v$ is the relative velocity ($\sim 100$ km/s), 
and where the cross section $\sigma$ is determined by the orbital 
radius of the planet ($\sigma \approx \pi R^2$).  Inserting these 
values, one finds
$$\tau = 1.3 \times 10^{15} {\rm yr} \Bigl( {R \over 1 {\rm AU} } 
\Bigr)^{-2} \, , \eqno(3.36)$$
$$\eta = 15.1 - 2 \log_{10} [ R/1 {\rm AU} ] \, , \eqno(3.37)$$
where $R$ is the radius of the planetary orbit. 

Comparing equation [3.33] with equation [3.36], we find that the 
time scale for gravitational radiation is equal to that of stellar 
encounters for planetary orbits of radius $R$ = 0.2 AU, which is 
about half the radius of the orbit of Mercury in our own solar 
system.  One might guess then, that very close planets, such as the
recently discovered companion to 51 Pegasus (Mayor \& Queloz, 1995;
Marcy, Butler, \& Williams, 1996), will eventually merge with their
parent stars as a result of radiative orbital decay, while planets
with larger initial orbits (e.g., the giant planets in our solar 
system) will be stripped away from their parent stars 
as a consequence of stellar encounters. However, since the
time scale for stellar evolution ($\eta_\ast < 14$) is much shorter
than the time scale for orbital decay, close-in planets around
solar-type stars will be destroyed during the red giant phases long
before their orbits are constricted by general relativity. Only the
inner planets of low mass M dwarfs (which experience no giant phases)
will find their fate sealed by gravitational radiation. 

\newpage 
\bigskip 
\noindent{\bf IV. LONG TERM FATE OF DEGENERATE STELLAR OBJECTS} 
\medskip 

Brown dwarfs, white dwarfs, neutron stars, and black holes have
lifetimes which are not only much longer than the current age of the
universe ($\eta = 10$), but also greatly exceed the expected lifetime
of the galaxy ($\eta = 20 - 25$).  Due to a general lack of urgency,
the ultimate fate of these objects has not yet been extensively
considered.  Nevertheless, these objects will not live forever. 
If the proton is unstable, then proton decay will drive the long term
evolution of degenerate stellar objects.  Black holes are essentially
unaffected by proton decay, but they gradually dissipate via the
emission of Hawking radiation. Both proton decay and Hawking radiation
yield many interesting astrophysical consequences.  In the following
discussion, we work out the details of these processes (see also Dicus
et al., 1982; Feinberg, 1981).

\bigskip 
\noindent{\bf A. Proton Decay} 
\medskip 

In Grand Unified Theories (GUTs), the proton is unstable and has 
a finite, albeit quite long, lifetime.  For example, the proton can 
decay through the process 
$$p \to e^{+} \, + \, \pi^0 \, , \eqno(4.1)$$ 
and the Feynman diagram for this decay process is shown in Figure 4. 
Many different additional decay channels are possible and the 
details ultimately depend on the particular theory (e.g., see 
the reviews of Langacker, 1981; Perkins, 1984). In particular, we note 
that many other decay products are possible, including neutrinos.
If protons are unstable, then neutrons will also be unstable over
a commensurate time scale. Free neutrons are of course unstable 
to $\beta$-decay on a very short time scale ($\sim 10$ minutes); 
however, bound neutrons will be unstable through processes 
analogous to the proton decay modes (e.g., see Figure 4). 
In the present context, the protons and neutrons of interest
are bound in ``heavy'' nuclei (mostly carbon and helium) 
within white dwarfs. 

For the simplest class of GUTs, as illustrated by the decay 
modes shown in Figure 4, the rate of nucleon decay $\gp$ 
is roughly given by 
$$\gp = \alpha_5^2 {m_P^5 \over M_X^4 } \, , \eqno(4.2)$$ 
where $m_P$ is the proton mass and $\alpha_5$ is a dimensionless 
coupling parameter (see, e.g., Langacker, 1981; Perkins, 1984; 
Kane, 1993). The mass scale $M_X$ is the mass of the particle 
which mediates the baryon number violating process.  The decay 
rate should also include an extra numerical factor which takes 
into account the probability that the interacting quarks 
(which participate in the decay) are in the same place at the 
same time; this numerical factor is less than unity so that the 
proton lifetime is larger by a corresponding factor. To a first 
approximation, the time scale for proton decay is thus given by 
$$\tau_P \approx 10^{37} {\rm yr} 
\Bigl[ {M_X \over 10^{16} {\rm GeV} } \Bigr]^4 \, , \eqno(4.3)$$ 
where we have taken into account the aforementioned numerical 
probability factor. The corresponding cosmological time scale is 
$$\eta_P = 37 + 4 \log_{10} [M_X/10^{16} {\rm GeV}] 
\, . \eqno(4.4)$$ 
Notice that this time scale has a very sensitive dependence on 
the mass scale $M_X$ of the mediating boson. 

We want to find the allowed range for the proton lifetime. This time
scale is constrained from below by current experimental limits on the
lifetime of the proton (e.g., Perkins, 1984). The proton lifetime must
be greater than $\eta \sim 32$ ($10^{32}$ yr), where the exact
limit depends on the particular mode of proton decay (Particle Data
Group, 1994).  Finding an upper bound is more difficult.  If we
restrict our attention to the class of proton decay processes for
which equation [4.4] is valid, then we must find an upper bound on 
the mass scale $M_X$.  Following cosmological tradition, we expect 
the scale $M_X$ to be smaller than the Planck scale 
$\mpl \approx 10^{19}$ GeV, which implies the following 
range for the proton lifetime,
$$32 < \eta_P < 49 \, . \eqno(4.5)$$ 
The lower bound is set by experimental data; the upper bound 
is more suggestive than definitive (see also \S IV.F). 

We can find a more restrictive range for the proton lifetime for the 
special case in which the decay mode from some GUT is responsible 
for baryogenesis in the early universe. (Note that some baryon 
number violating process is necessary for baryogenesis to take 
place -- see Sakharov, 1967). 
Let us suppose that the decay mode from some GUT is
valid and that baryogenesis takes place at an energy scale in the
early universe $E_B \sim M_X$.  This energy scale must be less than
the energy scale $E_I$ of the inflationary epoch (Guth, 1981).  
The inflationary energy scale is constrained to be less than 
$\sim 10^{-2}$ $\mpl$ in order to avoid overproducing scalar 
density perturbations and gravitational radiation perturbations 
(Lyth, 1984; Hawking, 1985; Krauss \& White, 1992; Adams \& Freese, 1995). 
Combining these two constraints, we obtain the following suggestive 
range for the time scale for proton decay, 
$$32 < \eta_P < 41 \, . \eqno(4.6)$$ 
Although a range of nine orders of magnitude in the relevant 
time scale seems rather severe, the general tenor of the following 
discussion does not depend critically on the exact value.  
For the sake of definiteness, we adopt $\eta_P = 37$ as a 
representative time scale. 

\bigskip 
\noindent{\bf B. White Dwarfs Powered by Proton Decay} 
\medskip 

On a sufficiently long time scale, the evolution of a white dwarf is
driven by proton decay.  When a proton decays inside a star, most of
the primary decay products (e.g., pions and positrons) quickly
interact and/or decay themselves to produce photons.  For example, the
neutral pion $\pi^0$ decays into a pair of photons with a lifetime of
$\sim 10^{-16}$ sec; positrons, $e^+$, last only $\sim 10^{-15}$ sec
before annihilating with an electron and producing gamma rays.
Therefore, one common net result of proton decay in a star is the
eventual production of four photons through the effective reaction
$$p + e^{-} \to \gamma + \gamma + \gamma + \gamma \, , \eqno(4.7)$$ 
where the typical energy of the photons is given by $E_\gamma \sim
m_P/4 \sim$ 235 MeV.  These photons have a relatively short mean free
path within the star and will thermalize and diffuse outwards through
a random walk process with a characteristic time scale of $\sim 10^5$
yr, much shorter than the evolutionary time scale of
the system.  Additionally, some fraction of the decay products are in
the form of neutrinos, which immediately leave the system.

When proton decay is a white dwarf's primary energy source, 
the luminosity is
$$L_\ast (t) = {\cal F} N_0 E \gp \, {\rm e}^{-\gp t} 
\, \approx  {\cal F} M(t) \gp , \eqno(4.8)$$
where $N_0 \sim 10^{57}$ is the initial number of protons in 
the star, $E \sim$ 1 GeV is the net energy produced per decay, 
and $\gp$ is the decay rate.  The factor $\cal F$ is an 
efficiency parameter which takes into account the fraction 
of energy lost in the form of neutrinos.  Very roughly, we 
expect $\sim 1/3$ of the energy in the decay products to be 
in neutrinos and hence ${\cal F} \sim 2/3$ (e.g., Dicus et al., 1982). 
The exact value of the fraction ${\cal F}$ depends on the branching 
ratios for a particular GUT and hence is model dependent. For a 
typical decay rate of $\gp \sim 10^{-37}$ yr$^{-1}$, the 
luminosity in solar units becomes 
$$L_\ast \sim 10^{-24} L_\odot \, . \eqno(4.9)$$
It is perhaps more illuminating to express this stellar luminosity 
in ordinary terrestrial units.  A white dwarf fueled by proton decay 
generates approximately 400 Watts, enough power to run a few light 
bulbs, or, alternately, about 1/2 horsepower. An entire galaxy of 
such stars has a total luminosity of $L_{\rm gal} \sim 10^{-13} L_\odot$, 
which is much smaller than that of a single hydrogen burning star.

The total possible lifetime for a star powered by proton decay 
is given by 
$$\tau = {1 \over \gp } \, \ln [N_0/N_{\rm min} ] \, , \eqno(4.10)$$
where $N_0 \sim 10^{57}$ is the initial number of nucleons in 
the star and $N_{\rm min}$ is the minimum number of nucleons 
required to consider the object a star. If, for example, one takes 
the extreme case of $N_{\rm min}$ = 1, the time 
required for the star to completely disappear is 
is $t \approx 130 / \gp$; in general we obtain 
$$\eta_\ast = \eta_P + \log_{10} \bigl[ \ln (N_0/N_{\rm min})
\bigl] \, . \eqno(4.11)$$
As we show in \S IV.D, the object ceases to be a star 
when $N_{\rm min} \sim$ 10$^{48}$ and hence 
$\eta_\ast \approx \eta_P + 1.3$. 

During the proton decay phase, the stellar surface temperature is 
given by 
$$T_\ast^4 = {{\cal F} N_0 E \, \gp \, \over 4 \pi \sigma_B R_\ast^2} 
{\rm e}^{-\gp t} \, , \eqno(4.12)$$
where we have assumed that the spectral energy distribution 
is simply a blackbody ($\sigma_B$ is the Stefan-Boltzmann constant). 
For a $1 M_\odot$ star and the typical decay rate $\gp$, 
the effective stellar temperature is $T_\ast \sim 0.06$ K. 
This temperature will be enormously hotter than the temperature 
of the universe's background radiation at the cosmological decade 
$\eta = 37$.

As a white dwarf loses mass via proton decay, the star expands 
according to the usual mass/radius relation 
$$R_\ast M_\ast^{1/3} = 0.114 {h^2 \over G m_e m_P^{5/3} } 
(Z/A)^{5/3} \, , \eqno(4.13)$$
where $Z$ and $A$ are the atomic number and atomic weight of 
the white dwarf material (e.g., Chandrasekhar, 1939; Shu, 1982; 
Shapiro \& Teukolsky, 1983). For simplicity, we will take typical 
values and use $A = 2 Z$.  If we also rewrite the white dwarf 
mass/radius relation in terms of natural units, 
we obtain the relation 
$$R_\ast = 1.42 \Bigl( {\mpl \over m_e} \Bigr)
\Bigl( {\mpl \over m_P} \Bigr)
\Bigl( {M_\ast \over m_P} \Bigr)^{-1/3} 
m_P^{-1} \, . \eqno(4.14)$$

While the white dwarf is in the proton decay phase of its 
evolution, the star follows a well defined track in the 
H-R diagram, i.e., 
$$L_\ast = L_0 (T_\ast / T_0)^{12/5} \, , \eqno(4.15)$$
or, in terms of numerical values, 
$$L_\ast = 10^{-24} L_\odot \Bigl[ {T_\ast \over 0.06 {\rm K} }
\Bigr]^{12/5} \, . \eqno(4.16)$$
We note that the white dwarf mass/radius relation depends on the star's 
chemical composition, which changes as the nucleons decay (see the 
following section).  This effect will cause the evolutionary tracks 
to depart slightly from the 12/5 power-law derived above. However, 
this modification is small and will not be considered here. 

\bigskip
\noindent{\bf C. Chemical Evolution in White Dwarfs}
\nobreak
\medskip
\nobreak

Over the duration of the proton decay phase, the chemical composition
of a white dwarf is entirely altered.  Several different effects 
contribute to the change in chemical composition.  The nucleon decay 
process itself directly alters the types of nuclei in the star 
and drives the chemical composition toward nuclei of increasingly 
lower atomic numbers.   However, pycnonuclear reactions can occur 
on the relevant (long) time scales and build nuclei back up to 
higher atomic numbers.  In addition, spallation interactions 
remove protons and neutrons from nuclei; these free nucleons 
then interact with other nuclei and lead to further changes in 
composition. 

In the absence of pycnonuclear reactions and spallation, the chemical
evolution of a white dwarf is a simple cascade toward lower atomic
numbers.  As protons and neutrons decay, the remaining nuclei become 
correspondingly smaller.  Some of the nuclear products are radioactive
and will subsequently decay.  Given the long time scale for proton
decay, these radioactive nuclei are extremely short-lived.  As a 
result, only the stable isotopes remain. At relatively late times, 
when the total mass of the star has decreased by a substantial factor
(roughly a factor of ten as we show below), almost all of the nuclei
left in the star will be in the form of hydrogen.

At high densities and low temperatures, nuclear reactions
can still take place, although at a slow rate. The quantum
mechanical zero point energy of the nuclei allows them to
overcome the Coulomb repulsion and fuse.  In natural units,
the nuclear reaction rate can be written in the form
$$W = 4 \Bigl( {2 \over \pi^3} \Bigr)^{1/2} S \,
(Z^2 \alpha \mu)^{3/4} \, R_0^{-5/4} \, \exp \Bigl[
- 4 Z (\alpha \mu R_0)^{1/2} \Bigr] \, , \eqno(4.17)$$
where $\mu$ is the reduced mass of the nucleus, $R_0$ is the
average spacing between nuclei, and $\alpha$ is the fine
structure constant (see Shapiro \& Teukolsky, 1983). A slightly
different form for this reaction rate can be derived by including
anisotropic and electron screening effects (Salpeter \& Van Horn, 
1969), but the basic form is similar. The parameter $S(E)$ is a
slowly varying function of energy which takes into account the
probability of two nuclei interacting given that tunneling
has occurred.  Specifically, the parameter $S$ is related to
the cross section $\sigma (E)$ through the relation
$$\sigma(E) = {S(E) \over E} {\cal T} \, , \eqno(4.18)$$
where $\cal T$ is the tunneling transition probability.
The parameter $S$ can be determined either from direct
experiments or from theoretical calculations (see
Shapiro \& Teukolsky, 1983; Bahcall, 1989).

In order to evaluate the time scale for pycnonuclear
reactions to occur, one needs to determine the spacing
$R_0$ of the nuclei, or, equivalently, the number density
of particles. Using the white dwarf mass/radius relation,
we obtain the result
$$\mu R_0 = 2.29 A \Bigl( {\mpl \over m_e} \Bigr)
\Bigl( {\mpl \over m_P} \Bigr)
\Bigl( {M_\ast \over m_P} \Bigr)^{-2/3}
\approx 4060 A \, {m_\ast}^{-2/3} \, , \eqno(4.19)$$
where $A$ is average the atomic weight of the nuclei
and where we have defined $m_\ast \equiv M_\ast / M_\odot$.

We can now obtain a rough estimate for the efficiency
of pycnonuclear reactions building larger nuclei within
white dwarfs.  As a reference point, we note that for a
density of $\rho \sim 10^6$ g cm$^{-3}$, the time scale
for hydrogen to fuse into helium is $\sim 10^5$ yr
(e.g., Shapiro \& Teukolsky, 1983; Salpeter \& van Horn, 1969),
which is much shorter than the proton decay time scale.
However, the form of equation [4.17] shows that the
rate of nuclear reactions becomes highly suppressed
as the reacting nuclei become larger.  The exponential
suppression factor roughly has the form
$\sim \exp [-\beta Z A^{1/2} ]$, where the numerical factor
$\beta \approx 22$.  Thus, as the quantity $Z A^{1/2}$
increases, the rate of nuclear reactions decreases exponentially.
For example, if $Z=6$ and $A=12$ (for carbon), this exponential
term is a factor of $\sim 10^{-190}$ smaller than that for hydrogen.
Because of this large exponential suppression, fusion reactions will
generally not proceed beyond helium during the late time chemical 
evolution considered here.  Thus, the net effect of pycnonuclear
reactions is to maintain the decaying dwarf with a predominantly
helium composition down to a lower mass scale.

Spallation is another important process that affects the chemical
evolution of white dwarf stars during the epoch of proton decay.  
The high energy photons produced through proton decay can interact
with nuclei in the star. The most common result of such an interaction
is the emission of a single free neutron, but charged particles
(protons), additional neutrons, and gamma rays can also result (e.g.,
Hubbell, Gimm, \& Overbo, 1980). The free neutrons will be promptly
captured by other nuclei in a type of late time $s$-process 
(the $r$-process is of course dramatically irrelevant). The free
protons can produce heavier nuclei through pycnonuclear reactions, as
described above. Both of these mechanisms thus allow heavier elements
to build up in the star, albeit at a very slow rate and a very low
abundance.  Thus, the process of spallation initially produces 
free neutrons and protons; but these nucleons are incorporated 
into other nuclei.  As a result, the net effect of spallation 
is to remove nucleons from some nuclei and then give them back 
to other nuclei within the star. The result of this redistribution 
process is to widen the distribution of the atomic numbers (and 
atomic weights) for the nuclei in the star. 

In order to assess the importance of spallation processes, we must
consider the interaction cross section.  To leading order, the 
cross section for nuclear absorption of photons is a single 
``giant resonance'' with a peak at about 24 MeV for light nuclei 
and a width in the range $\Gamma$ = 3 -- 9 MeV.  The relative 
magnitude of this resonance feature is $\sim 20$ mb (see, e.g.,
Hubbell, Gimm, \& Overbo, 1980; Brune \& Schmidt, 1974), roughly 
a factor of 30 smaller than the total interaction cross section 
(which is dominated by scattering and pair production). 
For each proton decay event, $\sim 940$ MeV of matter is converted
into photons, with some neutrino losses. When these photons 
cascade downward in energy through the resonance regime (at 
$\sim 24$ MeV), there will be 20 -- 40 photons and about one 
in 30 will produce a spallation event.  Hence, on average, 
each proton decay event leads to approximately one spallation 
event.  

Spallation products allow the interesting possibility that a CNO 
cycle can be set up within the star.  The time scale for pycnonuclear
reactions between protons (produced by spallation) and carbon nuclei
is short compared to the proton decay time scale.  The time scale for
pycnonuclear reactions between protons and nitrogen nuclei is
comparable to the proton decay time scale.  Thus, in principle, the
white dwarf can set up a CNO cycle analogous to that operating in
upper-main-sequence stars (see Clayton, 1983; Kippenhahn \& Weigert,
1990; Shu, 1982).  The energy produced by this cycle will be small
compared to that produced by proton decay and hence this process does
not actually affect the luminosity of the star.  However, this cycle
will affect the chemical composition and evolution of the star.  As
usual, the net effect of the CNO cycle is to build four free
protons into a helium nucleus and to maintain an equilibrium abundance
of the intermediate nitrogen and oxygen nuclei.

In order to obtain some understanding of the chemical evolution of
white dwarfs, we have performed a simple numerical simulation of the
process.  Figure 5 shows the results of this calculation for a 1
$M_{\odot}$ white dwarf with an initial chemical composition of pure
carbon $^{12}$C.  The simulation assumes that radioactive isotopes
decay immediately as they are formed through the preferred decay
modes.  For each proton decay event, a spallation event also occurs
(see above) and leads to the removal of a nucleon from a random
nucleus; the spallation products are then assumed to fuse immediately
and randomly with other nuclei through the $s$-process and
pycnonuclear reactions.  The spallation process builds up a small
abundance of nuclei heavier than the original $^{12}$C, particularly
$^{13}$C which has a substantial mass fraction at ``early'' times.
The white dwarf evolves through successive phases in which smaller and
smaller nuclei are the dominant elements by mass fraction. The star
never builds up a significant lithium fraction due to the immediate
fission of newly formed $^{8}$Be into $\alpha$ particles.  The star
has a broad phase during which $^{4}$He dominates the composition.
When the white dwarf has lost about 60\% of its original mass, the
hydrogen mass fraction begins to predominate.

\bigskip 
\noindent{\bf D. Final Phases of White Dwarf Evolution} 
\nobreak 
\medskip 

In the final phases in the life of a white dwarf, the star has lost
most of its mass through proton decay.  When the mass of the star
becomes sufficiently small, two important effects emerge: The first
effect is that degeneracy is lifted and the star ceases to be a white
dwarf. The second effect is that the object becomes optically thin to
its internal radiation produced by proton decay and thus ceases to be
a star.  In the following discussion, we present simple estimates of
the mass scales at which these events occur.

When the star has lost enough of its initial mass to become
nondegenerate, most of the nucleons in the star will be in the form 
of hydrogen (see the previous section).  A cold star composed of 
pure hydrogen will generally have a thick envelope of molecular
hydrogen surrounding a degenerate core of atomic hydrogen.  As 
the stellar mass continues to decline through the process of proton
decay, the degenerate core becomes increasingly smaller and finally
disappears altogether.  This transition occurs when the degeneracy 
energy, the Coulomb energy, and the self-gravitational energy of 
the star are all comparable in magnitude; this event, in turn, 
occurs when the central pressure $P_C$ drops below a critical 
value of roughly a few Megabars ($P_C \sim 10^{12}$ dyne/cm$^2$).  
The central pressure in a star can be written in the form 
$$P_C = \beta {G M_\ast^2 \over R_\ast^4} \, , \eqno(4.20)$$
where $\beta$ is a dimensionless number of order unity. 
Using the white dwarf mass/radius relation in the form of 
equation [4.13] and setting $Z=A=1$, we find the central 
pressure as a function of stellar mass, 
$$P_C \approx {\beta \over 410} \, \mpl^{-10} \, m_e^4 \, 
m_P^{20/3} \, \, M_\ast^{10/3} \, , \eqno(4.21)$$
or, equivalently (in cgs units), 
$$P_C \approx 2 \times 10^{21} \, {\rm dyne/cm^2} \, 
\Bigl( {M_\ast \over 1 M_\odot} \Bigr)^{10/3} \, . \eqno(4.22)$$ 
Combining these results, we find that the mass scale 
$M_{\ast {\rm nd}}$ at which the star becomes nondegenerate 
is given by 
$$M_{\ast {\rm nd}} \approx 10^{-3} \, M_\odot \, . \eqno(4.23)$$
This mass scale is roughly the mass of a giant planet
such as Jupiter (for more detailed discussion of this issue, 
see also Hamada \& Salpeter, 1961; Shu, 1982). 
At this point in its evolution, the star has a radius 
$R_\ast \sim 0.1 R_\odot \sim 7 \times 10^9$ cm and 
a mean density of roughly $\rho \sim 1$ g/cm$^3$; these 
properties are also comparable to those of Jupiter. 
As a reference point, notice also that neutral hydrogen 
atoms packed into a cubic array with sides equal to one 
Bohr radius would give a density of 1.4 g/cm$^3$. 
At this transition, a star powered by proton decay has
luminosity $L_\ast \approx 10^{-27}$ $L_\odot$ and 
effective surface temperature $T_\ast \approx 0.0034$ K. 

Once the star becomes nondegenerate, it follows new track in the H-R
diagram.  The expressions for the luminosity and surface temperature
(see equations [4.8] and [4.12]) remain valid, but the mass/radius 
relation changes. Since the density of matter is determined by Coulomb 
forces for the small mass scales of interest, the density is roughly 
constant with a value $\rho_0 \sim 1$ g/cm$^3$.  We can thus use 
the simple relationship $M_\ast = 4 \pi \rho_0 R_\ast^3 /3$.  
Combining these results, we obtain the relation 
$$L_\ast = {36 \pi \over {\cal F}^2 } 
{\sigma_B^3 \over \Gamma_P^2 \rho_0^2} T_\ast^{12} 
\, , \eqno(4.24)$$ 
or, in terms of numerical values, 
$$L_\ast \approx 10^{-27} L_\odot 
\Bigl[ {T_\ast \over 0.0034 {\rm K} } \Bigr]^{12} \, . 
\eqno(4.25)$$
This steep power-law implies that the effective temperature 
of the star does not change very much during the final phases of 
evolution (the mass has to decrease by 12 orders of magnitude 
in order for the temperature to change by a factor of 10). 
As a result, effective surface temperatures of order 
$T_\ast \sim 10^{-3}$ K 
characterize the final phases of stellar evolution. 

As the star loses mass, it also becomes increasingly optically 
thin to radiation. As an object becomes transparent, it becomes 
difficult to meaningfully consider the remnant as a star. 
An object becomes optically thin when 
$$R_\ast n \sigma < 1 \, , \eqno(4.26)$$
where $n$ is the number density of targets and $\sigma$ 
is the cross section of interaction between the radiation 
field and the stellar material. In this present context, 
we must consider whether the star is optically thin to 
both the gamma rays produced by proton decay and also to 
the internal radiation at longer wavelengths characteristic 
of its bolometric surface temperature.  This latter condition 
is required for the radiation field to be thermalized. 

We first consider the conditions for which the star becomes 
optically thin to the gamma rays (with energies $E_\gamma$ 
$\sim$ 250 MeV) produced by proton decay. 
Since we are considering the interaction of gamma rays with 
matter, we can write the cross section in the form 
$$\sigma = C \sigma_T = C {8 \pi \over 3} 
{\alpha^2 \over m_e^2} \, , \eqno(4.27)$$ 
where $C$ is a dimensionless number (of order unity) 
and $\sigma_T$ is the Thompson cross section.  To a rough 
approximation, the density will be $\rho \sim 1$ g/cm$^3$ and 
hence the number density will have a roughly constant value 
$n \sim 10^{24}$ cm$^{-3}$.  Using these values, we find that 
the ``star'' will be safely optically thick to gamma rays 
provided its characteristic size is larger than about one 
meter. In other words, the object must be as big as a large 
rock.  These rocks will not, however, look very much like 
stars. At the extremely low bolometric temperatures 
characteristic of the stellar photospheres at these late times, 
the wavelength of the photospheric photons will be macroscopic 
and hence will interact much less strongly than the gamma rays. 
As a result, the spectral energy distribution of these objects will
suffer severe departures from blackbody spectral shapes.

In order to consider the optical depth of the star to its 
internal radiation field, we rewrite the condition [4.26] 
using the relation $n \sigma$ = $\rho \kappa$, where 
$\kappa$ is the opacity.  As derived above (equation [4.24]), 
the surface temperature is a slowly varying function in this 
final phase of evolution; as a result, the wavelength of photons 
in the stellar photosphere will be of order $\lambda \sim 100$ cm. 
The interaction of this radiation with the star depends on the 
chemical purity and the crystal-grain structure of the stellar 
material.  We can obtain 
a very rough estimate of the opacity by scaling from known 
astrophysical quantities. For interstellar graphite, for example, 
the opacity at $\lambda$ = 100 $\mu$m is roughly 
$\kappa \sim 1$ cm$^2$/g and scales with wavelength 
according to $\kappa \propto \lambda^{-2}$ (see Draine \& Lee, 1984). 
We thus estimate that the opacity in the outer layers of the 
star/rock will be $\kappa \sim 10^{-8}$ cm$^2$/g. 
Thus, in order for the star to be optically thick to its 
internal radiation, its radius must be $R_\ast > 10^8$ cm, 
which corresponds to a mass scale of 
$$M_{\ast {\rm thin}} \sim \, 10^{24} \, {\rm g} \, . \eqno(4.28)$$ 
All of these values should be regarded as highly approximate. 

>From these results, the ultimate future of white dwarfs, and indeed
our own sun, becomes clear: A white dwarf emerges from degeneracy 
as a pure sphere of hydrogen when the mass drops below 
$M_\ast \sim 10^{-3} M_\odot$.  Finally, the remaining object becomes
transparent to its own internal radiation when its mass dwindles to
$M_\ast \sim 10^{24}$ g, and at this point it is no longer a star.  
Stellar evolution thus effectively comes to an end. 

Just prior to the conclusion of stellar evolution, the white dwarf 
experiences about 2000 proton decay events per second and hence has 
a luminosity of $L_\ast \sim 10^{-33} L_\odot$ $\sim 4$ erg/s, 
and a temperature $T_\ast \sim 10^{-3}$ K. The time at which this 
transition occurs is given by $\tau \sim 21 \gp^{-1}$.  

Given these results, we can now describe the complete evolution of a
1.0 $M_\odot$ star (e.g., the Sun), from its birth to its death.  The
entire evolution of the such a star in the Hertzsprung-Russell diagram
is plotted in Figure 6.  The star first appears on the stellar
birthline (Stahler, 1988) and then follows a pre-main sequence track
onto the main sequence.  After exhausting its available hydrogen, the
star follows conventional post-main sequence evolution, including red
giant, horizontal branch, red supergiant, and planetary nebula
phases. The star then becomes a white dwarf with mass $M_\ast$
$\approx 0.5 M_\odot$ and cools along a constant radius track.  
The white dwarf spends many cosmological decades $\eta = 11 - 25$ 
near the center of the diagram ($L_\ast = 10^{14}$ W; $T_\ast$ = 63 K), 
where the star is powered by annihilation of WIMPs accreted from the
galactic halo.  When the supply of WIMPs is exhausted, the star cools
relatively quickly and obtains its luminosity from proton decay
($L_\ast \approx$ 400 W). The star then follows the evolutionary 
track in the lower right part of the diagram (with 
$L_\ast \sim T_\ast^{12/5}$) until mass loss from proton decay causes
the star to become nondegenerate. The star then becomes a rock-like 
object supported by Coulomb forces and follows a steeper track 
(with $L_\ast \sim T_\ast^{12}$) in the H-R diagram until it 
becomes optically thin.  At this point, the object ceases
to be a star and stellar evolution effectively comes to an end.  
During its entire lifetime, the Sun will span roughly 33 orders of
magnitude in luminosity, 9 orders of magnitude in mass, and 8 orders
of magnitude in surface temperature.

\bigskip
\noindent{\bf E. Neutron Stars Powered by Proton Decay} 
\medskip

The evolution of neutron stars powered by proton decay is
qualitatively similar to that of white dwarfs.  Since neutron stars
are (roughly) the same mass as white dwarfs, and since proton decay
occurs on the size scale of an individual nucleon, the luminosity of
the neutron star is given by equations [4.8] and [4.9].  To leading
order, the mass/radius relation for a neutron star is the same as that
of white dwarfs with the electron mass $m_e$ replaced by the neutron
mass (see equations [4.13] and [4.14]). Neutron stars are thus
$\sim$2000 times smaller than white dwarfs of the same mass, and have
appropriately warmer surface temperatures. Neutron stars undergoing
nucleon decay follow a track in the H-R diagram given by 
$$L_\ast = 10^{-24} L_\odot \Bigl[ {T_\ast \over 3 {\rm K} } 
\Bigr]^{12/5} \, . \eqno(4.29)$$ 

The final phases of the life of a neutron star will differ from the
case of a white dwarf. In particular, the neutrons in a neutron star
come out of degeneracy in a somewhat different manner than the
electrons in a white dwarf. Within a neutron star, the neutrons exist
and do not $\beta$-decay (into protons, electrons, and anti-neutrinos)
because of the extremely high densities, which are close to nuclear
densities in the stellar interior.  On the exterior, however, every
neutron star has a solid crust composed of ordinary matter.  As a
neutron star squanders its mass through nucleon decay, the radius
swells and the density decreases.  The outer layers of the star are
less dense than the central regions and hence the outer region will
experience $\beta$-decay first.  Thus, as the mass decreases, neutrons
in the outer portion of the star begin to $\beta$-decay into their
constituent particles and the star must readjust itself accordingly;
the net effect is that the crust of ordinary matter thickens steadily
and moves inwards towards the center.  Once the stellar mass 
decreases below a critical value $M_{C\ast}$, the crust reaches the
center of the star and the transition becomes complete. At this point,
the star will resemble a white dwarf more than a neutron star.

This process thus defines a minimum mass neutron star (see Shapiro 
\& Teukolsky, 1983), which is roughly characterized by the parameters  
$$M_{C\ast} = 0.0925 \, M_\odot \, , \qquad 
\rho_C = 1.55 \times 10^{14} \, {\rm g} \, {\rm cm}^{-3} \, , 
\qquad R_\ast = 164 \, {\rm km} \, , \eqno(4.30)$$ 
where $\rho_C$ is the central density of the star. It is hard to 
imagine current-day astrophysical processes which produce stellar 
objects near this limit.  
The transformation from a neutron star to a white dwarf 
occurs with a time scale given by 
$$\tau = {1 \over \Gamma_P} \ln [ M_0 / M_{C\ast} ] 
\approx {2.7 \over \Gamma_P} \, , \eqno(4.31)$$  
where $M_0 \approx 1.4 M_\odot$ is the initial mass of the 
neutron star.  Notice that neutron stars have a possible mass range of 
only a factor of $\sim15$, considerably smaller than the mass range 
available to white dwarfs. 

\bigskip 
\noindent{\bf F. Higher Order Proton Decay} 
\medskip 

Not all particle physics theories predict proton decay through the
process described above with decay rate $\gp$ (equation [4.2] and 
Figure 4).  In theories which do not allow proton decay through 
this first order process, the proton can often decay through second
order processes and/or through gravitational effects. By a second
order process, we mean an interaction involving two protons and/or
neutrons, i.e., $\Delta B$ = 2, where $B$ is the baryon number.  The
decay rate for these alternate decay channels is typically much
smaller than that discussed above.  In this section, we discuss the
decay rates and time scales for these higher order processes (see also
Feinberg, Goldhaber, \& Steigman, 1978; Wilczek \& Zee, 1979; Weinberg, 
1980; Mohapatra \& Marshak, 1980).

We first consider a class of theories which allow baryon number
violation, but do not have the proper vertices for direct proton 
decay ($\Delta B$ = 1).  In such theories, proton decay can 
sometimes take place through higher order processes ($\Delta B > 1$). 
For example, if the quarks in two nucleons interact as shown in 
Figure 7, the decay rate is roughly given by 
$$\Gamma_2 \sim \alpha_5^4 {m_P^9 \over M_X^8 } \, . \eqno(4.32)$$ 
Even for this higher order example, the theory must have the 
proper vertices for this process to occur. We note that some 
theories forbid this class of decay channels and require 
$\Delta B$ = 3 reactions in order for nucleon decay to take place 
(e.g., Goity \& Sher, 1995; Castano \& Martin, 1994).  
For the example shown in Figure 7, the decay rate is 
suppressed by a factor of $(m_P/M_X)^4$ $\sim$ $10^{64}$ 
relative to the simplest GUT decay channel. As a result, 
the time scale for proton decay through this 
second order process is roughly given by
$$\tau_{P2} \approx 10^{101} {\rm yr} 
\Bigl[ {M_X \over 10^{16} {\rm GeV} } \Bigr]^8 \, , \eqno(4.33)$$ 
and the corresponding cosmological time scale is 
$$\eta_{P2} = 101 + 8 \log_{10} [M_X/10^{16} {\rm GeV}] 
\, . \eqno(4.34)$$ 
In order for this decay process to take place, the protons involved 
must be near each other.  For the case of interest, the protons in 
white dwarfs are (mostly) in carbon nuclei and hence meet this 
requirement.  Similarly, the neutrons in a neutron star are all 
essentially at nuclear densities. Notice, however, that free protons 
in interstellar or intergalactic space will generally not decay 
through this channel. 

The proton can also decay through virtual black hole processes in 
quantum gravity theories (e.g., Zel'dovich, 1976; Hawking, Page, 
\& Pope, 1979; Page, 1980; Hawking, 1987). Unfortunately, the time scale 
associated with this process is not very well determined, but it is 
estimated to lie in the range 
$$10^{46} {\rm yr} < \tau_{PBH} < 10^{169} {\rm yr} \, , \eqno(4.35)$$ 
with the corresponding range of cosmological decades 
$$46 < \eta_{PBH} < 169 \, . \eqno(4.36)$$ 
Thus, within the (very large) uncertainty, this time scale for 
proton decay is commensurate with the second order
GUT processes discussed above. 

We note that many other possible modes of nucleon decay exist. For
example, supersymmetric theories can give rise to a double neutron
decay process of the form shown in Figure 8a (see Goity \& Sher, 
1995).  In this case, two neutrons decay into two neutral kaons.
Within the context of standard GUTs, decay channels involving higher
order diagrams can also occur.  As another example, the process shown
in Figure 8b involves three intermediate vector bosons and thus leads
to a proton lifetime approximately given by
$$\eta_{P3} = 165 + 12 \log_{10} [M_X/10^{16} {\rm GeV}] 
\, . \eqno(4.37)$$ 
Other final states are possible (e.g., three pions), although the 
time scales should be comparable.  This process (Figure 8b) 
involves only the most elementary baryon number violating processes, 
which allow interactions of the general form $q q \to q {\bar q}$.  
As a result, this decay mode is likely to occur even when the lower 
order channels are not allowed.  

Finally, we mention the case of sphalerons, which provide yet another
mechanism that can lead to baryon number violation and hence proton
decay.  The vacuum structure of the electroweak theory allows for the
non-conservation of baryon number; tunneling events between the
different vacuum states in the theory give rise to a change in baryon
number (for further details, see Rajaraman, 1987; Kolb \& Turner, 1990).
Because these events require quantum tunneling, the rate for this
process is exponentially suppressed at zero temperature by the large
factor $f$ = $\exp [4 \pi / \alpha_W] \sim 10^{172}$, where $\alpha_W$
is the fine structure constant for weak interactions.  In terms of
cosmological decades, the time scale for proton decay through this
process has the form $\eta_P = \eta_{0} + 172$, where $\eta_0$ is the
natural time scale (for no suppression). Using the light crossing time
of the proton to determine the natural time scale (i.e., we
optimistically take $\eta_0 = -31$), we obtain the crude estimate
$\eta_P \approx 141$.  Since this time scale is much longer than the 
current age of the universe, this mode of proton decay has not been
fully explored. In addition, this process has associated selection 
rules (e.g., 't Hooft, 1976) that place further limits on the possible 
events which exhibit nonconservation of baryon number. 
However, this mode of baryon number violation could
play a role in the far future of the universe.

To summarize this discussion, we stress that many different mechanisms 
for baryon number violation and proton decay can be realized within 
modern theories of particle physics.  As a result, it seems likely 
that the proton must eventually decay with a lifetime somewhere 
in the range 
$$32 < \eta_P < 200 \, , \eqno(4.38)$$   
where the upper bound was obtained by using 
$M_X \sim \mpl$ in equation [4.37]. 

To put these very long time scales in perspective, we note that the
total number $N_N$ of nucleons in the observable universe (at the
present epoch) is roughly $N_N \sim 10^{78}$.  Thus, for a decay 
time of $\eta = 100$, the expected number $N_D$ of nucleons that have 
decayed within our observable universe during its entire history 
is far less than unity, $N_D \sim 10^{-12}$.  The experimental 
difficulties involved in detecting higher order proton decay 
processes thus become clear.  

If the proton decays with a lifetime corresponding to $\eta$ 
$\sim 100 - 200$, the evolution of white dwarfs will be qualitatively
the same as the scenario outlined above, but with a few differences.
Since the evolutionary time scale is much longer, pycnonuclear
reactions will be much more effective at building the chemical
composition of the stars back up to nuclei of high atomic number.
Thus, stars with a given mass will have higher atomic numbers for
their constituent nuclei.  However, the nuclear reaction rate
(equation [4.17]) has an exponential sensitivity to the density. 
As the star loses mass and becomes less dense (according to the white
dwarf mass/radius relation [4.13, 4.14]), pycnonuclear reactions will
shut down rather abruptly.  If these nuclear reactions stop entirely,
the star would quickly become pure hydrogen and proton decay through a
two body process would be highly suppressed.  However, hydrogen tends
to form molecules at these extremely low temperatures. The
pycnonuclear reaction between the two protons in a hydrogen molecule
proceeds at a fixed rate which is independent of the ambient
conditions and has a time scale of roughly $\eta \approx 60$ 
(see Dyson, 1979, Shapiro \& Teukolsky, 1983, and \S III.C 
for simple estimates of pycnonuclear reaction rates). 
This reaction will thus convert the star into deuterium and helium 
on a time scale significantly shorter than that of higher order 
proton decay.  The resulting larger nuclei can then still decay 
through a second or third order process.  We also note that this 
same mechanism allows for hydrogen molecules in intergalactic 
space to undergo proton decay through a two body process. 

\bigskip  
\noindent{\bf G. Hawking Radiation and the Decay of Black Holes} 
\medskip 

Black holes cannot live forever; they evaporate on long time scales 
through a quantum mechanical tunneling process that produces photons 
and other products (Hawking, 1975).  In particular, black holes radiate 
a thermal spectrum of particles with an effective temperature given by 
$$T_{BH} = {1 \over 8 \pi G M_{BH} } \, , \eqno(4.39)$$
where $M_{BH}$ is the mass of the black hole. 
The total life time of the black hole thus becomes 
$$\tau_{BH} = {2560 \pi \over g_\ast} G^2 M_{BH}^3 \, , \eqno(4.40)$$ 
where $g_\ast$ determines the total number of effective 
degrees of freedom in the radiation field.  Inserting 
numerical values and scaling to a reference black hole 
mass of $10^6$ $M_\odot$, we find the time scale 
$$\tau_{BH} = 10^{83} {\rm yr} \, \Bigl[ M_{BH} / 10^6 M_\odot 
\Bigr]^3 \, , \eqno(4.41)$$
or, equivalently, 
$$\eta_{BH} = 83 + 3 \log_{10} [M_{BH} /10^6 M_\odot] 
\, . \eqno(4.42)$$
Thus, even a black hole with a mass comparable to a galaxy 
($M_{BH} \sim 10^{11} M_\odot$) will evaporate through this 
process on the time scale $\eta_{BH} \sim 98$.  One important 
consequence of this result is that for $\eta > 100$, a large 
fraction of the universe will be in the form of radiation, 
electrons, positrons, and other decay products.  

\bigskip 
\noindent{\bf H. Proton Decay in Planets} 
\medskip 

Planets will also eventually disintegrate through the process 
of proton decay.  Since nuclear reactions have a time scale 
($\eta \sim 1500$) much longer than that of proton decay and hence 
are unimportant (see Dyson, 1979), the chemical evolution of the 
planet is well described by a simple proton decay cascade scenario
(see \S IV.C).  In particular, this cascade will convert a planet
initially composed of iron into a hydrogen lattice in $\sim 6$ proton
half lives, or equivalently, on a time scale given by
$$\tau_{\rm planet} \approx {6 \ln 2 \over \gp } 
\approx 10^{38} {\rm yr} \, ; \qquad \eta_{\rm planet} 
\approx 38 . \eqno(4.43)$$ 
This time scale also represents the time at which the planet is 
effectively destroyed. 

During the epoch of proton decay, planets radiate energy 
with an effective luminosity given by 
$$L_{\rm planet} = {\cal F} M_{\rm planet} (t) \, 
\gp \approx 10^{-30} L_\odot \Bigl[ 
{M_{\rm planet} \over M_E } \Bigr] \, , \eqno(4.44)$$ 
where $M_E$ is the mass of the Earth and where we have 
used a proton decay lifetime of $10^{37}$ yr.  The efficiency 
factor $\cal F$ is expected to be of order unity. Thus, the 
luminosity corresponds to $\sim 0.4$ mW.

\newpage 
\bigskip 
\noindent{\bf V. LONG TERM EVOLUTION OF THE UNIVERSE} 
\medskip 
\nobreak 

In spite of the wealth of recent progress in our understanding 
of cosmology, the future evolution of the universe cannot be 
unambiguously predicted.  In particular, the geometry of the 
universe as a whole remains unspecified.  The universe can be 
closed ($k=+1$; $\Omega > 1$), flat ($k=0$; $\Omega=1$), or open
($k=-1$; $\Omega < 1$).  In addition, the contribution of vacuum
energy density remains uncertain and can have important implications
for the long term evolution of the universe.

\bigskip 
\noindent{\bf A. Future Expansion of a Closed Universe}
\medskip 

If the universe is closed, then the total lifetime of the universe,
from Big Bang to Big Crunch, can be relatively short in comparison 
with the characteristic time scales of many of the physical processes 
considered in this paper.  For a closed universe with density parameter 
$\Omega_0 > 1$, the total lifetime $\tau_U$ of the universe can be 
written in the form 
$$\tau_U = \Omega_0 (\Omega_0 - 1)^{-3/2} \pi H_0^{-1} 
\, , \eqno(5.1)$$ 
where $H_0$ is the present value of the Hubble parameter
(see, e.g., Peebles, 1993). 
Notice that, by definition, the age $\tau_U \to \infty$ as 
$\Omega_0 \to 1$. Current cosmological observations suggest that 
the Hubble constant is restricted to lie in the range 50 -- 100 
km s$^{-1}$ Mpc$^{-1}$ (e.g., Riess, Press, \& Kirshner, 1995), 
and hence the time scale $H_0^{-1}$ is restricted to be greater than 
$\sim 10$ Gyr. Additional observations (e.g., Loh \& Spillar, 1986) 
suggest that $\Omega_0 < 2$.  Using these results, we thus obtain 
a lower bound on the total lifetime of the universe, 
$$\tau_U > 20 \pi \, \, {\rm Gyr} \, . \eqno(5.2)$$
In terms of the time variable $\eta$, this limit takes the form 
$$\eta_U > 10.8 \, . \eqno(5.3)$$ 
This limit is not very strong -- if the universe is indeed 
closed, then there will be insufficient time to allow for many 
of the processes we describe in this paper. 

We also note that a closed universe model can in principle be
generalized to give rise to an oscillating universe. In this case, 
the Big Crunch occurring at the end of the universe is really a  
``Big Bounce'' and produces a new universe of the next generation.
This idea originated with ${\rm Lema{\hat i}tre}$ (1933) and has been
subsequently considered in many different contexts (from Tolman, 1934
to Peebles, 1993).

\bigskip 
\noindent
{\bf B. Density Fluctuations and the Expansion of a Flat or Open Universe}
\medskip 

The universe will either continue expanding forever or will collapse
back in on itself, but it is not commonly acknowledged that
observations are unable to provide a definitive answer to this important
question.  The goal of many present day astronomical observations is
to measure the density parameter $\Omega$, which is the ratio of the
density of the universe to that required to close the
universe. However, measurements of $\Omega$ do not necessarily
determine the long term fate of the universe.

Suppose, for example, that we can ultimately measure $\Omega$ to be
some value $\Omega_0$ (either less than or greater than unity).  This
value of $\Omega_0$ means that the density within the current horizon
volume has a given ratio to the critical density.  If we could view the
universe (today) on a much larger size scale (we can't because of
causality), then the mean density of the universe of that larger size
scale need not be the same as that which we measure within our horizon
today.  Let $\Omega_{\rm big}$ denote the ratio of the density of the
universe to the critical density on the aforementioned larger size
scale.  In particular, we could measure a value $\Omega_0 < 1$ and
have $\Omega_{\rm big} > 1$, or, alternately, we could measure 
$\Omega_0 > 1$ and have $\Omega_{\rm big} < 1$.  This possibility has 
been discussed at some length by Linde (1988, 1989, 1990). 

To fix ideas, consider the case in which the local value of the
density parameter is $\Omega_0 \approx 1$  and the larger scale value is
$\Omega_{\rm big}$ = 2 $>$ 1. (Note that $\Omega$ is not constant in time
and hence this value refers to the time when the larger scale enters
the horizon.) In other words, we live in an apparently flat universe, 
which is actually closed on a larger scale.  This state of affairs 
requires that our currently observable universe lies within
a large scale density fluctuation of amplitude 
$${\Delta \rho \over \rho} = 
{\Omega_0 - \Omega_{\rm big} \over \Omega_{\rm big}} 
= - {1 \over 2} \, , \eqno(5.4)$$
where the minus sign indicates that we live in a locally underdense
region. Thus, a density perturbation with amplitude of order unity 
is required; furthermore, as we discuss below, the size scale of 
the perturbation must greatly exceed the current horizon size. 

On size scales comparable to that of our current horizon, 
density fluctuations are constrained to be quite small 
($\Delta \rho / \rho \sim 10^{-5}$) because of measurements of
temperature fluctuations in the cosmic microwave background radiation
(Smoot et al., 1992; Wright et al., 1992).  On smaller size scales,
additional measurements indicate that density fluctuations 
are similarly small in amplitude (e.g., Meyer, Cheng, \& Page, 1991; 
Gaier et al., 1992; Schuster et al., 1993). The microwave background 
also constrains density fluctuations on scales {\it larger than the 
horizon} (e.g., Grischuk \& Zel'dovich, 1978), although the sensitivity
of the constraint decreases with increasing size scale $\lambda$
according to the relation $\sim (\lambda_{\rm hor} / \lambda)^2$, 
where $\lambda_{\rm hor}$ is the horizon size.  Given that density 
fluctuations have amplitudes of roughly $\sim 10^{-5}$ on the size
scale of the horizon today, the smallest size scale $\lambda_1$ for 
which fluctuations can be of order unity is estimated to be 
$$\lambda_1 \sim 300 \lambda_{\rm hor} \, \approx 10^6 \, 
{\rm Mpc} \, . \eqno(5.5)$$ 
For a locally flat universe ($\Omega_0 \approx 1$), density 
fluctuations with this size scale will enter the horizon at a time 
$t_1 \approx 3 \times 10^7 t_0 \approx 3 \times 10^{17}$ yr, 
or, equivalently, at the cosmological decade 
$$\eta_1 \approx 17.5 \, . \eqno(5.6)$$ 
This time scale represents a lower bound on the (final) age of the 
universe if the present geometry is spatially flat. In practice, 
the newly closed universe will require some additional time to 
re-collapse (see equation [5.1]) and hence the lower bound on 
the total age becomes approximately $\eta > 18$. 

The situation is somewhat different for the case of an open universe
with $\Omega_0 < 1$.  If the universe is open, then the expansion
velocity will (relatively) quickly approach the speed of light, i.e.,
the scale factor will expand according to $R \propto t$ (for this
discussion, we do not include the possibility that $\Omega_0$ = $1 -
\epsilon$, where $\epsilon \ll 1$, i.e., we consider only manifestly
open cases).  In this limit, the (comoving) particle horizon expands
logarithmically with time and hence continues to grow.  However, the
speed of light sphere -- the distance out to which particles in the
universe are receding at the speed of light -- approaches a constant
in comoving coordinates. As a result, density perturbations on very
large scales will remain effectively ``frozen out'' and are thus
prevented from further growth as long as the universe remains open.
Because the comoving horizon continues to grow, albeit quite slowly,
the possibility remains for the universe to become closed at some
future time.  The logarithmic growth of the horizon implies that the
time scale for the universe to become closed depends exponentially on
the size scale $\lambda_1$ for which density perturbations are of
order unity. The resulting time scale is quite long ($\eta \gg 100$),
even compared to the time scales considered in this paper.

To summarize, if the universe currently has a nearly flat spatial 
geometry, then microwave background constraints imply a lower bound 
on the total age of universe, $\eta > 18$. The evolution of the universe 
at later times depends on the spectrum of density perturbations.  
If large amplitude perturbations ($\Delta \rho/\rho > 1$) enter 
the horizon at late times, then the universe could end in a big 
crunch at some time $\eta > \eta_1 = 17.5$. On the other hand, if 
the very large scale density perturbations have small amplitude 
($\Delta \rho/\rho \ll 1$), then the universe can continue to 
expand for much longer time scales.  If the universe is currently 
open, then large scale density perturbations are essentially 
frozen out. 

\bigskip 
\noindent{\bf C. Inflation and the Future of the Universe} 
\medskip 

The inflationary universe scenario was originally invented (Guth, 1981)
to solve the horizon problem and the flatness problem faced by
standard Big Bang cosmology (see also Albrecht \& Steinhardt, 1982;
Linde, 1982).  The problem of magnetic monopoles was also a motivation,
but will not be discussed here.  In addition, inflationary models
which utilize ``slowly rolling'' scalar fields can produce density
fluctuations which later grow into the galaxies, clusters, and
super-clusters that we see today (e.g., Bardeen, Steinhardt, \& 
Turner, 1983; Starobinsky, 1982; Guth \& Pi, 1982; Hawking, 1982). 

During the inflationary epoch, the scale factor of the universe grows
superluminally (usually exponentially with time).  During this period
of rapid expansion, a small causally connected region of the universe
inflates to become large enough to contain the presently observable
universe.  As a result, the observed homogeneity and isotropy of the
universe can be explained, as well as the observed flatness.  In order
to achieve this resolution of the horizon and flatness problems, 
the scale factor of the universe must inflate by a factor of 
${\rm e}^{N_I}$, where the number of e-foldings $N_I \sim 60$.  
At the end of this period of rapid expansion, the universe must be
re-thermalized in order to become radiation dominated and recover 
the successes of standard Big Bang theory. 

Since the conception of inflation, many models have been produced and
many treatments of the requirements for sufficient inflation have been
given (e.g., Steinhardt \& Turner, 1984; Kolb \& Turner, 1990; Linde, 
1990).  These constraints are generally written in terms of explaining
the flatness and causality of the universe at the present epoch.
However, it is possible, or even quite likely, that inflation will
solve the horizon and flatness problems far into the future.  In this
discussion, we find the number $N_I$ of inflationary e-foldings
required to solve the horizon and flatness problems until a future
cosmological decade $\eta$.

Since the number of e-foldings required to solve the flatness 
problem is (usually) almost the same as that required to solve 
the horizon problem, it is sufficient to consider only the latter
(for further discussion of this issue, see, e.g., Kolb \& Turner, 
1990; Linde, 1990).  The condition for sufficient inflation can 
be written in the form 
$${1 \over (H R)_\eta } < {1 \over (H R)_B} \, , \eqno(5.7)$$
where the left hand side of the inequality refers to the inverse of the
product of the Hubble
parameter and the scale factor evaluated at the future cosmological 
decade $\eta$ and the right hand side refers to the same quantity
evaluated at the beginning of the inflationary epoch.

The Hubble parameter at the beginning of inflation takes
the form 
$$H_B^2 = {8 \pi \over 3} {M_I^4 \over M_{\rm Pl}^2} 
\, , \eqno(5.8)$$
where $M_I$ is the energy scale at the start of inflation 
(typically, the energy scale $M_I \sim 10^{16}$ GeV, which 
corresponds to cosmological decade $\eta_I \sim -44.5$). 
Similarly, the Hubble parameter at some future time $\eta$ 
can be written in the form 
$$H_\eta^2 = {8 \pi \over 3} {M_\eta^4 \over M_{\rm Pl}^2} 
\, , \eqno(5.9)$$
where the energy scale $M_\eta$ is defined by 
$$\rho(\eta) \equiv M_\eta^4 = \rho_0 R_\eta^{-3} \, . \eqno(5.10)$$
In the second equality, we have written the energy density in 
terms of its value $\rho_0$ at the present epoch and we assume 
that the universe remains matter dominated. 
We also assume that the evolution of the universe is 
essentially adiabatic from the end of inflation
(scale factor $R_{end}$) until the future epoch of interest 
(scale factor $R_\eta$), i.e., 
$${R_{end} \over R_\eta} = {T_\eta \over f M_I} \, , \eqno(5.11)$$
where $T_\eta = T_0/R_\eta$ is the CMB temperature at time $\eta$
and $T_0 \approx$ 2.7 K is the CMB temperature today. The quantity 
$f M_I$ is the CMB temperature at the end of inflation, after 
thermalization, and we have introduced the dimensionless 
factor $f < 1$. 

Combining all of the above results, we obtain the following 
constraint for sufficient inflation, 
$${\rm e}^{N_I} = {R_{end} \over R_B} > 
{M_I \, T_0 \, R_\eta^{1/2} \over f \sqrt{\rho_0} } 
\, . \eqno(5.12)$$
Next, we write the present day energy density $\rho_0$ in 
terms of the present day CMB temperature $T_0$, 
$$\rho_0 = \beta^2 T_0^4 \, , \eqno(5.13)$$
where $\beta \approx 100$.  The number of e-foldings 
is thus given by 
$$N_I = \ln [R_{end} / R_B ] = \ln [M_I/ \beta T_0] + 
{1 \over 2} \ln R_\eta - \ln f \, . \eqno(5.14{\rm a})$$ 
Inserting numerical values and using the definition [1.1] of 
cosmological decades, we can write this constraint in the form 
$$N_I \approx 61 + \ln \bigl[M_I/ (10^{16} {\rm GeV}) \bigr] 
+ {1 \over 3} (\eta - 10) \, \ln 10 \, . \eqno(5.14{\rm b})$$
For example, in order to have enough inflation for the universe 
to be smooth and flat up to the cosmological decade $\eta = 100$, 
we require $N_I \approx 130$ e-foldings of inflation. This value 
is not unreasonable in that $N_I$ = 130 is just as natural from 
the point of view of particle physics as the $N_I$ = 61 value 
required by standard inflation. 

We must also consider the density perturbations produced 
by inflation. All known models of inflation produce density
fluctuations and most models predict that the amplitudes 
are given by 
$${\Delta \rho \over \rho} \approx {1 \over 10} 
{H^2 \over {\dot \Phi} } \, , \eqno(5.15)$$
where $H$ is the Hubble parameter and $\Phi$ is the scalar 
field responsible for inflation (Starobinsky, 1982; Guth \& 
Pi, 1982; Hawking, 1982; Bardeen, Steinhardt, \& Turner, 1983). 
In models of inflation with more than one scalar field
(e.g., La \& Steinhardt, 1989; Adams \& Freese, 1991), the 
additional fields can also produce density fluctuations in 
accordance with equation [5.15].

In order for these density fluctuations to be sufficiently small, 
as required by measurements of the cosmic microwave background, 
the potential $V(\Phi)$ for the inflation field must be very 
flat. This statement can be quantified by defining a 
``fine-tuning parameter'' $\lambda_{FT}$ through the relation 
$$\lambda_{FT} \equiv {\Delta V \over (\Delta \Phi)^4 } \, , 
\eqno(5.16)$$
where $\Delta V$ is the change in the potential during 
a given portion of the inflationary epoch and $\Delta \Phi$ 
is the change in the scalar field over the same period
(Adams, Freese, \& Guth, 1991).  The parameter $\lambda_{FT}$ 
is constrained to less than $\sim 10^{-8}$ for all models 
of inflation of this class and is typically much smaller, 
$\lambda_{FT} \sim 10^{-12}$, for specific models. 
The required smallness of this parameter places tight 
constraints on models of inflation. 

The aforementioned constraints were derived by demanding that the
density fluctuations (equation [5.15]) are sufficiently small in
amplitude over the size scales of current cosmological interest, 
i.e., from the horizon size (today) down to the size scale of
galaxies.  These density perturbations are generated over 
$N_\delta \approx 8$ e-foldings during the inflationary epoch.
However, as discussed in \S V.B, large amplitude density fluctuations
can come across the horizon in the future and effectively close the
universe (see also Linde, 1988, 1989, 1990).  In order for the universe
to survive (not become closed) up until some future cosmological
decade $\eta$, density fluctuations must be small in amplitude for 
all size scales up to the horizon size at time $\eta$ (within an 
order of magnitude -- see equation [5.1]). As a result, inflation 
must produce small amplitude density fluctuations over many
more e-foldings of the inflationary epoch, namely 
$$N_\delta \approx 8 + 
{1 \over 3} (\eta - 10) \ln 10 \, , \eqno(5.17)$$
where $\eta$ is the future cosmological decade of interest.  
For example, for $\eta = 100$ we would require $N_\delta \approx 77$.  
Although this larger value of $N_\delta$ places a tighter bound 
on the fine-tuning parameter $\lambda_{FT}$, and hence a tighter 
constraint on the inflationary potential, such bounds can be 
accommodated by inflationary models (see Adams, Freese, \& 
Guth, 1991 for further discussion).  Loosely speaking, once 
the potential is flat over the usual $N_\delta = 8$ e-foldings 
required for standard inflationary models, it is not that 
difficult to make it flat for $N_\delta = 80$. 

\bigskip 
\noindent{\bf D. Background Radiation Fields} 
\medskip 

Many of the processes discussed in this paper will produce background
radiation fields, which can be important components of the universe
(see, e.g., Bond, Carr, \& Hogan, 1991 for a discussion of present day
backgrounds).  Stars produce radiation fields and low mass stars will
continue to shine for several more cosmological decades (\S II).  The
net effect of WIMP capture and annihilation in white dwarfs (\S III.E)
will be to convert a substantial portion of the mass energy of
galactic halos into radiation.  Similarly, the net effect of proton
decay (\S IV) will convert the mass energy of the baryons in the
universe into radiation.  Finally, black holes will evaporate as well,
(\S IV.H), ultimately converting their rest mass into radiation
fields.  As we show below, each of these radiation fields will
dominate the radiation background of the universe for a range of
cosmological decades, before being successively redshifted to
insignificance.

The overall evolution of a radiation field in an expanding universe 
can be described by the simple differential equation, 
$${d \rho_{\rm rad} \over dt} + 4 \, { {\dot R} \over R} \, 
\rho_{\rm rad} = S(t) \, , \eqno(5.18)$$ 
where $\rho_{\rm rad}$ is the energy density of the radiation field 
and $S(t)$ is a source term (see, e.g., Kolb \& Turner, 1990). 

Low mass stars will continue to shine far into the future. The 
source term for this stellar radiation can be written in the form 
$$S_\ast (t) = n_\ast L_\ast = \epsilon_\ast \Omega_\ast \, 
\rho_0 R^{-3} \, {1 \over t_\ast} \, , \eqno(5.19)$$ 
where $L_\ast$ and $n_\ast$ are the luminosity and number density 
of the low mass stars.  In the second equality, we have introduced 
the present day mass fraction of low mass stars $\Omega_\ast$, 
the nuclear burning efficiency $\epsilon_\ast \sim 0.007$, 
the effective stellar lifetime $t_\ast$, and the present day energy 
density of the universe $\rho_0$.  For this example, we have written 
these expressions for a population of stars with only a single mass; 
in general, one should of course consider a distribution of stellar 
masses and then integrate over the distribution. As a further 
refinement, one could also include the time dependence of the 
stellar luminosity $L_\ast$ (see \S II). 

For a given geometry of the universe, we find the solution for 
the background radiation field from low mass stars, 
$$\rho_{{\rm rad}\ast} = \epsilon_\ast \, \Omega_\ast \, 
\rho(R) \, f \, {t \over t_\ast} \, , \eqno(5.20)$$ 
where the dimensionless factor $f=1/2$ for an open universe 
and $f=3/5$ for a flat universe.  This form is valid until the 
stars burn out at time $t = t_\ast$. 
After that time, the radiation field simply redshifts in 
the usual manner, $\rho_{{\rm rad}\ast} \sim R^{-4}$. 

For the case of WIMP annihilation in white dwarfs, the source 
term is given by 
$$S_W (t) = L_\ast n_\ast = \Omega_W \rho_0 R^{-3} \Gamma 
\, , \eqno(5.21)$$ 
where $L_\ast$ and $n_\ast$ are the luminosity and number density 
of the white dwarfs. In the second equality, we have written the 
source in terms of the energy density in WIMPs, where $\Omega_{W}$ 
is the present day mass fraction of WIMPs and $\Gamma$ is the 
effective annihilation rate.  The solution for the background 
radiation field from WIMP annihilation can be found, 
$$\rho_{\rm wrb} (t) = f \, \Omega_{W} \, \rho(R) \, 
\Gamma \, t  \,  , \eqno(5.22)$$ 
where the dimensionless factor $f$ is defined above. 
This form is valid until the galactic halos begin to run out of WIMP
dark matter at time $t \sim \Gamma^{-1} \sim 10^{25}$ yr, or until 
the galactic halo ejects nearly all of its white dwarfs. 
We note that direct annihilation of dark matter will also 
contribute to the background radiation field of the universe. 
However, this radiation will be highly nonthermal; the annihilation 
products will include gamma rays with characteristic 
energy $E_\gamma \sim 1$ GeV. 

For the case of proton decay, the effective source term for the 
resulting radiation field can be written  
$$S_P (t) = {\cal F} \, \Omega_B \, \rho_0 \, R^{-3} \, 
\gp \, {\rm e}^{-\gp t} \, , \eqno(5.23)$$ 
where $\Omega_B$ is the present day contribution of baryons to the 
total energy density $\rho_0$, $\gp$ is the proton decay rate, and 
$\cal F$ is an efficiency factor of order unity.  For a given 
geometry of the universe, we obtain the solution for the background 
radiation field from proton decay, 
$$\rho_{\rm prb} (t) = {\cal F} \, \Omega_B \, \rho(R) \, 
F(\xi) \, , \eqno(5.24)$$ 
where $F(\xi)$ is a dimensionless function of the 
dimensionless time variable $\xi \equiv \gp t$. 
For an open universe, 
$$F(\xi) = {1 - (1 + \xi) {\rm e}^{-\xi} \over \xi } 
\, , \eqno(5.25)$$ 
whereas for a flat universe, 
$$F(\xi) = \xi^{-2/3} \, \int_0^\xi \, 
x^{2/3} {\rm e}^{-x} \, dx \, \, 
= \xi^{-2/3} \, \gamma(5/3, \xi) \, , \eqno(5.26)$$ 
where $\gamma(5/3, \xi)$ is the incomplete gamma function 
(Abramowitz \& Stegun, 1972). 

For black hole evaporation, the calculation of the radiation 
field is more complicated because the result depends on the 
mass distribution of black holes in the universe.  For simplicity, 
we will consider a population of black holes with a single mass 
$M$ and mass fraction $\Omega_{BH}$ (scaled to the present epoch). 
The source term for black hole evaporation can be written 
in the form 
$$S_{BH} (t) = \Omega_{BH} \, \rho_0 \, R^{-3} \, 
{1 \over 3 \tau_{BH}} \, {1 \over 1 - t/\tau_{BH} } 
\, , \eqno(5.27)$$ 
where $\tau_{BH}$ is the total lifetime of a black hole 
of the given mass $M$ (see equation [4.37]).  For an open 
universe, we obtain the solution for the background radiation 
field from black hole evaporation 
$$\rho_{\rm bhr} (t) = \Omega_{BH} \, \rho(R) \, 
F(\xi) \, , \eqno(5.28)$$  
where the dimensionless time variable $\xi = t/\tau_{BH}$. 
For an open universe, the dimensionless function $F(\xi)$ 
is given by 
$$F(\xi) = {1 \over 3 \xi} \, \Bigl\{ \ln \bigl[ 
{1 \over 1 - \xi} \bigr] - \xi \Bigr\} \, , \eqno(5.29)$$ 
whereas for a flat universe, 
$$F(\xi) = {1 \over 3 \xi^{2/3} } \, \int_0^\xi \, 
{x^{2/3} \, dx \over 1 - x } \, . \eqno(5.30)$$ 

Each of the four radiation fields discussed here has the 
same general time dependence.  For times short compared to 
the depletion times, the radiation fields have the form 
$$\rho(t) \approx \Omega_X \, \rho(R) \, 
\Gamma_X \, t \, , \eqno(5.31)$$ 
where $\Omega_X$ is the present day abundance of the raw material 
and $\Gamma_X$ is the effective decay rate (notice that we have 
neglected dimensionless factors of order unity). After the sources 
(stars, WIMPs, protons, black holes) have been successively exhausted, 
the remaining radiation fields simply redshift away, i.e., 
$$\rho(t) = \rho (t_{\rm end}) \, 
(R/R_{\rm end})^{-4} \, , \eqno(5.32)$$ 
where the subscript refers to the end of the time 
period during which the ambient radiation was produced. 

Due to the gross mismatch in the characteristic time scales, 
each of the radiation fields will provide the dominate contribution 
to the radiation content of the universe over a given time period.
This trend is illustrated in Figure 9, which shows the relative
contribution of each radiation field as a function of cosmological
time $\eta$.  For purposes of illustration, we have assumed an open
universe and the following source abundances: low mass stars
$\Omega_\ast$ = $10^{-3}$, weakly interacting massive particles
$\Omega_W$ = 0.2, baryons $\Omega_B$ = 0.05, and black holes
$\Omega_{BH}$ = 0.1.  At present, the cosmic microwave background
(left over from the big bang itself) provides the dominant radiation
component.  The radiation field from star light will dominate the
background for the next several cosmological decades.  At cosmological
decade $\eta \sim 16$, the radiation field resulting from WIMP
annihilation will overtake the starlight background and become the
dominant component.  At the cosmological decade $\eta \sim 30$, the
WIMP annihilation radiation field will have redshifted away and the
radiation field from proton decay will begin to dominate.  At much
longer time scales, $\eta \sim 60$, the radiation field from black
hole evaporation provides the dominant contribution (where we have 
used $10^6$ $M_\odot$ black holes for this example). 

The discussion thus far has focused on the total energy density 
$\rho_{\rm rad}$ of the background radiation fields.  One can 
also determine the spectrum of the background fields as a function 
of cosmological time, i.e., one could follow the time evolution of 
the radiation energy density per unit frequency.  In general, the 
spectra of the background radiation fields will be non-thermal for 
two reasons: 

\item{[1]} The source terms are not necessarily perfect blackbodies.
The stars and black holes themselves produce nearly thermal spectra,
but objects of different masses will radiate like blackbodies of
different temperatures.  One must therefore integrate over the mass
distribution of the source population.  It is interesting that this
statement applies to all of the above sources.  For the first three
sources (low mass stars, white dwarfs radiating WIMP annihilation
products, and white dwarfs powered by proton decay), the mass 
distribution is not very wide and the resulting composite spectrum 
is close to that of a blackbody.  For the case of black holes, 
the spectrum is potentially much wider, but the mass distribution 
is far more uncertain. 

\item{[2]} The expansion of the universe redshifts the radiation 
field as it is produced and thereby makes the resultant spectrum 
wider than a thermal distribution.  However, due to the linear
time dependence of the emission (equation [5.31]), most of the 
radiation is emitted in the final cosmological decade of the 
source's life.  The redshift effect is thus not as large as 
one might naively think. 

\noindent 
To summarize, the radiation fields will experience departures 
from a purely thermal distribution. However, we expect that the 
departures are not overly severe.

The above results, taken in conjunction with our current cosmological
understanding, imply that it is unlikely that the universe will become
radiation dominated in the far future.  The majority of the energy
density at the present epoch is (most likely) in the form of
non-baryonic dark matter of some kind.  A substantial fraction of 
this dark matter resides in galactic halos, and some fraction of these
halos can be annihilated and hence converted into radiation through
the white dwarf capture process outlined in \S III.E.  However, an
equal or larger fraction of this dark matter resides outside of
galaxies and/or can escape destruction through evaporation from
galactic halos.  Thus, unless the dark matter particles themselves
decay into radiation, it seems that enough non-baryonic dark matter
should survive to keep the universe matter dominated at all future 
epochs; in addition, the leftover electrons and positrons will 
help prevent the universe from becoming radiation dominated 
(see also Page \& McKee, 1981ab). 

\bigskip 
\noindent{\bf E. Possible Effects of Vacuum Energy Density} 
\medskip 

If the universe contains a nonvanishing contribution of vacuum energy
to the total energy density, then two interesting long term effects
can arise. The universe can enter a second inflationary phase,
in which the universe expands superluminally (Guth, 1981; see also
Albrecht \& Steinhardt, 1983; Linde, 1982). Alternately, the
vacuum can, in principle, be unstable and the universe can tunnel 
into an entirely new state (e.g., Coleman, 1977, 1985).  
Unfortunately, the contribution of the vacuum to the energy density 
of the universe remains unknown.  In fact, the ``natural value'' of 
the vacuum energy density appears to be larger than the 
cosmologically allowed value by many orders of magnitude. 
This discrepancy is generally known as the ``cosmological constant 
problem'' and has no currently accepted resolution (see the 
reviews of Weinberg, 1989; Carroll, Press, \& Turner, 1992). 

\bigskip 
\noindent 
{1. Future Inflationary Epochs} 
\medskip 

We first consider the possibility of a future inflationary epoch.  
The evolution equation for the universe can be written in the form 
$$\Bigl( { {\dot R} \over R} \Bigr)^2 = 
{8 \pi G \over 3} \bigl( \rho_M + \rhovac \bigr) \, , \eqno(5.33)$$
where $R$ is the scale factor, $\rho_M$ is the energy density 
in matter, and $\rhovac$ is the vacuum energy density. 
We have assumed a spatially flat universe for simplicity. 
The matter density varies with the scale factor according to 
$\rho_M \sim R^{-3}$, whereas the vacuum energy density is 
constant.  We can define the ratio 
$$\nu \equiv \rhovac/\rho_0 \, , \eqno(5.34)$$ 
i.e., the ratio of the vacuum energy density to that of the matter 
density $\rho_0$ at the present epoch.  We can then integrate equation 
[5.6] into the future and solve for the time $t_{\rm vac}$ at
which the universe becomes vacuum dominated.  We find the result 
$$t_{\rm vac} = t_0 \, + \, \tau \, \, 
{ \sinh^{-1}[1] - \sinh^{-1}[\nu^{1/2}] \over \nu^{1/2} } 
\, ,  \eqno(5.35)$$
where $t_0$ is the present age of the universe and we have defined 
$\tau \equiv (6 \pi G \rho_0)^{-1/2}$; both time scales 
$t_0$ and $\tau$ are approximately $10^{10}$ yr. 

Several results are immediately apparent from equation [5.35]. If the
vacuum energy density provides any appreciable fraction of the total
energy density at the present epoch (in other words, if $\nu$ is not
too small), then the universe will enter an inflationary phase in the
very near future.  Furthermore, almost any nonvanishing value of the
present day vacuum energy will lead the universe into an inflationary
phase on the long time scales considered in this paper. For small
values of the ratio $\nu$, the future inflationary epoch occurs 
at the cosmological decade given by 
$$\eta_{\rm inflate} \approx 10 + {1\over2} 
\log_{10} \bigl[ {1 \over \nu} \bigr] \, . \eqno(5.36)$$
For example, even for a present day vacuum contribution as small 
as $\nu \sim 10^{-40}$, the universe will enter an inflationary 
phase at the cosmological decade $\eta_{\rm inflate} \approx 30$, 
long before protons begin to decay.  In other words, the 
traditional cosmological constant problem becomes even more 
severe when we consider future cosmological decades. 

If the universe enters into a future inflationary epoch, several
interesting consequences arise.  After a transition time comparable 
to the age of the universe at the epoch [5.36], the scale factor 
of the universe will begin to grow superluminally. Because of this 
rapid expansion, all of the astrophysical objects in the universe
become isolated and eventually become out of causal contact.  
In other words, every given co-moving observer will see an effectively  
{\it shrinking horizon} (the particle horizon does not actually get 
smaller, but this language has become common in cosmology -- see 
Ellis \& Rothman, 1993 for further discussion of horizons in this 
context). In particular, astrophysical objects, such as galaxies 
and stars, will cross outside the speed-of-light sphere and hence 
disappear from view. For these same astrophysical objects, the 
velocity relative to the observer becomes larger than the speed 
of light and their emitted photons are redshifted to infinity. 

\bigskip 
\noindent 
{2. Tunneling Processes} 
\nobreak 
\medskip 
\nobreak 

We next consider the possibility that the universe is 
currently in a false vacuum state.  In other words, a lower 
energy vacuum state exists and the universe can someday tunnel 
to that lower energy state.  This problem, the fate of the 
false vacuum, was first explored quantitatively by Voloshin 
et al. (1974) and by Coleman (1977). Additional effects have
been studied subsequently, including gravity (Coleman \& De 
Luccia, 1980) and finite temperature effects (e.g., Linde, 1983). 

To obtain quantitative results, we consider an illustrative 
example in which the vacuum energy density of the universe 
can be described by the dynamics of a single scalar field. 
Once a field configuration becomes trapped in a metastable state 
(the false vacuum), bubbles of the true vacuum state nucleate in
the sea of false vacuum and begin growing spherically.  The speed of 
the bubble walls quickly approaches the speed of light. The basic
problem is to calculate the tunneling rate (the decay probability)
from the false vacuum state to the true vacuum state, i.e., the bubble
nucleation rate ${\cal P}$ per unit time per unit volume.  For tunneling 
of scalar fields at zero temperature (generally called quantum 
tunneling), the four-dimensional Euclidean action $S_4$ of the 
theory largely determines this tunneling rate. The decay probability 
$\cal P$ can be written in the form  
$${\cal P} = K {\rm e}^{-S_4} \, , \eqno(5.37)$$
where $K$ is a determinental factor (see Coleman, 1977, 1985). 
For purposes of illustration, we assume a generic quartic 
potential of the form 
$$V (\Phi) = \lambda \Phi^4 - a \Phi^3 + b \Phi^2 + c \Phi + d 
\, . \eqno(5.38)$$
We can then write the action $S_4$ in the form  
$$S_4 = {\pi^2 \over 3 \lambda} \, (2 - \delta)^{-3} \, 
{\cal R}(\delta) \, , \eqno(5.39)$$
where $\delta \equiv 8 \lambda b/a^2$ and where $\cal R$ is a 
slowly varying function which has a value near unity for most 
of the range of possible quartic potentials (Adams, 1993). 
The composite shape parameter $\delta$ varies from 0 to 2 as 
the potential $V (\Phi)$ varies from having no barrier height 
to having nearly degenerate vacua (see Figure 10). 

Even though equations [5.37 -- 5.39] describe the tunneling rate, 
we unfortunately do not know what potential (if any) describes our
universe and hence it is difficult to obtain a precise numerical 
estimate for this time scale. To get some quantitative feeling for 
this problem, we consider the following example.  For the case of 
no tunneling barrier (i.e., for $S_4 = 0$), the characteristic 
decay probability is given by ${\cal P}_0 \sim K \sim M_V^4$, 
where $M_V$ is the characteristic energy scale for the scalar 
field.  For $M_V = 10^{16}$ GeV (roughly the GUT scale), 
${\cal P}_0 \sim 10^{129}$ s$^{-1}$ cm$^{-3}$.  
With this decay rate, the universe within a characteristic volume 
$M_V^{-3}$ would convert from false vacuum to true vacuum on a 
time scale of $\sim 10^{-24}$ s. 
Clearly, however, the actual decay time scale must be long enough 
that the universe has not decayed by the present epoch.  In order 
to ensure that the universe has survived, we require that no 
nucleation events have occurred within the present horizon volume 
($\sim [3000 \, {\rm Mpc}]^3$) during the current age of the 
universe ($\sim 10^{10}$ yr). This constraint implies that the 
action $S_4$ must be sufficiently large in order to suppress 
nucleation, in particular, 
$$S_4 > 231 \ln 10 \approx 532 \, . \eqno(5.40)$$
The question then becomes: is this value for $S_4$ reasonable? 
For the parameter $\lambda$, a reasonable range of values is 
$0.1 < \lambda <  1$; similarly, for $\delta$, we take the range 
$0.1 < \delta < 1.9$. Using the form [5.39] for the action and 
setting $\cal R$ = 1, we find the approximate range 
$$ 0.5 < S_4 < 3 \times 10^4 \, . \eqno(5.41)$$ 
Thus, the value required for the universe to survive to the present
epoch (equation [5.40]) can be easily realized within this simple 
model. In the future, however, the universe could tunnel into its 
false vacuum state at virtually any time, as soon as tomorrow, or 
as late as $\eta$ = $10^4$.  If and when this tunneling effect occurs,
the universe will change its character almost completely. The physical
laws of the universe, or at least the values of all of the physical
constants, would change as the phase transition completes (see Sher, 
1989 and Crone \& Sher, 1990 for a discussion of changing laws of
physics during a future phase transition). The universe, as we know
it, would simply cease to exist.

Vacuum tunneling of the entire universe
is certainly one of the more speculative topics considered in this
paper.  Nevertheless, its inclusion is appropriate since the act of
tunneling from a false vacuum into a true vacuum would change the
nature of the universe more dramatically than just about any other
physical process. 

It is also possible for the universe to spontaneously create 
``child universes'' through a quantum tunneling process roughly
analogous to that considered above (e.g., Sato et al., 1982; 
Hawking, 1987; Blau, Guendelman, \& Guth, 1987).  In this situation, 
a bubble of false vacuum energy nucleates in an otherwise empty
space-time.  If this bubble is sufficiently large, it will grow
exponentially and will eventually become causally disconnected
from the original space-time.  In this sense, the newly created bubble
becomes a separate ``child universe''.  The newly created universe
appears quite different to observers inside and outside the bubble.
Observers inside the bubble see the local universe in a state of
exponential expansion. Observers outside the bubble, in the empty
space-time background, see the newly created universe as a black hole
that collapses and becomes causally disconnected.  As a result, 
these child universes will not greatly affect the future evolution 
of our universe because they (relatively) quickly become out of 
causal contact.  

One potentially interesting effect of these child universes is that
they can, in principle, receive information from our universe. Before
the newly created universe grows out of causal contact with our own
universe, it is connected through a relativistic wormhole, which can
provide a conduit for information transfer and perhaps even the 
transfer of matter (see Visser, 1995 for further discussion of 
wormholes and transferability).  The implications of this possibility 
are the subject of current debate (for varying points of view, see, 
e.g., Linde, 1988, 1989; Tipler, 1992; Davies, 1994). 

\bigskip 
\noindent{\bf F. Speculations about Energy and 
Entropy Production in the Far Future} 
\medskip 

Thus far in this paper, we have shown that entropy can be 
generated (and hence work can be done) up to cosmological decades
$\eta \sim 100$.  For very long time scales $\eta \gg 100$, the 
future evolution of the universe becomes highly uncertain, but 
the possibility of continued entropy production is very important 
(see \S VI.D).  Here, we briefly assess some of the possible ways 
for energy and entropy to be generated in the far future. 

\bigskip 
\noindent 
{1. Continued Formation and Decay of Black Holes}
\medskip 

For the case of a flat spatial geometry for the universe, future
density perturbations can provide a mechanism to produce entropy.
These density perturbations create large structures which can
eventually collapse to form black holes. The resulting black holes, 
in turn, evaporate by emitting Hawking radiation and thus represent 
entropy (and energy) sources (e.g., see also Page \& McKee, 1981a; 
Frautschi, 1982).  Density perturbations of increasingly larger 
size scale $\lambda$ will enter the horizon as the universe 
continues to expand. The corresponding mass scale $M_\lambda$ 
of these perturbations is given by 
$$M_\lambda = M_0 \Bigl( {t_\lambda \over t_0} 
\Bigr) \, , \eqno(5.42)$$
where $t_\lambda$ is the time at which the perturbation 
enters the horizon and $M_0 \approx 10^{22} M_\odot$ is the 
total mass within the present day horizon (at time $t_0$). 

The time $t_\lambda$ represents the time at which a given 
perturbation enters the horizon and begins to grow; a large 
structure (such as a black hole) can only form at some later
time after the perturbation becomes nonlinear. 
Suppose that a density perturbation has an initial amplitude 
$\delta_\lambda$ when it enters the horizon.  In the 
linear regime, the perturbation will grow according 
to the usual relation
$$\delta = \delta_\lambda \Bigl( {t \over t_\lambda} 
\Bigr)^{2/3} \, , \eqno(5.43)$$
where $\delta \equiv \Delta \rho / \rho$ and $t > t_\lambda$ 
(see Peebles, 1993). Using this growth law, the epoch 
$\eta_{\rm nl}$ at which the perturbation becomes nonlinear 
can be written in the form 
$$\eta_{\rm nl} = \eta_\lambda - {3 \over 2} \, 
\log_{10} \delta_\lambda \, . \eqno(5.44)$$
For example, if the perturbation has an amplitude 
$\delta_\lambda = 10^{-4}$, then it becomes nonlinear at time 
$\eta_{\rm nl}$ = $\eta_\lambda$ + 6.  Since we are interested 
in very long time scales $\eta > 100$, the difference between 
the horizon crossing time $\eta_\lambda$ and the time $\eta_{\rm nl}$ 
of nonlinearity is not overly large. 

One possible result of this process is the production of a large black
hole with a mass $M_{BH} \sim$ $M_\lambda$. The time scale for such a
black hole to evaporate through the Hawking process is given by
$$\eta_{BH} = 101 + 3 \eta_\lambda \, , \eqno(5.45)$$ 
where we have combined equations [4.42] and [5.42]. 
Since $\eta_{BH} \gg \eta_\lambda \sim \eta_{\rm nl}$, 
the universe can form black holes faster than they can 
evaporate. Thus, for the case of a geometrically flat universe, 
future density perturbations can, in principle, continue to 
produce black holes of increasingly larger mass. In this case, 
the universe will always have a source of entropy -- the Hawking 
radiation from these black holes.  

We note that these bound perturbations need not necessarily form black
holes.  The material is (most likely) almost entirely non-dissipative
and collisionless, and will thus have a tendency to form virialized
clumps with binding energy per unit mass of order $\sim \delta c^2$.
Thus, unless the perturbation spectrum is tilted so that $\delta$ is
of order unity on these much larger scales, the ensuing dynamics is
probably roughly analogous to that of a a cluster-mass clump of cold
dark matter in our present universe.  However, even if the mass of the
entire perturbation does not form a single large black hole, smaller
scale structures can in principle form black holes, in analogy to
those currently in the centers of present-day galaxies.  In addition,
it is possible that the existing black holes can merge faster than
they evaporate through the Hawking process (see also \S III.D).
Thus, the possibility remains for the continued existence of
black holes in the universe.

The process outlined here, the formation of larger and larger black
holes, can continue as long as the universe remains spatially flat 
and the density perturbations that enter the horizon are not overly
large. The inflationary universe scenario provides a mechanism to
achieve this state of affairs, at least up to some future epoch 
(see \S V.C and in particular equation [5.14]).  Thus, the nature 
of the universe in the far future $\eta \gg 100$ may be determined 
by the physics of the early universe (in particular, inflation) at 
the cosmological decade $\eta \sim -45$. 

Notice that at these very late times, $\eta \gg 100$, 
the matter entering the horizon will already be ``processed'' 
by the physical mechanisms described earlier in the paper. 
Thus, the nucleons will have (most likely) already decayed 
and the matter content of the universe will be mostly 
electrons, positrons, and non-baryonic dark matter particles. 
Annihilation of both $e^+$--$e^-$ pairs and dark matter will 
occur simultaneously with perturbation growth and hence the 
final mass of the black hole will be less than $M_\lambda$.  
This issue must be studied in further depth. 

\bigskip 
\noindent 
{2. Particle Annihilation in an Open Universe}
\medskip

If the universe is open, however, then future density perturbations 
are effectively frozen out (see \S V.B) and the hierarchy of black holes 
described above cannot be produced.  For an open universe, 
continued energy and entropy production is more difficult to 
achieve. One process that can continue far into the future, albeit 
at a very low level, is the continued annihilation of particles. 
Electrons and positrons represent one type of particle that can
annihilate (see also Page \& McKee, 1981ab), but the discussion 
given below applies to a general population of particles. 

Consider a collection of particles with number density $n$. 
The time evolution of the particle population is governed 
by the simple differential equation 
$${dn \over dt} + 3 H n = - \langle \sigma v \rangle n^2 
\, , \eqno(5.46)$$
where $H = {\dot R}/R$ is the Hubble parameter and 
$\langle \sigma v \rangle$ is the appropriate average of 
interaction cross section times the speed 
(e.g., see Kolb \& Turner, 1990).  Since we are interested in 
the case for which the expansion rate is much larger than the 
interaction rate, the particles are very far from thermal 
equilibrium and we can neglect any back reactions that produce
particles. For this example, we consider the universe to be open, 
independent of the activity of this particle population. 
As a result, we can write $R \propto t$ and hence $H = 1/t$.  
We also take the quantity $\langle \sigma v \rangle$ to be 
a constant in time (corresponding to $s$-wave annihilation). 
With these approximations, the differential equation [5.46] 
can be integrated to obtain the solution 
$$n(t) = n_1 \, \Bigl( {t_1 \over t} \Bigr)^3 \, 
\Bigl[ 1 + \Delta_\infty [1 - (t_1/t)^2 ] \Bigr]^{-1} 
\, , \eqno(5.47)$$
where we have defined the quantity 
$$\Delta_\infty \equiv {1 \over 2} n_1 t_1 
\langle \sigma v \rangle \, , \eqno(5.48)$$
and where we have invoked the boundary condition 
$$n(t_1) = n_1 = {\sl constant} \, . \eqno(5.49)$$ 
Analogous solutions for particle annihilation can be 
found for the case of a flat universe ($H = 2/3t$) 
and an inflating universe ($H$ = {\sl constant}). 

The difference between the solution [5.47] and the simple 
adiabatic scaling solution $n(t) = n_1 (t_1/t)^3$ is due to particle 
annihilation, which is extremely small but non-zero. This statement 
can be quantified by defining the fractional difference $\Delta$ 
between the solution [5.47] and the adiabatic solution, i.e., 
$$\Delta(t) \equiv {\Delta n \over n} (t) = 
\Delta_\infty [1 - (t_1/t)^2 ] \, . \eqno(5.50)$$ 
Over the entire (future) lifetime of the universe, the 
comoving fraction of particles that annihilate is given 
by the quantity $\Delta_\infty$, which is both finite and 
typically much less than unity.  For example, if we consider 
the largest possible values at the present epoch 
($\sigma \approx \sigma_T \approx 10^{-24}$ cm$^2$, 
$n_1 \approx 10^{-6}$ cm$^{-3}$, 
$t_1 \approx 3 \times 10^{17} s$, and $v=c$), 
then $\Delta_\infty \approx 10^{-2}$. 
The fraction $\Delta_\infty$ will generally be much smaller 
than this example.  The fact that the fraction $\Delta_\infty$ 
is finite implies that the process of particle annihilation 
can provide only a finite amount of energy over the infinite 
time interval $\eta_1 < \eta < \infty$. 

\bigskip 
\noindent 
{3. Formation and Decay of Positronium}
\medskip 

Another related process that will occur on long time scales
is the formation and eventual decay of positronium.  This 
process has been studied in some detail by Page \& McKee 
(1981ab; see also the discussion of Barrow \& Tipler, 1986); 
here we briefly summarize their results. The time scale for 
the formation of positronium in a flat universe is given by 
$$\eta_{\rm form} \approx 85 + 2 (\eta_P - 37) 
- {2 \over 3} \log_{10} [ \Omega_e ] \, , \eqno(5.51)$$
where $\eta_P$ is the proton lifetime (see \S IV) and where 
$\Omega_e$ is the mass fraction of $e^\pm$ after proton decay.  
For a flat or nearly flat universe, most of the electrons 
and positrons become bound into positronium. 
In an open universe, some positronium formation occurs, 
but most electrons and positrons remain unattached. 

At the time of formation, the positronium atoms are generally 
in states of very high quantum number (and have radii larger 
than the current horizon size).  The atoms emit a cascade of 
low energy photons until they reach their ground state; once 
this occurs, the positronium rapidly annihilates. The relevant 
time scale for this decay process is estimated to be 
$$\eta_{\rm decay} \approx 141 + 4 (\eta_P - 37) 
- {8 \over 3} \log_{10} [ \Omega_e ] \, . \eqno(5.52)$$

\newpage 
\noindent{\bf VI. SUMMARY AND DISCUSSION} 
\medskip 

Our goal has been to present a plausible and quantitative description
of the future of the Universe.  Table I outlines the most important
events in the overall flow of time, as well as the cosmological
decades at which they occur (see equation [1.1]).  In constructing
this table, representative values for the (often uncertain) parameters
have been assumed; the stated time scales must therefore be viewed as
approximate. Furthermore, as a general rule, both the overall future
of the universe, as well as the time line suggested in Table I, 
become more and more uncertain in the face of successively deeper
extrapolations into time. Some of the effects we have described will
compete with one another, and hence not all the relevant physical
processes can proceed to completion. Almost certainly, parts of our
current time line will undergo dramatic revision as physical
understanding improves. We have been struck by the remarkable natural
utility of the logarithmic ``clock'', $\eta$, in organizing the
passage of time. Global processes which can characterize the entire
universe rarely span more than a few cosmological decades, and the 
ebb and flow of events is dispersed quite evenly across a hundred 
and fifty orders of magnitude in time, i.e., $-50 < \eta < 100$. 

\bigskip 
\noindent{\bf A. Summary of Results} 
\medskip 
\nobreak 

Our specific contributions to physical eschatology can be 
summarized as follows: 

\item{\bf [1]} We have presented new stellar evolution calculations 
which show the long term behavior of very low mass stars (see Figure 1). 
Stars with very small mass ($\sim 0.1 M_\odot$) do not experience 
any red giant phases.  As they evolve, these stars become steadily
brighter and bluer, reaching first a maximum luminosity, and second, a 
maximum temperature, prior to fading away as helium white dwarfs.

\item{\bf [2]} Both stellar evolution and conventional star formation 
come to an end at the cosmological decade $\eta \sim$ 14.  This time 
scale only slightly exceeds the longest evolution time for a low mass 
star. It also corresponds to the time at which the galaxy runs out of 
raw material (gas) for producing new stars.  The era of conventional 
stars in the universe is confined to the range $6 < \eta < 14$.

\item{\bf [3]}  We have introduced the final mass function (FMF), 
i.e., the distribution of masses for the degenerate stellar objects
left over from stellar evolution (see Figure 2). 
Roughly half of these objects will be white dwarfs, with most of the 
remainder being brown dwarfs. Most of the mass, however, will 
be in the form of white dwarfs (see equations [2.22] and [2.23]).  

\item{\bf [4]} We have explored a new mode of continued
star formation through the collisions of substellar objects (see
Figure 3).  Although the time scale for this process is quite long,
this mode of star formation will be the leading source of new stars for
cosmological decades in the range $15 < \eta < 23$.

\item{\bf [5]} We have presented a scenario for the future evolution 
of the galaxy.  The galaxy lives in its present state until a time of 
$\eta \sim 14$ when both conventional star formation ceases and the 
smallest ordinary stars leave the main sequence. For times $\eta > 14$, 
the principle mode of additional star formation is through the 
collisions and mergers of brown dwarfs (substellar objects). 
The galaxy itself evolves through the competing processes of 
orbital decay of orbits via gravitational radiation and the 
evaporation of stars into the intergalactic medium via stellar 
encounters.  Stellar evaporation is the dominant process and 
most of the stars will leave the system at a time $\eta \sim 19$. 
Some fraction (we roughly estimate $\sim$0.01--0.10) of the 
galaxy is left behind in its central black hole. 

\item{\bf [6]} We have considered the annhilation and capture of
weakly interacting massive particles (WIMPs) in the galactic halo.  
In the absence of other evolutionary processes, the WIMPs in the halo
annihilate on the time scale $\eta \sim 23$.  On the other hand, white
dwarfs can capture WIMPs and thereby deplete the halo on the somewhat
longer time scale $\eta \sim 25$.  The phenomenon of WIMP capture
indicates that white dwarf cooling will be arrested rather shortly 
at a luminosity $L_\ast \sim 10^{-12} L_\odot$. 

\item{\bf [7]} Depending on the amount of mass loss suffered by the Sun 
when it becomes a red giant, the Earth may be vaporized by the Sun during 
its asymptotic giant phase of evolution; in this case, the Earth will be 
converted to a small (0.01 \%) increase in the solar metallicity. 
In general, however, planets can end their lives in a variety of ways. 
They can be vaporized by their parent stars, ejected into interstellar 
space through close stellar encounters, merge with their parent stars 
through gravitational radiation, and can eventually disappear 
as their protons decay. 

\item{\bf [8]} We have discussed the allowed range for the proton 
lifetime.  A firm lower bound on the lifetime arises from current 
experimental searches. Although no definitive upper limit exists, 
we can obtain a suggestive upper ``bound'' on the proton lifetime 
by using decay rates suggested by GUTs and by invoking the constraint 
the mass of the mediating boson, $M_X < \mpl \sim 10^{19}$ GeV.  
We thus obtain the following expected range for the proton lifetime
$$32 < \eta_P < 49 + 76 (N-1) \, , \eqno(6.1)$$ 
where the integer $N$ is order of the process, i.e., the number of 
mediating bosons required for the decay to take place. Even for the 
third order case, we have $\eta_P < 201$. Quantum gravity effects 
also lead to proton decay with time scales in the range 
$46 < \eta_P < 169$. Finally, sphalerons imply $\eta_P \sim 140$. 

\item{\bf [9]} We have presented a scenario for the future evolution
of sun-like stars (see Figure 6). In this case, stars evolve into
white dwarf configurations as in conventional stellar evolution. 
On sufficiently long time scales, however, proton decay becomes 
important.  For cosmological decades in the range $20 < \eta < 35$,
the mass of the star does not change appreciably, but the luminosity
is dominated by the energy generated by proton decay.  In the
following cosmological decades, $\eta = 35 - 37$, mass loss plays a
large role in determining the stellar structure.  The star expands as
it loses mass and follows the usual mass/radius relation for white
dwarfs. The chemical composition changes as well (see Figure 5). 
Proton decay by itself quickly reduces the star to a state of pure
hydrogen.  However, pycnonuclear reactions will be sufficient to
maintain substantial amounts of helium ($^3$He and $^4$He) until the
mass of the star decreases below $\sim 0.01 M_\odot$. During the
proton decay phase of evolution, a white dwarf follows a well-defined
track in the H-R Diagram given by $L_\ast \propto T_\ast^{12/5}$.
After the stellar mass decreases to $M_\ast \approx 10^{-3} M_\odot$, 
the star is lifted out of degeneracy and follows a steeper track 
$L_\ast \propto T_\ast^{12}$ in the H-R Diagram. 

\item{\bf [10]} If proton decay does not take place through the 
first order process assumed above, then white dwarfs and other 
degenerate objects will still evolve, but on a much longer time 
scale.  The relevant physical process is likely to be proton decay 
through higher order effects. The time scales for the 
destruction and decay of degenerate stars obey the ordering 
$$\eta_P \ll \eta_{BH} \ll \eta_{P2} \, , \eqno(6.2)$$
where $\eta_P \sim 37$ is the time scale for first order proton decay,
$\eta_{BH} \sim 65$ is the time scale for a stellar-sized black hole 
to evaporate, and $\eta_{P2} \sim 100 - 200$ is the time scale for 
proton decay through higher order processes. 

\item{\bf [11]} In the future, the universe as a whole can evolve in 
a variety of different possible ways.  Future density perturbations
can come across the horizon and close the universe; this effect would
ultimately lead (locally) to a big crunch.  Alternately, the universe
could contain a small amount of vacuum energy (a cosmological constant
term) and could enter a late time inflationary epoch. Finally, the
universe could be currently in a false vacuum state and hence 
kevorking on the brink of instability. In this case, when the 
universe eventually tunnels into the true vacuum state, the
laws of physics and hence the universe as we know it would 
change completely. 

\item{\bf [12]} As the cosmic microwave background redshifts away,
several different radiation fields will dominate the background.  
In the near term, stellar radiation will overtake the cosmic 
background. Later on, the radiation produced by dark matter
annihilation (both direct and in white dwarfs) will provide the
dominant contribution.  This radiation field will be replaced by that
arising from proton decay, and then, eventually, by the radiation
field arising from evaporation of black holes (see Figure 9).

\bigskip
\noindent{\bf B. Eras of the Future Universe} 
\medskip

Our current understanding of the universe suggests that we can
organize the future into distinct eras, somewhat analogous to
geological eras: 

\item{\bf [A]} {\sl The Radiation Dominated Era}. 
$-\infty < \eta < 4$.  This era corresponds to the usual 
time period in which most of the energy density of the 
universe is in the form of radiation. 

\item{\bf [B]} {\sl The Stelliferous Era}. $6 < \eta < 14$.  
Most of the energy generated in the universe arises from 
nuclear processes in conventional stellar evolution. 

\item{\bf [C]} {\sl The Degenerate Era}. $15 < \eta < 37$. 
Most of the (baryonic) mass in the universe is locked up in 
degenerate stellar objects: brown dwarfs, white dwarfs, and 
neutron stars. Energy is generated through proton decay 
and particle annihilation. 

\item{\bf [D]} {\sl The Black Hole Era}. $38 < \eta < 100$. 
After the epoch of proton decay, the only stellar-like objects 
remaining are black holes of widely disparate masses, which are 
actively evaporating during this era. 

\item{\bf [E]} {\sl The Dark Era}. $\eta > 100$. 
At this late time, protons have decayed and black holes have evaporated.  
Only the waste products from these processes remain: mostly photons of 
colossal wavelength, neutrinos, electrons, and positrons. The seeming 
poverty of this distant epoch is perhaps more due to the difficulties 
inherent in extrapolating far enough into the future, rather 
than an actual dearth of physical processes.

\bigskip 
\noindent{\bf C. Experimental and Theoretical Implications}
\medskip

Almost by definition, direct experiments that test theoretical
predictions of the very long term fate of the universe cannot be 
made in our lifetimes.  However, this topic in general and this 
paper in particular have interesting implications for present 
day experimental and theoretical work.  If we want to gain more 
certainty regarding the future of the universe and the astrophysical 
objects within it, then several issues must be resolved. 
The most important of these are as follows:  

\item{\bf [A]} Does the proton decay? What is the lifetime?  
This issue largely determines the fate stellar objects in the universe
for time scales longer than $\eta \sim 35$.  If the proton is stable
to first order decay processes, then stellar objects in general and
white dwarfs in particular can live in the range of cosmological
decades $\eta < 100$.  If the proton is also stable to second order
decay processes, then degenerate stellar objects can live for a much 
longer time.  On the other hand, if the proton does decay, a large 
fraction of the universe will be in the form of proton decay products 
(neutrinos, photons, positrons, etc.) for times $\eta > 35$. 

\item{\bf [B]} What is the vacuum state of the universe? 
This issue plays an important role in determining the ultimate fate 
of the universe itself. If the vacuum energy density of the universe
is nonzero, then the universe might ultimately experience a future 
epoch of inflation. On the other hand, if the vacuum energy density 
is strictly zero, then future (large) densities perturbations can,  
in principle, enter our horizon and lead (locally) to a closed 
universe and hence a big crunch.

\item{\bf [C]} What is the nature of the dark matter? 
Of particular importance is the nature of the dark matter 
that makes up galactic halos.  The lifetime of the dark matter 
particles is also of great interest. 

\item{\bf [D]} What fraction of the stars in a galaxy 
are evaporated out of the system and what fraction are
accreted by the central black hole (or black holes)? 
This issue is important because black holes dominate 
the energy and entropy production in the universe in the 
time range $36 < \eta < 100$ and the mass of a black hole 
determines its lifetime. 

\item{\bf [E]} Does new physics occur at extremely low temperatures?
As the universe evolves and continues to expand, the relevant
temperatures become increasingly small. In the scenario outlined here,
photons from the cosmic microwave background and other radiation
fields, which permeate all of space, can redshift indefinitely in
accordance with the classical theory of radiation.  It seems possible
that classical theory will break down at some point.  For example, in
an open universe, the CMB photons will have a wavelength longer than
the current horizon size ($\sim 3000$ Mpc) at a time $\eta \sim 40$,
just after proton decay.  Some preliminary models for future phase 
transitions have been proposed (Primack \& Sher, 1980; Suzuki, 1988; 
Sher, 1989), but this issue calls out for further exploration. 

\newpage 
\noindent{\bf D. Entropy and Heat Death} 
\medskip

The concept of the heat death of the universe has troubled many
philosophers and scientists since the mid-nineteenth century when 
the second law of thermodynamics was first understood (e.g., Helmholz, 
1854; Clausius, 1865, 1868).  Very roughly, {\sl classical heat death}
occurs when the universe as a whole reaches thermodynamic equilibrium;
in such a state, the entire universe has a constant temperature at all
points in space and hence no heat engine can operate.  Without the
ability to do physical work, the universe ``runs down'' and becomes 
a rather lifeless place.  Within the context of modern Big Bang 
cosmology, however, the temperature of the universe is continually
changing and the issue shifts substantially; many authors have
grappled with this problem, from the inception of Big Bang theory
(e.g., Eddington, 1931) to more recent times (Barrow \& Tipler, 1978,
1986; Frautschi, 1982).  A continually expanding universe never reaches
true thermodynamic equilibrium and hence never reaches a constant
temperature.  Classical heat death is thus manifestly avoided.
However, the expansion can, in principle, become purely adiabatic so
that the entropy in a given comoving volume of the universe approaches
(or attains) a constant value.  In this case, the universe can still
become a dull and lifeless place with no ability to do physical work.
We denote this latter possibility as {\sl cosmological heat death}.

Long term entropy production in the universe is constrained in 
fairly general terms for a given class of systems (Bekenstein, 1981). 
For a spatially bounded physical system with effective radius $R$, 
the entropy $S$ of the system has a well defined maximum value. 
This upper bound is given by 
$$S \le {2 \pi R E \over \hbar c} \, , \eqno(6.3)$$
where $E$ is the total energy of the system.  Thus, for a bounded 
system (with finite size $R$), the ratio $S/E$ of entropy to energy 
has a firm upper bound. Furthermore, this bound can be actually 
attained for black holes (see Bekenstein, 1981 for further discussion). 

The results of this paper show that cosmological events continue 
to produce energy and entropy in the universe, at least until the 
cosmological decade $\eta \sim 100$.  As a result, cosmological heat
death is postponed until after that epoch, i.e., until the Dark Era.
After that time, however, it remains possible in principle for the
universe to become nearly adiabatic and hence dull and lifeless.
The energy and entropy generating mechanisms available to the universe
depend on the mode of long term evolution, as we discuss below. 

If the universe is closed (\S V.A) or becomes closed at some future
time (\S V.B), then the universe will end in a big crunch and long
term entropy production will not be an issue.  For the case in which
the universe remains nearly flat, density perturbations of larger and
larger size scales can enter the horizon, grow to nonlinearity, and
lead to continued production of energy and entropy through the
evaporation of black holes (see \S V.F.1).  These black holes saturate
the Bekenstein bound and maximize entropy production.  Cosmological
heat death can thus be avoided as long as the universe remains nearly 
flat. 

On the other hand, if the universe is open, then density fluctuations
become frozen out at some finite length scale (\S V.B).  The energy
contained within the horizon thus becomes a finite quantity.  However,
the Bekenstein bound does not directly constrain entropy production in
this case because the effective size $R$ grows without limit.  For an
open universe, the question of cosmological heat death thus remains
open.  For a universe experiencing a future inflationary phase 
(\S V.E.1), the situation is similar.  Here, the horizon is 
effectively shrinking with time.  However, perturbations that have
grown to nonlinearity will be decoupled from the Hubble flow. The
largest nonlinear perturbation will thus define a largest length scale
$\lambda$ and hence a largest mass scale in the universe; this mass
scale once again implies a (finite) maximum possible amount of energy
available to a local region of space.  However, the system is not
bounded spatially and the questions of entropy production and
cosmological heat death again remain open.

To close this paper, we put forth the point of view that the universe
should obey a type of {\sl Copernican Time Principle} which applies to
considerations of the future. This principle holds that the current
cosmological epoch ($\eta = 10$) has no special place in time.  
In other words, interesting things can continue to happen at the
increasingly low levels of energy and entropy available in the
universe of the future.

\newpage 
\noindent{\bf Acknowledgments}
\nobreak 

This paper grew out of a special course taught at the University of
Michigan for the theme semester ``Death, Extinction, and the Future of
Humanity'' (Winter 1996).  We would like to thank Roy Rappaport for
providing the initial stimulation for this course and hence this
paper.  We also thank R. Akhoury, M. Einhorn, T. Gherghetta, G. Kane,
and E. Yao for useful discussions regarding proton decay and other
particle physics issues.  We thank P. Bodenheimer, G. Evrard,
J. Jijina, J. Mohr, M. Rees, D. Spergel, F. X. Timmes, and R. Watkins 
for many interesting astrophysical discussions and for critical commentary 
on the manuscript. This work was supported by an NSF Young Investigator 
Award, NASA Grant No. NAG 5-2869, and by funds from the Physics 
Department at the University of Michigan.

\newpage 
\par\pp
{\bf REFERENCES} 

\medskip 

\par\pp 
Abramowitz, M., and I. A. Stegun, 1972, 
{\sl Handbook of Mathematical Functions} (New York: Dover). 

\par\pp
Adams, F. C. 1993, {\sl Phys. Rev.} D {\bf 48}, 2800. 

\par\pp 
Adams, F. C., and M. Fatuzzo, 1996, {\sl Astrophys. J.} {\bf 464}, 256. 

\par\pp
Adams, F. C., and K. Freese, 1991, {\sl Phys. Rev.} D {\bf 43}, 353. 

\par\pp
Adams, F. C., and K. Freese, 1995, {\sl Phys. Rev.} D {\bf 51}, 6722. 

\par\pp
Adams, F. C., K. Freese, and A. H. Guth, 1991, 
{\sl Phys. Rev.} D {\bf 43}, 965. 

\par\pp
Albrecht, A., and P. J. Steinhardt, 1982, {\sl Phys. Rev. Lett.} 
{\bf 48}, 1220. 

\par\pp
Alcock, C. et al. 1993, {\sl Nature} {\bf 365}, 621. 

\par\pp
Aubourg, E. et al. 1993, {\sl Nature} {\bf 365}, 623. 

\par\pp
Bahcall, J. N. 1989, {\sl Neutrino Astrophysics} 
(Cambridge: Cambridge Univ. Press). 

\par\pp
Bardeen, J. M., P. J. Steinhardt, and M. S. Turner, 1983, 
{\sl Phys. Rev.} D {\bf 28}, 679.  

\par\pp
Barrow, J. D., and F. J. Tipler, 1978, {\sl Nature} {\bf 276}, 453. 

\par\pp
Barrow, J. D., and F. J. Tipler, 1986, {\sl The Anthropic Cosmological 
Principle} (Oxford: Oxford Univ. Press). 

\par\pp
Bekenstein, J. D. 1981, {\sl Phys. Rev.} D {\bf 23}, 287. 

\par\pp
Binney, J., and S. Tremaine, 1987, {\sl Galactic Dynamics} 
(Princeton: Princeton Univ. Press).

\par\pp 
Blau, S. K., E. I. Guendelman, and A. H. Guth 1987, 
{\sl Phys. Rev.} D {\bf 35}, 1747.  

\par\pp 
Laughlin, G., P. Bodenheimer, and F. C. Adams, 1996, submitted to 
{\sl Astrophys. J.} 

\par\pp
Bond, J. R., B. J. Carr, and C. J. Hogan, 1991, 
{\sl Astrophys. J.} {\bf 367}, 420. 

\par\pp
Brune, D., and J. J. Schmidt, 1974, editors, 
{\sl Handbook on Nuclear Activation Cross-Sections} 
(Vienna: International Atomic Energy Agency).  

\par\pp
Burrows, A., W. B. Hubbard, D. Saumon, and J. I. Lunine, 1993, 
{\sl Astrophys. J.} {\bf 406}, 158. 

\par\pp
Burrows, A., and J. Liebert, 1993, {\sl Rev. Mod. Phys.} {\bf 65}, 301.

\par\pp
Carroll, S. M., W. H. Press, and E. L. Turner, 1992, 
{\sl Ann. Rev. Astron. Astrophys.} {\bf 30}, 499. 

\par\pp
Castano, D. J., and S. P. Martin, 1994, {\sl Phys. Lett.} 
{\bf 340 B}, 67. 

\par\pp
Chandrasekhar, S. 1939, {\sl Stellar Structure} (New York: Dover). 

\par\pp
Clausius, R. 1865, {\sl Ann. Physik}, {\bf 125}, 353. 

\par\pp
Clausius, R. 1868, {\sl Phil. Mag.}, {\bf 35}, 405.  

\par\pp
Clayton, D. D. 1983, {\sl Principles of Stellar Evolution and
Nucleosynthesis} (Chicago: Univ. Chicago Press). 

\par\pp
Coleman, S. 1977, {\sl Phys. Rev.} D {\bf 15}, 2929. 

\par\pp
Coleman, S. 1985, {\sl Aspects of Symmetry} 
(Cambridge, England: Cambridge Univ. Press). 

\par\pp
Coleman, S., and F. De Luccia, 1980, {\sl Phys. Rev.} D {\bf 21}, 3305. 

\par\pp
Copeland, H., J. O. Jensen, and H. E. Jorgensen, 1970, 
{\sl Astron. Astrophys.} {\bf 5}, 12. 

\par\pp
Crone, M. M., and M. Sher, 1990, {\sl Am. J. Phys.} {\bf 59}, 25. 

\par\pp
D'Antona, F., and I. Mazzitelli, 1985, {\sl Astrophys. J.} {\bf 296}, 502. 

\par\pp
Davies, P.C.W. 1994, {\sl The Last Three Minutes} 
(New York: BasicBooks). 

\par\pp
Dicus, D. A., J. R. Letaw, D. C. Teplitz, and V. L. Teplitz, 
1982, {\sl Astrophys. J.} {\bf 252}, 1.  

\par\pp
Diehl, E., G. L. Kane, C. Kolda, and J. D. Wells, 1995, 
{\sl Phys. Rev.} D {\bf 52}, 4223. 

\par\pp
Dorman, B., L. A. Nelson, and W. Y. Chan, 1989, 
{\sl Astrophys. J.} {\bf 342}, 1003.  

\par\pp 
Draine, B. T., and H. M. Lee, 1984, {\sl Astrophys. J.} {\bf 285}, 89. 

\par\pp
Dyson, F. J. 1979, {\sl Rev. Mod. Phys.} {\bf 51}, 447. 

\par\pp
Dyson, F. J. 1988, {\sl Infinite in All Directions}
(New York: Harper and Row). 

\par\pp
Eddington, A. S. 1931, {\sl Nature} {\bf 127}, 447. 

\par\pp
Ellis, G.F.R., and T. Rothman, 1993, {\sl Am. J. Phys.} 
{\bf 61}, 883. 

\par\pp
Ellis, G.F.R., and D. H. Coule, 1994, {\sl Gen. Rel. and Grav.} 
{\bf 26}, 731. 

\par\pp
Elmegreen, B. G., and R. D. Mathieu, 1983, 
{\sl Mon. Not. R. Astron. Soc.} {\bf 203}, 305. 

\par\pp
Faulkner, J., and R. L. Gilliland, 1985, {\sl Astrophys. J.} 
{\bf 299}, 994. 

\par\pp
Feinberg, G., 1981, {\sl Phys. Rev.} D {\bf 23}, 3075. 

\par\pp
Feinberg, G., M. Goldhaber, and G. Steigman, 1978, 
{\sl Phys. Rev.} D {\bf 18}, 1602. 

\par\pp
Frautschi, S. 1982, {\sl Science} {\bf 217}, 593.  

\par\pp
Freese, K. 1986, {\sl Phys. Lett.} {\bf 167 B}, 295.  

\par\pp
Gaier, T., et al. 1992, {\sl Astrophys. J. Lett.} {\bf 398}, L1. 

\par\pp
Goity, J. L., and M. Sher, 1995, {\sl Phys. Lett.} {\bf 346 B}, 69. 

\par\pp 
Golimowski, D. A., T. Nakajima, S. R. Kulkarni, and B. R. Oppenheimer, 
1995, {\sl Astrophys. J. Lett.} {\bf 444}, L101. 

\par\pp
Gott, J. R. III 1993, {\sl Nature} {\bf 363}, 315.  

\par\pp
Gould, A. 1987, {\sl Astrophys. J.} {\bf 321}, 571. 

\par\pp
Gould, A. 1991, {\sl Astrophys. J.} {\bf 388}, 338. 

\par\pp
Grischuk, L. P., and Ya. B. Zel'dovich, 1978, {\sl Sov. Astron.} 
{\bf 22}, 125. 

\par\pp
Grossman, A. S., and H. C. Graboske, 1971, {\sl Astrophys. J.} 
{\bf 164}, 475. 

\par\pp
Guth, A. 1981, {\sl Phys. Rev.} D {\bf 23}, 347. 

\par\pp
Guth, A. H., and S.-Y. Pi, 1982, {\sl Phys. Rev. Lett.} {\bf 49}, 1110. 

\par\pp
Hamada, T., and E. E. Salpeter, 1961, {\sl Astrophys. J.} {\bf 134}, 683.  

\par\pp
Hawking, S. W. 1975, {\sl Comm. Math. Phys.} {\bf 43}, 199. 

\par\pp
Hawking, S. W. 1982, {\sl Phys. Lett.} {\bf 115 B}, 295. 

\par\pp
Hawking, S. W. 1985, {\sl Phys. Lett.} {\bf 150 B}, 339. 

\par\pp 
Hawking, S. W. 1987, {\sl Phys. Lett.} {\bf 195 B}, 337. 

\par\pp
Hawking, S. W., D. N. Page, and C. N. Pope, 1979, 
{\sl Phys. Lett.} {\bf 86 B}, 175. 

\par\pp
Helmholz, H. von 1854, {\sl On the Interaction of Natural Forces}. 

\par\pp
Henry, T. J., J. D. Kirkpatrick, and D. A. Simons, 1994, 
{\sl Astron. J.} {\bf 108}, 1437. 

\par\pp
't Hooft, G. 1976, {\sl Phys. Rev. Lett.} {\bf 37}, 8. 

\par\pp
Hubbell, J. H., H. A. Grimm, and I. Overbo, 1980, 
{\sl J. Phys. Chem. Ref. Data} {\bf 9}, 1023. 

\par\pp
Islam, J. N. 1977, {\sl Quart. J. R. Astron. Soc.} {\bf 18}, 3. 

\par\pp
Islam, J. N. 1979, {\sl Sky and Telescope} {\bf 57}, 13. 

\par\pp
Jungman, G., M. Kamionkowski, and K. Griest, 1996,
{\sl Physics Reports} in press. 

\par\pp
Jura, M. 1986, {\sl Astrophys. J.} {\bf 301}, 624. 

\par\pp
Kane, G. L. 1993, {\sl Modern Elementary Particle Physics}
(Reading MA: Addison--Wesley). 

\par\pp
Kane, G. L., and J. D. Wells, 1996, {\sl Phys. Rev. Lett.} in press. 

\par\pp
Kennicutt, R. C., P. Tamblyn, and C. W. Congdon, 1995, 
{\sl Astrophys. J.} {\bf 435}, 22. 

\par\pp
{\obeylines 
Kippenhahn, R., and A. Weigert, 1990, {\sl Stellar Structure and Evolution}
(Berlin: Springer)}. 

\par\pp
Kolb, E. W., and M. S. Turner, 1990, {\sl The Early Universe} 
(Redwood City CA: Addison--Wesley). 

\par\pp
Krauss, L. M., M. Srednicki, and F. Wilczek, 1986, 
{\sl Phys. Rev.} D {\bf 33}, 2206.  

\par\pp
Krauss, L. M., and M. White, 1992, {\sl Phys. Rev. Lett.} 
{\bf 69}, 869. 

\par\pp
Kumar, S. S. 1963, {\sl Astrophys. J.} {\bf 137}, 1121. 

\par\pp
La, D., and P. J. Steinhardt, 1989, {\sl Phys. Rev. Lett.} 
{\bf 62}, 376. 

\par\pp
Langacker, P. 1981, {\sl Physics Reports} {\bf 72}, 186. 

\par\pp
Larson, R. B. 1973, {\sl Mon. Not. R. Astron. Soc.} {\bf 161}, 133. 

\par\pp
Larson, R. B., and B. M. Tinsley, 1978, {\sl Astrophys. J.} {\bf 219}, 46. 

\par\pp
Laughlin, G., and P. Bodenheimer, 1993, {\sl Astrophys. J.} {\bf 403}, 303.

\par\pp
Lee, H. M., and J. P. Ostriker, 1986, {\sl Astrophys. J.} {\bf 310}, 176. 

\par\pp
${\rm Lema{\hat i}tre}$, G. 1993, {\sl Ann. Soc. Sci. Bruxelles} 
A{\bf 53}, 51. 

\par\pp
Lightman, A. P., and S. L. Shapiro, 1978, {\sl Rev. Mod. Phys.} 
{\bf 50}, 437. 

\par\pp
Linde, A. D. 1982, {\sl Phys. Lett.} {\bf 108 B}, 389. 

\par\pp
Linde, A. D. 1983, {\sl Nucl. Phys.} {\bf B216}, 421. 

\par\pp
Linde, A. D. 1988, {\sl Phys. Lett.} {\bf 211 B}, 29. 

\par\pp
Linde, A. D. 1989, {\sl Phys. Lett.} {\bf 227 B}, 352. 

\par\pp
Linde, A. D. 1990, {\sl Particle Physics and Inflationary Cosmology} 
(New York: Harwood Academic). 

\par\pp
Loh, E., and E. Spillar, 1986, {\sl Astrophys. J. Lett.} {\bf 307}, L1. 

\par\pp
Lyth, D. H. 1984, {\sl Phys. Lett.} {\bf 147 B}, 403. 

\par\pp
Manchester, R. N., and J. H. Taylor, 1977, {\sl Pulsars} 
(San Francisco: W. H. Freeman). 

\par\pp
Marcy, G. W., R. P. Butler, and E. Williams, 1996, 
submitted to {\sl Astrophys. J.} 

\par\pp
Mayor, M., and D. Queloz, 1995, {\sl Nature} {\bf 378}, 355. 

\par\pp
Meyer, S. S., E. S. Cheng, and L. A. Page, 1991, 
{\sl Astrophys. J. Lett.} {\bf 410}, L57. 

\par\pp
Mihalas, D., and J. Binney, 1981, {\sl Galactic Astronomy: Structure 
and Kinematics} (New York: W. H. Freeman). 

\par\pp
Miller, G. E., and J. M. Scalo, 1979, 
{\sl Astrophys. J. Suppl.} {\bf 41}, 513. 

\par\pp
Misner, C. W., K. S. Thorne, and J. A. Wheeler, 1973, 
{\sl Gravitation} (San Francisco: W. H. Freeman). 

\par\pp
Mohapatra, R. N., and R. E. Marshak, 1980, 
{\sl Phys. Rev. Lett.} {\bf 44}, 1316. 

\par\pp 
Ohanian, H. C., and R. Ruffini, 1994, {\sl Gravitation and Spacetime} 
(New York: W. W. Norton). 

\par\pp
Oppenheimer, B. R., S. R. Kulkarni, K. Matthews, and T. Nakajima, 
1995, {\sl Science} {\bf 270}, 1478. 

\par\pp
Page, D. N. 1980, {\sl Phys. Lett.} {\bf 95 B}, 244. 

\par\pp
Page, D. N., and M. R. McKee, 1981a, {\sl Phys. Rev.} D {\bf 24}, 1458. 

\par\pp
Page, D. N., and M. R. McKee, 1981b, {\sl Nature} {\bf 291}, 44. 

\par\pp
Particle Data Group, 1994, {\sl Phys. Rev.} D {\bf 50}, 1173. 

\par\pp
Peccei, R. D., and H. R. Quinn 1977a, {\sl Phys. Rev. Lett.} 
{\bf 38}, 1440. 

\par\pp
Peccei, R. D., and H. R. Quinn 1977b, {\sl Phys. Rev.} D 
{\bf 16}, 1791. 

\par\pp
Peebles, P.J.E. 1993, {\sl Principles of Physical Cosmology} 
(Princeton: Princeton Univ. Press). 

\par\pp
Peebles, P.J.E. 1994, {\sl Astrophys. J.} {\bf 429}, 43. 

\par\pp
Perkins, D. 1984, {\sl Ann. Rev. Nucl. Part. Sci.} {\bf 34}, 1. 

\par\pp
Phillips, A. C. 1994, {\sl The Physics of Stars}
(Chichester: Wiley). 

\par\pp
Poundstone, W. 1985, {\sl The Recursive Universe} 
(New York: Morrow). 

\par\pp
Press, W. H., and D. N. Spergel, 1985, {\sl Astrophys. J.} {\bf 296}, 679.

\par\pp
Press, W. H., and S. A. Teukolsky, 1977, {\sl Astrophys. J.} 
{\bf 213}, 183. 

\par\pp
Primack, J. R., and M. Sher, 1980, {\sl Nature} {\bf 288}, 680. 

\par\pp
Rajaraman, R. 1987, {\sl Solitons and Instantons}
(Amsterdam: North-Holland). 

\par\pp
Rana, N. C. 1991, {\sl Ann. Rev. Astron. Astrophys.} {\bf 29}, 129. 

\par\pp
Rees, M. J. 1969, {\sl Observatory} {\bf 89}, 193. 

\par\pp
Rees, M. J. 1981, {\sl Quart. J. R. Astron. Soc.} {\bf 22}, 109. 

\par\pp
Rees, M. J. 1984, {\sl Ann. Rev. Astron. Astrophys.} {\bf 22}, 471.  

\par\pp
Reimers, D. 1975, in {\sl Problems in Stellar Astrophysics}, eds. 
B. Bascheck, W. H. Kegel, and G. Traving (New York: Springer), p. 229. 

\par\pp
Riess, A. G., W. H. Press, and R. P. Kirshner, 1995, 
{\sl Astrophys. J. Lett.} {\bf 438}, L17. 

\par\pp
Roberts, M. S. 1963, {\sl Ann. Rev. Astron. Astrophys.} {\bf 1}, 149. 

\par\pp
Sackmann, I.-J., A. I. Boothroyd, and K. E. Kraemer, 1993, 
{\sl Astrophys. J.} {\bf 418}, 457. 

\par\pp
Sakharov, A. D. 1967, {\sl JETP Letters} {\bf 5}, 24. 

\par\pp
Salpeter, E. E. 1955, {\sl Astrophys. J.} {\bf 121}, 161. 

\par\pp
Salpeter, E. E. 1982, in {\sl Essays in Nuclear Astrophysics} 
(Cambridge: Cambridge Univ. Press). 

\par\pp 
Salpeter, E. E., and H. M. Van Horn, 1969, 
{\sl Astrophys. J.} {\bf 155}, 183. 

\par\pp
Sato, K., H. Kodama, M. Sasaki, and K. Maeda, 1982, 
{\sl Phys. Lett.} {\bf 108 B}, 103. 

\par\pp
Scalo, J. M. 1986, {\sl Fund. Cos. Phys.} {\bf 11}, 1. 

\par\pp
Schuster, J., et al. 1993, {\sl Astrophys. J. Lett.} {\bf 412}, L47. 

\par\pp
Shapiro, S. L., and S. A. Teukolsky, 1983, {\sl Black Holes, 
White Dwarfs, and Neutron Stars: The Physics of Compact Objects} 
(New York: Wiley). 

\par\pp
Sher, M. 1989, {\sl Physics Reports} {\bf 179}, 273. 

\par\pp
Shu, F. H. 1982, {\sl The Physical Universe} 
(Mill Valley: University Science Books). 

\par\pp
Shu, F. H., F. C. Adams, and S. Lizano, 1987, 
{\sl Ann. Rev. Astron. Astrophys.} {\bf 25}, 23.   

\par\pp
Smoot, G., et al. 1992, {\sl Astrophys. J. Lett.} {\bf 396}, L1. 

\par\pp
Stahler, S. W. 1988, {\sl Astrophys. J.} {\bf 332}, 804. 

\par\pp
Starobinsky, A. A. 1982, {\sl Phys. Lett.} {\bf 117 B}, 175.  

\par\pp
Steinhardt, P. J., and M. S. Turner, 1984, 
{\sl Phys. Rev.} D {\bf 29}, 2162. 

\par\pp
Stevenson, D. J. 1991, {\sl Ann. Rev. Astron. Astrophys.} {\bf 29}, 163.   

\par\pp
Suzuki, M. 1988, {\sl Phys. Rev.} D {\bf 38}, 1544. 

\par\pp
Timmes, F. X. 1996, unpublished. 

\par\pp
Tinney, C. G. 1995, Editor, {\sl The Bottom of the Main Sequence 
and Beyond} (Berlin: Springer). 

\par\pp
Tipler, F. J. 1992, {\sl Phys. Lett.} {\bf 286 B}, 36.  

\par\pp
Tolman, R. C. 1934, {\sl Relativity, Thermodynamics, and 
Cosmology}, (Oxford: Clarendon Press). 

\par\pp
Turner, M. S. 1983, {\sl Nature} {\bf 306}, 161. 

\par\pp
Visser, M. 1995, {\sl Lorentzian Wormholes: From Einstein to Hawking} 
(Woodbury NY: AIP Press). 

\par\pp 
Voloshin, M. B., I. Yu. Kobzarev, and L. B. Okun, 1975, 
{\sl Sov. J. Nucl. Phys.} {\bf 20}, 644. 

\par\pp
Weinberg, M. D. 1989, {\sl Mon. Not. R. Astron. Soc.} {\bf 239}, 549. 

\par\pp
Weinberg, S. 1972, {\sl Gravitation and Cosmology} 
(New York: Wiley). 

\par\pp
Weinberg, S. 1977, {\sl The First Three Minutes} 
(New York: Basic). 

\par\pp
Weinberg, S. 1978, {\sl Phys. Rev. Lett.} {\bf 40}, 223. 

\par\pp
Weinberg, S. 1980, {\sl Phys. Rev.} D {\bf 22}, 1694. 

\par\pp
Weinberg, S. 1989, {\sl Rev. Mod. Phys.} {\bf 61}, 1. 

\par\pp
Wilczek, F. 1978, {\sl Phys. Rev. Lett.} {\bf 40}, 279. 

\par\pp
Wilczek, F., and A. Zee, 1979, {\sl Phys. Lett.} {\bf 88 B}, 311. 

\par\pp
Wood, M. A. 1992, {\sl Astrophys. J.} {\bf 386}, 529. 

\par\pp
Wright, E. L., et al. 1992, {\sl Astrophys. J. Lett.} {\bf 396}, L3. 

\par\pp
Zel'dovich, Ya. B. 1976, {\sl Phys. Lett.} {\bf 59 A}, 254. 

\par\pp
Zinnecker, H. 1984, {\sl Mon. Not. R. Astron. Soc.} {\bf 210}, 43. 

\par\pp
Zuckerman, B., and M. A. Malkan, 1996, {\sl The Origin and Evolution 
of the Universe} (Sudbury, MA: Jones and Bartlett).

\newpage 
\noindent{\bf FIGURE CAPTIONS} 
\nobreak 
\medskip 

\bigskip 
\noindent 
Figure 1. The Hertzsprung-Russell diagram for low mass stars 
for time scales much longer than the current age of the universe. 
The labeled curves show the evolutionary tracks for stars of 
varying masses, from 0.08 $M_\odot$ to $0.25 M_\odot$, 
as well as the brown dwarf track for a substellar object 
with mass $M_\ast = 0.06 M_\odot$.  The inset figure shows the 
main sequence lifetimes as a function of stellar mass. 

\bigskip 
\noindent 
Figure 2. The Final Mass Function (FMF) for stars. Solid curve 
shows the predicted distribution $m(dN/dm)$ for the masses of 
the degenerate stellar objects (brown dwarfs, white dwarfs, 
and neutron stars) remaining at the cosmological epoch when 
conventional star formation has ceased.  The dashed curve 
shows the mass distribution of the initial progenitor 
population (the initial mass function). 

\bigskip 
\noindent 
Figure 3. Numerical simulation of a collision between two 
brown dwarfs.  The two initial objects have masses less than 
that required for hydrogen burning; the final product of the 
collision is a true star and is capable of sustained hydrogen fusion. 
The two stars collide with a relative velocity of 200 km/s 
and an impact parameter of $\sim 1$ stellar radius.
The top series of panels shows the collision from a side view; 
the bottom series of panels shows the top view.  

\bigskip  
\noindent 
Figure 4. Representative Feynman diagrams for proton decay 
(top diagram) and neutron decay (bottom diagram) shown in terms 
of the constituent quarks ($u$, $d$, $\bar d$). These processes 
are the form expected for the simplest Grand Unified Theories. 
The particles $X$ and $Y$ are the intermediate vector bosons 
which mediate the baryon number violating process and are 
expected to have masses comparable to the GUT scale 
$\sim 10^{16}$ GeV. 

\bigskip 
\noindent 
Figure 5. Chemical evolution of a white dwarf star during proton 
decay.  The curves show the mass fractions of the major component 
nuclei in the star as a function of time, which is measured 
here in terms of the stellar mass.  The initial state is a 
1.0 $M_\odot$ white dwarf made of pure $^{12}$C.  This simulation 
includes the effects of spallation and radioactivity (see text). 

\bigskip 
\noindent 
Figure 6. The the complete evolution of the Sun 
(or any $1 M_\odot$ star) in the H-R Diagram.  The track shows the
overall evolution of a star, from birth to final death.  The star
first appears in the H-R diagram on the stellar birthline and then
follows a pre-main sequence track onto the main sequence.  After its
post-main sequence evolution (red giant, horizontal branch, red
supergiant, and planetary nebula phases), the star becomes a white
dwarf and cools along a constant radius track.  The star spends 
many cosmological decades $\eta = 11 - 25$ at a point near the center
of the diagram ($L_\ast = 10^{14}$ W; $T_\ast$ = 63 K), where the star 
is powered by annihilation of WIMPs accreted from the galactic halo. 
When the supply of WIMPs is exhausted, the star cools relatively 
quickly and obtains its luminosity from proton decay
($L_\ast \approx$ 400 W). The star then follows the evolutionary 
track in the lower right part of the diagram (with 
$L_\ast \sim T_\ast^{12/5}$) until mass loss from proton 
decay causes the star to become optically thin.  At this 
point, the object ceases to be a star and stellar evolution 
comes to an end. 

\bigskip 
\noindent  
Figure 7. Representative Feynman diagram for nucleon decay 
for a $\Delta B$ = 2 process, i.e., a decay involving two 
nucleons. The net result of this interaction (shown here 
in terms of the constituent quarks) is the decay of a neutron 
and a proton into two pions, $n + p \to \pi^0 + \pi^+$. 
The $Y$ particle mediates the baryon number violating process.
Similar diagrams for neutron-neutron decay and for 
proton-proton decay can be obtained by changing the 
type of spectator quarks. 

\bigskip 
\noindent  
Figure 8. Representative Feynman diagram for higher order nucleon 
decay processes, shown here in terms of the constituent quarks. 
(a) Double neutron decay for a supersymmetric theory. The net 
reaction converts two neutrons $n$ into two neutral kaons $K^0$. 
The tildes denote the supersymmetric partners of the particles. 
(b) Double nucleon decay involving three intermediate vector 
bosons $Y$.  Other final states are possible (e.g., three pions), 
but the overall decay rate is comparable and implies a decay time 
scale $\eta_P \sim 165$ + $12 \log_{10} [M_Y / 10^{16} {\rm GeV}]$. 
 
\bigskip 
\noindent 
Figure 9. Background radiation fields in the universe.  The vertical
axis represents the ratio of the energy density in radiation to the
total energy density (assuming the universe remains matter dominated). 
The horizontal axis is given in terms of cosmological decades
$\eta$. The various curves represent the radiation fields from the
cosmic microwave background (CMB), light from low mass stars (S),
radiation from WIMP annihilation in white dwarfs (WIMPs), radiation
from proton decay (p decay), and black hole evaporation (black 
holes).

\bigskip 
\noindent 
Figure 10. Potential $V(\Phi)$ of a scalar field which determines the 
vacuum state of the universe. This potential has both a false vacuum 
state (labeled $F$) and a true vacuum state (labeled $T$). 
As illustrated by the dashed curve, the universe can tunnel from the 
false vacuum state into the true vacuum state at some future time. 

\newpage

{\bf Table I: Important Events in the History and Future of the Universe} 
\bigskip

The Big Bang \dotfill $\eta = -\infty$

Planck Epoch \dotfill --50.5 

GUT Epoch \dotfill --44.5 

Electroweak Phase Transition \dotfill --17.5 

Quarks become confined into Hadrons \dotfill --12.5 

Nucleosynthesis \dotfill --6 

\bline 

Matter Domination \dotfill 4 

Recombination \dotfill 5.5

First possible Stellar Generation \dotfill 6

Formation of the Galaxy \dotfill 9 

Formation of the Solar System \dotfill 9.5 

{\bf Today: The Present Epoch} \dotfill 10 

Our Sun dies \dotfill 10.2 

Close Encounter of Milky Way with Andromeda (M31) \dotfill 10.2 

Lower Bound on the Age of closed Universe \dotfill 10.8    

Lifetime of Main Sequence Stars with Lowest Mass \dotfill 13

End of conventional Star Formation \dotfill 14 

\bline 

Planets become detached from Stars \dotfill 15

Star Formation via Brown Dwarf Collisions \dotfill 16 

Lower Bound on Age of flat Universe 
(with future $\Delta \rho / \rho > 1$) \dotfill 18  

Stars evaporate from the Galaxy \dotfill 19 

Planetary Orbits decay via Gravitational Radiation \dotfill 20 

WIMPs in the Galactic Halo annihilate \dotfill 22.5 

Star Formation via Orbital Decay of Brown Dwarf Binaries \dotfill 23 

Stellar Orbits in the Galaxy decay via Gravitational Radiation \dotfill 24 

White Dwarfs deplete WIMPs from the Galactic Halo \dotfill 25 

Black Holes accrete Stars on Galactic Size Scale \dotfill 30  

Black Holes accrete Stars on Cluster Size Scale \dotfill 33 

Protons decay \dotfill 37 

Neutron Stars $\beta$-decay \dotfill 38 

Planets destroyed by Proton Decay \dotfill 38 

White Dwarfs destroyed by Proton Decay \dotfill 39 

\bline 

Axions decay into Photons \dotfill 42 

Hydrogen Molecules experience Pycnonuclear Reactions \dotfill 60 

Stellar-sized Black Holes evaporate \dotfill 65 

Black Holes with $M=10^6 M_\odot$ evaporate \dotfill 83 

Positronium formation in a Flat Universe \dotfill 85 

Galaxy-sized Black Holes evaporate \dotfill 98 

Black Hole with Mass of current Horizon Scale evaporates \dotfill 131 

Positronium decay in a Flat Universe \dotfill 141 

Higher order Proton Decay Processes \dotfill $\sim$100 -- 200 

\bline 

\bye